\documentclass[twocolumn,times]{aastex701}
\usepackage{amsmath}
\usepackage{bm}
\usepackage[normalem]{ulem}
\usepackage{lipsum}
\usepackage{float}

\newcommand{\alphasat}{\alpha_{\rm sat}}

\newcommand{\ds}{\Delta\Sigma}

\newcommand{\fsat}{f_{\rm sat}}
\newcommand{\hmpc}{h^{-1}\,\mathrm{Mpc}}
\newcommand{\hi}{H\,{\sc i}}

\newcommand{\msun}{\mathrm{M_\odot}}
\newcommand{\hmsun}{h^{-1}\,\mathrm{M_\odot}}
\newcommand{\hsqmsun}{h^{-2}\,\mathrm{M_\odot}}
\newcommand{\avgmh}{\langle M_h \rangle}
\newcommand{\mhwl}{M_h^{\rm WL}}
\newcommand{\ngobs}{n_g^{\rm obs}}
\newcommand{\hsmr}{\langle M_h(M_\star) \rangle}
\newcommand{\Ncen}{\langle N_{\rm cen}\rangle}
\newcommand{\Ncenmh}{\langle N_{\rm cen}(M_h)\rangle}
\newcommand{\Nsat}{\langle N_{\rm sat}\rangle}
\newcommand{\Ns}{\textsc{Nsat}}
\newcommand{\Nsatmh}{\langle N_{\rm sat}(M_h)\rangle}
\newcommand{\Nsatmhwl}{\langle N_{\rm sat}(M_h^{\rm WL})\rangle}

\newcommand{\sigmalogm}{\sigma_{\log M_\star}}
\newcommand{\vmax}{V_{\rm max}}
\newcommand{\wpgg}{w_p^{gg}}
\newcommand{\wphg}{w_p^{hg}}
\newcommand{\xiggrs}{\xi^{rs}_{gg}}
\newcommand{\xigg}{\xi_{gg}}
\newcommand{\xihg}{\xi_{hg}}

\newcommand{\diff}{\mathrm{d}}

\defcitealias{ZM2015}{ZM15}

\begin{document}

\title{Extending the Stellar-to-Halo Mass Relation to Dwarf Galaxies with DESI DR1}

\correspondingauthor{Ying Zu}
\email{yingzu@sjtu.edu.cn}

\author[0000-0002-4585-3985]{Zhiwei Shao}
\affiliation{Department of Astronomy, School of Physics and Astronomy, Shanghai Jiao Tong University, Shanghai 200240, China}
\affiliation{State Key Laboratory of Dark Matter Physics, \& Tsung-Dao Lee Institute, Shanghai Jiao Tong University, Shanghai 200240, China}
\email{zwshao@sjtu.edu.cn}

\author[0000-0001-6966-6925]{Ying Zu}
\affiliation{Department of Astronomy, School of Physics and Astronomy, Shanghai Jiao Tong University, Shanghai 200240, China}
\affiliation{State Key Laboratory of Dark Matter Physics, \& Tsung-Dao Lee Institute, Shanghai Jiao Tong University, Shanghai 200240, China}
\email{yingzu@sjtu.edu.cn}

\author[0000-0003-1420-527X]{Andr{\'e}s N. Salcedo}
\affiliation{Department of Astronomy/Steward Observatory, University of Arizona, 933 North Cherry Avenue, Tucson, AZ 85721-0065, USA}
\affiliation{Department of Physics, University of Arizona, 1118 East Fourth Street, Tucson, AZ 85721, USA}
\email{ansalcedo@arizona.edu}

\author[0000-0002-5551-483X]{Yuanye Lin}
\affiliation{Department of Astronomy, School of Physics and Astronomy, Shanghai Jiao Tong University, Shanghai 200240, China}
\affiliation{State Key Laboratory of Dark Matter Physics, \& Tsung-Dao Lee Institute, Shanghai Jiao Tong University, Shanghai 200240, China}
\email{ythylyy@sjtu.edu.cn}

\author[0000-0002-5551-483X]{Zhao Chen}
\affiliation{Department of Astronomy, School of Physics and Astronomy, Shanghai Jiao Tong University, Shanghai 200240, China}
\email{chyiru@sjtu.edu.cn}

\author{Xiaoju Xu}
\affiliation{Department of Astronomy, School of Physics and Astronomy, Shanghai Jiao Tong University, Shanghai 200240, China}
\affiliation{Shanghai Key Lab for Astrophysics, Shanghai Normal University, Shanghai 200234, China}
\email{xiaojuxu@shnu.edu.cn}

\author[0009-0007-7882-7832]{Junyu Hua}
\affiliation{Department of Astronomy, School of Physics and Astronomy, Shanghai Jiao Tong University, Shanghai 200240, China}
\email{hjy157hjy@sjtu.edu.cn}

\author{Zhongxu Zhai}
\affiliation{Department of Astronomy, School of Physics and Astronomy, Shanghai Jiao Tong University, Shanghai 200240, China}
\email{zhongxuzhai@sjtu.edu.cn}

\author{Jessica Nicole Aguilar}
\affiliation{Lawrence Berkeley National Laboratory, 1 Cyclotron Road, Berkeley, CA 94720, USA}
\email{jaguilar@lbl.gov}

\author[0000-0001-6098-7247]{Steven Ahlen}
\affiliation{Department of Physics, Boston University, 590 Commonwealth Avenue, Boston, MA 02215 USA}
\email{ahlen@bu.edu}

\author[0000-0003-0467-5438]{Florian Beutler}
\affiliation{Institute for Astronomy, University of Edinburgh, Royal Observatory, Blackford Hill, Edinburgh EH9 3HJ, UK}
\email{florian.beutler@ed.ac.uk}

\author[0000-0001-9712-0006]{Davide Bianchi}
\affiliation{Dipartimento di Fisica ``Aldo Pontremoli'', Universit\`a degli Studi di Milano, Via Celoria 16, I-20133 Milano, Italy}
\affiliation{INAF-Osservatorio Astronomico di Brera, Via Brera 28, 20122 Milano, Italy}
\email{davide.bianchi1@unimi.it}

\author{David Brooks}
\affiliation{Department of Physics \& Astronomy, University College London, Gower Street, London, WC1E 6BT, UK}
\email{david.brooks@ucl.ac.uk}

\author[0000-0003-3044-5150]{Aurelio Carnero Rosell}
\affiliation{Departamento de Astrof\'{\i}sica, Universidad de La Laguna (ULL), E-38206, La Laguna, Tenerife, Spain}
\affiliation{Instituto de Astrof\'{\i}sica de Canarias, C/ V\'{\i}a L\'{a}ctea, s/n, E-38205 La Laguna, Tenerife, Spain}
\email{acarnero@iac.es}

\author[0000-0001-7316-4573]{Francisco Javier Castander}
\affiliation{Institut d'Estudis Espacials de Catalunya (IEEC), c/ Esteve Terradas 1, Edifici RDIT, Campus PMT-UPC, 08860 Castelldefels, Spain}
\affiliation{Institute of Space Sciences, ICE-CSIC, Campus UAB, Carrer de Can Magrans s/n, 08913 Bellaterra, Barcelona, Spain}
\email{fjc@ice.csic.es}

\author{Todd Claybaugh}
\affiliation{Lawrence Berkeley National Laboratory, 1 Cyclotron Road, Berkeley, CA 94720, USA}
\email{tmclaybaugh@lbl.gov}

\author[0000-0002-1769-1640]{Axel de la Macorra}
\affiliation{Instituto de F\'{\i}sica, Universidad Nacional Aut\'{o}noma de M\'{e}xico,  Circuito de la Investigaci\'{o}n Cient\'{\i}fica, Ciudad Universitaria, Cd. de M\'{e}xico  C.~P.~04510,  M\'{e}xico}
\email{macorra@fisica.unam.mx}

\author[0000-0002-5665-7912]{Biprateep Dey}
\affiliation{Department of Astronomy \& Astrophysics, University of Toronto, Toronto, ON M5S 3H4, Canada}
\affiliation{Department of Physics \& Astronomy and Pittsburgh Particle Physics, Astrophysics, and Cosmology Center (PITT PACC), University of Pittsburgh, 3941 O'Hara Street, Pittsburgh, PA 15260, USA}
\email{b.dey@utoronto.ca}

\author[0000-0002-3369-3718]{Zhejie Ding}
\affiliation{University of Chinese Academy of Sciences, Nanjing 211135, People's Republic of China.}
\email{dingzhejie@ucas.ac.cn}

\author[0000-0002-2890-3725]{Jaime E. Forero-Romero}
\affiliation{Departamento de F\'isica, Universidad de los Andes, Cra. 1 No. 18A-10, Edificio Ip, CP 111711, Bogot\'a, Colombia}
\affiliation{Observatorio Astron\'omico, Universidad de los Andes, Cra. 1 No. 18A-10, Edificio H, CP 111711 Bogot\'a, Colombia}
\email{je.forero@uniandes.edu.co}

\author[0000-0001-9632-0815]{Enrique Gazta\~naga}
\affiliation{Institut d'Estudis Espacials de Catalunya (IEEC), c/ Esteve Terradas 1, Edifici RDIT, Campus PMT-UPC, 08860 Castelldefels, Spain}
\affiliation{Institute of Cosmology and Gravitation, University of Portsmouth, Dennis Sciama Building, Portsmouth, PO1 3FX, UK}
\affiliation{Institute of Space Sciences, ICE-CSIC, Campus UAB, Carrer de Can Magrans s/n, 08913 Bellaterra, Barcelona, Spain}
\email{gaztanaga@gmail.com}

\author[0000-0003-3142-233X]{Satya Gontcho A Gontcho}
\affiliation{University of Virginia, Department of Astronomy, Charlottesville, VA 22904, USA}
\email{satya@virginia.edu}

\author{Gaston Gutierrez}
\affiliation{Fermi National Accelerator Laboratory, PO Box 500, Batavia, IL 60510, USA}
\email{gaston@fnal.gov}

\author[0000-0003-1197-0902]{ChangHoon Hahn}
\affiliation{Department of Astronomy, University of Texas at Austin, 2515 Speedway, TX 78712, USA}
\email{changhoon.hahn@utexas.edu}

\author[0000-0002-0000-2394]{Stephanie Juneau}
\affiliation{NSF NOIRLab, 950 N. Cherry Ave., Tucson, AZ 85719, USA}
\email{stephanie.juneau@noirlab.edu}

\author{Robert Kehoe}
\affiliation{Department of Physics, Southern Methodist University, 3215 Daniel Avenue, Dallas, TX 75275, USA}
\email{kehoe@physics.smu.edu}

\author[0000-0001-6356-7424]{Anthony Kremin}
\affiliation{Lawrence Berkeley National Laboratory, 1 Cyclotron Road, Berkeley, CA 94720, USA}
\email{akremin@lbl.gov}

\author[0000-0002-1134-9035]{Ofer Lahav}
\affiliation{Department of Physics \& Astronomy, University College London, Gower Street, London, WC1E 6BT, UK}
\email{o.lahav@ucl.ac.uk}

\author{Andrew Lambert}
\affiliation{Lawrence Berkeley National Laboratory, 1 Cyclotron Road, Berkeley, CA 94720, USA}
\email{arlambert@lbl.gov}

\author[0000-0003-1838-8528]{Martin Landriau}
\affiliation{Lawrence Berkeley National Laboratory, 1 Cyclotron Road, Berkeley, CA 94720, USA}
\email{mlandriau@lbl.gov}

\author[0000-0001-7178-8868]{Laurent Le Guillou}
\affiliation{Sorbonne Universit\'{e}, CNRS/IN2P3, Laboratoire de Physique Nucl\'{e}aire et de Hautes Energies (LPNHE), FR-75005 Paris, France}
\email{llg@lpnhe.in2p3.fr}

\author[0000-0003-4962-8934]{Marc Manera}
\affiliation{Departament de F\'{i}sica, Serra H\'{u}nter, Universitat Aut\`{o}noma de Barcelona, 08193 Bellaterra (Barcelona), Spain}
\affiliation{Institut de F\'{i}sica d’Altes Energies (IFAE), The Barcelona Institute of Science and Technology, Edifici Cn, Campus UAB, 08193, Bellaterra (Barcelona), Spain}
\email{mmanera@ifae.es}

\author[0000-0002-4279-4182]{Paul Martini}
\affiliation{Center for Cosmology and AstroParticle Physics, The Ohio State University, 191 West Woodruff Avenue, Columbus, OH 43210, USA}
\affiliation{Department of Astronomy, The Ohio State University, 4055 McPherson Laboratory, 140 W 18th Avenue, Columbus, OH 43210, USA}
\affiliation{The Ohio State University, Columbus, 43210 OH, USA}
\email{martini.10@osu.edu}

\author[0000-0002-1125-7384]{Aaron Meisner}
\affiliation{NSF NOIRLab, 950 N. Cherry Ave., Tucson, AZ 85719, USA}
\email{aaron.meisner@noirlab.edu}

\author{Ramon Miquel}
\affiliation{Instituci\'{o} Catalana de Recerca i Estudis Avan\c{c}ats, Passeig de Llu\'{\i}s Companys, 23, 08010 Barcelona, Spain}
\affiliation{Institut de F\'{i}sica d’Altes Energies (IFAE), The Barcelona Institute of Science and Technology, Edifici Cn, Campus UAB, 08193, Bellaterra (Barcelona), Spain}
\email{rmiquel@ifae.es}

\author[0000-0002-2733-4559]{John Moustakas}
\affiliation{Department of Physics and Astronomy, Siena University, 515 Loudon Road, Loudonville, NY 12211, USA}
\email{jmoustakas@siena.edu}

\author[0000-0001-9070-3102]{Seshadri Nadathur}
\affiliation{Institute of Cosmology and Gravitation, University of Portsmouth, Dennis Sciama Building, Portsmouth, PO1 3FX, UK}
\email{seshadri.nadathur@port.ac.uk}

\author[0000-0002-4637-2868]{Enrique Paillas}
\affiliation{Instituto de Estudios Astrof\'isicos, Facultad de Ingenier\'ia y Ciencias, Universidad Diego Portales, Av. Ej\'ercito Libertador 441, Santiago, Chile}
\affiliation{Steward Observatory, University of Arizona, 933 N. Cherry Avenue, Tucson, AZ 85721, USA}
\email{enrique.paillas@udp.cl}

\author[0000-0002-0644-5727]{Will Percival}
\affiliation{Department of Physics and Astronomy, University of Waterloo, 200 University Ave W, Waterloo, ON N2L 3G1, Canada}
\affiliation{Perimeter Institute for Theoretical Physics, 31 Caroline St. North, Waterloo, ON N2L 2Y5, Canada}
\affiliation{Waterloo Centre for Astrophysics, University of Waterloo, 200 University Ave W, Waterloo, ON N2L 3G1, Canada}
\email{will.percival@uwaterloo.ca}

\author[0000-0001-7145-8674]{Francisco Prada}
\affiliation{Instituto de Astrof\'{i}sica de Andaluc\'{i}a (CSIC), Glorieta de la Astronom\'{i}a, s/n, E-18008 Granada, Spain}
\email{fprada@iaa.es}

\author[0000-0001-6979-0125]{Ignasi P\'erez-R\`afols}
\affiliation{Departament de F\'isica, EEBE, Universitat Polit\`ecnica de Catalunya, c/Eduard Maristany 10, 08930 Barcelona, Spain}
\email{ignasi.perez.rafols@upc.edu}

\author[0000-0002-3500-6635]{Corentin Ravoux}
\affiliation{Universit\'{e} Clermont-Auvergne, CNRS, LPCA, 63000 Clermont-Ferrand, France}
\email{corentin.ravoux@clermont.in2p3.fr}

\author{Graziano Rossi}
\affiliation{Department of Physics and Astronomy, Sejong University, 209 Neungdong-ro, Gwangjin-gu, Seoul 05006, Republic of Korea}
\email{graziano@sejong.ac.kr}

\author[0000-0002-0394-0896]{Rossana Ruggeri}
\affiliation{Queensland University of Technology,  School of Chemistry \& Physics, George St, Brisbane 4001, Australia}
\email{rossana.ruggeri@qut.edu.au}

\author[0000-0003-4755-6404]{Manasvee Saraf}
\affiliation{Department of Physics \& Astronomy, University College London, Gower Street, London, WC1E 6BT, UK}
\email{manasvee.saraf.16@ucl.ac.uk}

\author[0000-0002-1609-5687]{Lado Samushia}
\affiliation{Abastumani Astrophysical Observatory, Tbilisi, GE-0179, Georgia}
\affiliation{Department of Physics, Kansas State University, 116 Cardwell Hall, Manhattan, KS 66506, USA}
\email{lado@phys.ksu.edu}

\author[0000-0002-9646-8198]{Eusebio Sanchez}
\affiliation{CIEMAT, Avenida Complutense 40, E-28040 Madrid, Spain}
\email{eusebio.sanchez@ciemat.es}

\author[0000-0002-0408-5633]{Christoph Saulder}
\affiliation{Max Planck Institute for Extraterrestrial Physics, Gie\ss enbachstra\ss e 1, 85748 Garching, Germany}
\email{csaulder@mpe.mpg.de}

\author{David Schlegel}
\affiliation{Lawrence Berkeley National Laboratory, 1 Cyclotron Road, Berkeley, CA 94720, USA}
\email{djschlegel@lbl.gov}

\author[0000-0002-3461-0320]{Joseph Harry Silber}
\affiliation{Lawrence Berkeley National Laboratory, 1 Cyclotron Road, Berkeley, CA 94720, USA}
\email{jhsilber@lbl.gov}

\author[0000-0002-2949-2155]{Ma{\l}gorzata Siudek}
\affiliation{Institute of Space Sciences, ICE-CSIC, Campus UAB, Carrer de Can Magrans s/n, 08913 Bellaterra, Barcelona, Spain}
\affiliation{Instituto de Astrof\'{\i}sica de Canarias, C/ V\'{\i}a L\'{a}ctea, s/n, E-38205 La Laguna, Tenerife, Spain}
\email{msiudek@iac.es}

\author[0000-0003-1704-0781]{Gregory Tarl\'e}
\affiliation{University of Michigan, 500 S. State Street, Ann Arbor, MI 48109, USA}
\email{gtarle@umich.edu}

\author{Benjamin Alan Weaver}
\affiliation{NSF NOIRLab, 950 N. Cherry Ave., Tucson, AZ 85719, USA}
\email{benjamin.weaver@noirlab.edu}

\author[0000-0002-6684-3997]{Hu Zou}
\affiliation{National Astronomical Observatories, Chinese Academy of Sciences, A20 Datun Road, Chaoyang District, Beijing, 100101, P.~R.~China}
\email{zouhu@nao.cas.cn}




\begin{abstract}
Constraining the dark matter halos of the smallest galaxies offers fundamental
insights into the nature of dark matter and stellar feedback. Using the Dark
Energy Spectroscopic Instrument (DESI) Data Release 1, we infer the
stellar-to-halo mass relation (SHMR) down to the dwarf scale
($M_\star<10^9\,\msun$), without extrapolation from the higher mass range.
Leveraging the unprecedented depth of the DESI Bright Galaxy Survey at
$0.01{<}z{<}0.2$, we construct 12 samples spanning nearly four orders of
magnitude in stellar mass, and measure their projected clustering $w_p$,
galaxy--galaxy lensing $\Delta\Sigma$, as well as a novel observable: satellite
occupation number \textsc{Nsat}.  The addition of \textsc{Nsat} enables robust
subtraction of satellite contributions to both $w_p$ and $\Delta\Sigma$ across
the 12 individual halo occupation distribution analyses, yielding an average
halo-to-stellar mass relation (HSMR) of $\log \langle M_h(M_\star) \rangle =
12.06+0.58\log \left(M_\star/10^{11}\right)+
\left(M_\star/10^{11}\right)^{0.73}$.  Combining this HSMR with an observed
stellar mass function, we constrain the SHMR across five orders of magnitude in
halo mass, with the power-law slope steepening from $0.32{\pm}0.06$ above the
Milky Way mass to $2.08{\pm}0.21$ in the dwarf regime. Interestingly, the
scatter about the SHMR grows from $0.17{\pm}0.02$ dex at Milky Way-like scales
to $0.68_{-0.33}^{+0.21}$ dex for systems comparable to the Large Magellanic
Cloud, suggesting that smaller galaxies follow increasingly diverse evolutionary
paths. Our work highlights the power of DESI in probing the galaxy--halo
connection within the dwarf regime, offering an exciting avenue to bridge the
gap between large-scale and near-field cosmologies in the future.
\end{abstract}



\section{Introduction}
\label{sec:intro}


The stellar-to-halo mass relation~(SHMR) of dwarf
galaxies~($M_\star<10^{9}\,\msun$) provides a critical test of both
galaxy formation physics~\citep{Bullock2017} and the nature of dark
matter~\citep{Buckley2018,Nadler2021}. However, constraints in this
low-mass regime have remained elusive, as near-field cosmology studies are
limited by small sample sizes~\citep{Monzon2024}. While large-scale
structure~(LSS) surveys have yielded consistent constraints on the SHMR for
galaxies above the Milky Way~(MW)
mass~\citep[e.g.,][]{Yang2012,Leauthaud2012,Coupon2015,ZM2015,Rodriguez2017,Moster2018,Behroozi2019,Dvornik2020,Zacharegkas2025},
they generally lack the depth to robustly probe the galaxy-halo
connection~\citep{Wechsler2018} for Large Magellanic Cloud~(LMC)-like
systems. In this paper, we bridge this gap using the unprecedented depth
and volume of the Dark Energy Spectroscopic
Instrument~\citep[DESI;][]{DESI2016a,DESI2016b,DESI2022Overview} survey.
Using a novel combination of LSS observables, we derive a robust constraint
on the SHMR extended into the dwarf galaxy regime, connecting near-field
observations with the statistical power of cosmological surveys.

The SHMR is formally defined by the approximately log-normal distribution
of stellar mass for {\it central} galaxies in halos of mass $M_h$,
$P(M_\star|M_h)$.  It is therefore fully characterized by a mean relation,
$f_{\rm SHMR}{\equiv}\langle \log M_\star(M_h)\rangle$, and a logarithmic
scatter $\sigma_{\log M_\star}$.  However, the inverted relation,
$P(M_h|M_\star)$, which we refer to as the halo-to-stellar mass
relation~(HSMR), is more accessible through observations, because galaxy
samples are usually selected by galaxy stellar mass or luminosity rather
than $M_h$.  Since the two relations are connected by Bayes' theorem, the
key to constraining SHMR is to acquire robust measurements of the average
halo mass as a function of the stellar mass of the centrals, i.e., the mean
HSMR $\langle M_h(M_\star) \rangle$.

However, surveys of dwarf galaxies in the local Universe~($z<0.01$)
predominantly target satellites of nearby massive
galaxies~\citep{Geha2017,Drlica-Wagner2020,Carlsten2022}. To connect theory with
these observations, near-field studies often employ the subhalo abundance
matching (SHAM) technique.\footnote{A more comprehensive matching can be
obtained with more advanced empirical models such as
UniverseMachine~\citep{Behroozi2019}, EMERGE~\citep{Moster2018}, and
GRUMPY~\citep{Kravtsov2022}. We refer to them collectively as empirical galaxy
formation models based on SHAM.} This method assigns stellar masses to
halos~(including subhalos) by matching them to the observed dwarf satellite
stellar mass functions~(SMFs) in systems like the
MW~\citep{Koposov2009,Jethwa2018,Nadler2020,Manwadkar2022}, the Local
Group~\citep{Brook2014,Garrison-Kimmel2017}, or the broader Local
Volume~\citep{Danieli2023,Wang2024SAGAUM,Kado-Fong2025}. Despite the substantial
success of SHAM-based methods, halo masses are not directly measured but are
instead inferred indirectly from galaxy abundance.

Alternatively, the halo mass of individual dwarf spiral galaxies can be
measured dynamically in the local Universe using their rotation
curves~\citep[e.g.,][]{Oh2015,Read2017,Posti2019,ManceraPina2025}.
Although this method can reveal exquisite details of the inner density
profile, determining the {\it total} halo mass requires extrapolating the
profile from tens to hundreds of kiloparsecs. The systematic uncertainty
from this extrapolation is irreducible by larger sample sizes.  A more
fundamental limitation is that the method is inapplicable to
dispersion-supported dwarf galaxies, which constitute a significant
fraction of the dwarf population~\citep{Wheeler2017} and do not permit
rotation curve analysis for mass estimation.

Beyond the local Universe, weak gravitational lensing provides a direct
probe of dark matter surrounding galaxies~\citep{Mandelbaum2018}, and the
halo occupation
distribution~\citep[HOD;][]{Jing1998,Scoccimarro2001,Peacock2000,Berlind2002,Zheng2005}
model offers a powerful statistical framework for inferring the SHMR from
such measurements~(a.k.a., galaxy--galaxy lensing) with cosmological
surveys. An HOD is characterized by the mean occupation functions $\Ncenmh$
and $\Nsatmh$, which prescribe the expected numbers of central and
satellite galaxies, respectively, at fixed host halo mass $M_h$. In
particular, the HOD of centrals determines their average halo mass
$\avgmh$, which is measurable through galaxy--galaxy lensing if the
satellite HOD can be correctly inferred. Applying this scheme to the main
sample of the Sloan Digital Sky Survey~\citep[SDSS;][]{York2000},
\citet{Mandelbaum2006} measured the mean HSMR through a simple HOD modeling
of the galaxy--galaxy lensing signals. Adopting the more comprehensive HOD
framework of~\citet{Leauthaud2011}, \citet{ZM2015}~(hereafter
\citetalias{ZM2015}) combined galaxy clustering and galaxy--galaxy lensing
from the final SDSS data release to place stringent constraints on the SHMR
down to $M_\star\sim10^9\,\msun$.  They found the best-fitting mean
SHMR has a much steeper slope for MW-like galaxies than for massive
galaxies, with a logarithmic scatter of $\sim0.2\,{\rm dex}$. However, it
has remained unclear whether dwarf galaxies land on this relation, due to
the relatively shallow coverage of the SDSS main sample~(with r-band
magnitude $m_r<17.77$).

The advent of DESI now enables a critical constraint on the dwarf SHMR using
weak lensing under the HOD framework. Aiming at determining the nature of dark
energy~\citep{DESIDR1cosmology,DESIDR2cosmology}, DESI is designed
to effectively measure the redshifts of 63 million galaxies and quasars with
5000 fibers on the 4-meter Mayall telescope at Kitt Peak National
Observatory~\citep{Guy2023DESI,Schlafly2023DESI,Miller2024DESI,Poppett2024DESI}.
As part of the program, the DESI Bright Galaxy Survey~\citep[BGS;][]{Hahn2023}
is a flux-limited survey of galaxies with $m_r<19.5$, which also provides a
large, homogeneous, and complete spectroscopic sample at the dwarf mass scale.
In light of this new dataset, we develop a novel HOD framework to extend the
SHMR constraint to the dwarf regime, while significantly improving upon previous
constraints for MW-like and massive galaxies. In particular, we employ the new
method developed recently by \citet{Shao2025} to directly measure the satellite
HOD $\Nsatmh$~(hereafter shortened to \Ns) using the halo-based DESI group
catalog~\citep{Yang2021}. The addition of \Ns\ significantly reduces the
systematic uncertainties in measuring central halo masses, thereby sharpening
our SHMR constraints into the dwarf galaxy regime.

The methodology of this paper is explicitly divided into two parts to
elucidate the procedures that yield the SHMR constraint. In the first
part~(\S\ref{sec:method}--\S\ref{sec:results}), we infer the mean HSMR by
performing a comprehensive HOD analysis of the DESI LSS statistics after
describing the data sets~(\S\ref{sec:data}) and
observables~(\S\ref{sec:statistics}). The second part~(\S\ref{sec:shmr})
leverages an independent SMF measurement from~\citet{Xu2025} to convert the
HSMR measurement into our final SHMR constraint by virtue of Bayes'
theorem~(\S\ref{sec:shmr}). We conclude by summarizing our results
in~\S\ref{sec:conclusion}.

Throughout this paper, we adopt a flat $\Lambda$-dominated cold dark
matter~(CDM) cosmology with matter density~(in units of the critical density)
$\Omega_m{=}0.3$, matter clustering amplitude $\sigma_8=0.8$, and Hubble
parameter~(in units of $100\,\mathrm{km/s/Mpc}$) $h{=}0.7$. This vanilla CDM
model is a conservative choice amid the various tensions in cosmology
today~\citep{DiValentino2026}. Unless otherwise specified, we adopt
$M_h{\equiv}M_{\rm 200m}{=}200\bar\rho_m(4\pi/3)r_{\rm 200m}^3$ as our
definition of halo mass, where $\bar\rho_m$ is the mean matter density of the
universe and $r_h{\equiv}r_{\rm 200m}$ is the radius within which the mean
density is 200 times $\bar\rho_m$.  The units of stellar mass and halo mass are
$\hsqmsun$ and $\hmsun$, respectively, and $\log$
always denotes the base-10 logarithm.

\section{Data}
\label{sec:data}

\subsection{Galaxies}
\subsubsection{DESI BGS DR1}
\label{subsubsec:bgs_gals}


We employ galaxies from the Bright Galaxy Survey \citep[BGS;][]{Hahn2023}
of the DESI Data Release~1 \citep[DR1;][]{DESIDR1}. The BGS sample consists
of the main BGS, a flux-limited sample selected with $m_r{<}19.5$ (i.e.,
BGS Bright), and the BGS Faint, a fainter ($19.5{<}m_r{<}20.175$) sample
with additional selections in surface brightness and color.  We only use
the BGS Bright sample, because its simple target selection allows for the
robust construction of volume-limited subsamples.  As noted in the
Introduction, the main BGS sample is analogous to the SDSS main galaxy
sample, but extends roughly 0.8 dex ($\Delta m_r{=}1.7\,{\rm mag}$) lower
in $M_\star$, thereby probing further into the dwarf mass range.

In particular, we adopt the \texttt{BGS\_BRIGHT} large-scale structure
(LSS) catalog constructed by
\citet{Ross2025}.\footnote{\url{https://data.desi.lbl.gov/public/dr1/survey/catalogs/dr1/LSS/iron/LSScats/v1.5pip/}}
This catalog includes additional quality cuts for selecting reliable redshifts
\citep[\texttt{ZWARN}=0, $\Delta\chi^2{>40}$;][]{DESIDR1Sample}
and provides the necessary weights for correcting various observational
systematics, including spectroscopic incompleteness ($w_{\rm comp}$), imaging
systematics ($w_{\rm sys}$), and redshift failures ($w_{\rm zfail}$). The total
LSS weight of each galaxy is computed as $w_{\rm tot} = w_{\rm comp} \times
w_{\rm sys} \times w_{\rm zfail}$. We further adopt the corresponding LSS random
catalogs that match the survey geometry and redshift distribution of the
\texttt{BGS\_BRIGHT} sample. The total area covered by the
\texttt{BGS\_BRIGHT} DR1 sample is $\sim7400$ deg$^2$. Throughout this pilot
study using DR1, we focus exclusively on the redshift range $0.01{<}z{<}0.2$,
within which the redshift evolution of SHMR is negligible \citep{Moster2018}.
The star-forming clumps within nearby massive galaxies could be misidentified as
individual dwarf galaxies in the \texttt{BGS\_BRIGHT} LSS catalog. We remove
these ``shredded'' objects from our dwarf galaxy samples $(\log M_\star < 9.6)$
by applying the $\mathtt{FRACFLUX}{\leq}0.2$ cut suggested by
\citep{Manwadkar2026}, where $\mathtt{FRACFLUX}$ is the profile-weighted
fraction of the flux from other sources. The fraction of removed objects
decreases from 12\% at $\log M_\star{=}7.8$ to 2\% at $\log M_\star{=}9.6$.

\subsubsection{Stellar Mass Estimate}
\label{subsubsec:stellar_mass}

We make use of the stellar mass measurements from \citet{Zou2024}. Briefly,
they used the stellar population synthesis code \texttt{CIGALE}
\citep{Boquien2019cigale,Yang2020cigale,Yang2022cigale} to fit to the
multi-band photometry from the DESI Legacy Imaging Surveys \citep{Dey2019}
as well as ten artificial medium-band photometry generated from the
observed BGS spectra. To facilitate the comparison with previous studies
based on SDSS, we apply an overall calibration to the \citet{Zou2024}
stellar mass estimates $\log M_\star {=} 0.99\log M_\star^{\rm
zou}{+}0.01$, so that the rescaled masses $M_\star$ are consistent with the
widely used MPA/JHU stellar mass estimates of SDSS galaxies
\citep{Kauffmann2003,Salim2007}. The calibration is derived from the 23127
BGS galaxies that were also present in the SDSS MPA/JHU catalog. We adopt
this rescaled stellar mass $M_\star$ for the DESI BGS galaxies throughout
the paper.

\subsubsection{Selection of galaxy samples binned by stellar mass}
\label{subsubsec:complete_sample}

Before selecting galaxies into stellar mass bins, we first define the stellar
mass limit at each redshift, $M_\star^{\rm lim}(z)$, above which the BGS sample
is volume-complete at that redshift. In other words, there exists a maximum
redshift $z_{\rm max}(M_\star)$ below which the sample of galaxies with stellar
mass $M_\star$ is complete: $M_\star^{\rm lim}(z)$ is the inverse function of
$z_{\rm max}(M_\star)$.  In essence, $z_{\rm max}$ corresponds to where the
galaxy with the highest mass-to-light ratio
($\Upsilon_\star{\equiv}M_\star/L_r$) has the same apparent magnitude as the BGS
flux limit, $m_r{=}19.5$.  To locate $z_{\rm max}(M_\star)$, we first measure
the complete mass-to-light ratio distribution $P(\Upsilon_\star|M_\star)$ within
narrow stellar mass bins ($\Delta \log M_\star=0.1$), using the standard
$1/\vmax$ weights to correct for Malmquist bias (\citealt{Schmidt1968}; see
Appendix~\ref{appendix:vmax} for more details).  With the measured
$P(\Upsilon_\star|M_\star)$, we compute the 98th percentile of the distribution
as a function of stellar mass, $\Upsilon^{98}_\star(M_\star)$, and use it to
derive the maximum redshift $z_{\rm max}(M_\star)$. Finally, the corresponding
98\% completeness stellar mass limit can be well described by an analytic
fitting formula
\begin{equation}
    \log M_\star^{\rm lim}(z) = 7.66\times(z-0.01)^{0.28}+5.60.
    \label{eq:mlim}
\end{equation}

Above the completeness limit, we construct 12 {\it volume-limited} samples
spanning almost four orders of magnitude in stellar mass.\footnote{We
note that for the lowest stellar mass bin at $z\leq0.021$, the peculiar
velocity of galaxies can influence distance measurements by up to 20\%. This
will also influence the stellar mass estimates, but the effect is sub-dominant
compared to the uncertainties in the stellar mass estimates themselves ($\sim
0.3\,{\rm dex}$) and we treat it as is.} We divide the 12 samples into three
mass scales, each consisting of four stellar mass bins:
\begin{itemize}
     \item Dwarf galaxies: $7.8{\leq}\log M_\star{<}9.6$;
     \item MW-like galaxies: $9.6{\leq}\log M_\star{<}10.6$;
     \item Massive galaxies: $10.6{\leq}\log M_\star{<}11.4$.
\end{itemize}
These {\it volume-limited} samples will be directly used to measure the
auto-correlation functions for their respective stellar mass bins.

In contrast, when measuring the galaxy--galaxy lensing and cluster--galaxy
cross-correlations, we make use of all the BGS galaxies within each stellar
mass bin, regardless of whether they are above or below $M_\star^{\rm
lim}(z)$.  We hereafter refer to those samples as the {\it total} galaxy
samples, which follow the same stellar mass binning as the volume-limited
samples.  Although these total samples are biased against systems with high
$\Upsilon_\star$, we are able to correct for this bias by applying
additional $1/\vmax$ weights to individual galaxies, so that the effective
mass-to-light ratio distribution of the total sample matches that of the
corresponding volume-limited sample (see \S\ref{subsec:ggl} and
Appendix~\ref{appendix:vmax} for more details).  This $1/\vmax$-method
increases the number of dwarf lenses (664175) by approximately 900\%
compared to the volume-limited samples (75338), leading to a significantly
improved signal-to-noise ratio (S/N) in the lensing and cross-correlation
measurements.

\begin{figure}[!t]
     \centering
     \includegraphics[width=.99\linewidth]{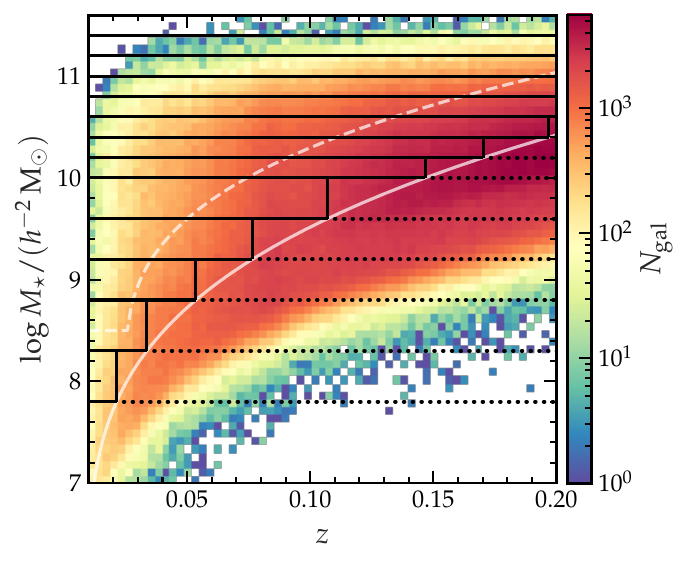}
     \caption{Observed number counts of DESI BGS galaxies on the stellar
     mass versus redshift plane. The white solid curve indicates the
     stellar mass completeness limit in DESI BGS~(Equation~\ref{eq:mlim}),
     while the dashed white curve is the limit for the SDSS main galaxy
     survey~\citepalias{ZM2015}. Black solid rectangles delineate the 12
     volume-limited $M_\star$-binned samples, used exclusively for the
     $w_g$ measurements. Total galaxy samples selected by the dotted
     horizontal lines are used for the $\ds$ and $\Ns$ measurements~(see
     text for details).}
     \label{fig:sample}
\end{figure}

Our sample selection is summarized in Table~\ref{tab:sample} and
illustrated in Figure~\ref{fig:sample}.  The background 2D histogram of
Figure~\ref{fig:sample} shows the galaxy number distribution on the
$M_\star$ vs. $z$ plane, color-coded according to the color bar on the
right. Black solid boxes and dotted horizontal lines indicate the
volume-limited and total galaxy samples, respectively, of the 12 stellar
mass bins. White solid and dashed curves mark the completeness limit of our
Equation~\ref{eq:mlim} for DESI and that used by \citetalias{ZM2015} (their
Equation~1) for SDSS.

\begin{deluxetable}{lccccc}
\tablecaption{Summary of volume-limited galaxy samples in this study. Numbers
    in the parentheses are for the ``total'' samples.}
    \label{tab:sample}
\tablewidth{0pt}
\tablehead{
\colhead{$\log M_\star$} & \colhead{$z_{\rm min}$} &
\colhead{$z_{\rm max}$} & \colhead{$N_g$} & \colhead{S/N of $\ds$}
}
\startdata
\multicolumn{5}{l}{Dwarf} \\
7.8--8.3 & 0.01 & 0.021 & 2735 (36927) & 2.4 (5.9) \\
8.3--8.8 & 0.01 & 0.034 & 9871 (108842) & 6.5 (10.2) \\
8.8--9.2 & 0.01 & 0.054 & 20378 (198097) & 9.2 (16.3) \\
9.2--9.6 & 0.01 & 0.077 & 41751 (327328) & 13.2 (20.6) \\
\hline
\multicolumn{5}{l}{MW-like} \\
9.6--10.0 & 0.01 & 0.107 & 96903 (428497) & 18.3 (26.9) \\
10.0--10.2 & 0.01 & 0.147 & 110025 (240671) & 21.0 (26.2) \\
10.2--10.4 & 0.01 & 0.170 & 143982 (210585) & 23.4 (27.1) \\
10.4--10.6 & 0.01 & 0.200 & 158445  & 24.0 \\
\hline
\multicolumn{5}{l}{Massive} \\
10.6--10.8 & 0.01 & 0.200 & 98992 & 24.3  \\
10.8--11.0 & 0.01 & 0.200 & 45904 & 20.6  \\
11.0--11.2 & 0.01 & 0.200 & 14435 & 17.4  \\
11.2--11.4 & 0.01 & 0.200 & 2637 & 13.1  \\
\enddata
\end{deluxetable}

\subsection{Halo-Based DESI Groups}
\label{subsec:group_sample}

The main advantage of our HOD modeling technique is the inclusion of direct
satellite HOD measurements $(\Ns)$ using galaxy groups.\footnote{We use
clusters and groups interchangeably in the paper, loosely referring to all
dark matter halos with mass above a few $\times10^{13}\,\hmsun$.}
The $\Ns$ measurement method was recently proposed by \citet{Shao2025}, who
demonstrated that the satellite occupation numbers $\Nsat$ can be robustly
determined from cluster--galaxy cross-correlation, while cluster weak
lensing constrains the average halo mass $M_h$.

Following \citet{Shao2025}, we use the extended halo-based group catalog
constructed by \citet{Yang2021} (hereafter Y21) from the DESI Legacy
Imaging Surveys.  They demonstrated that the group catalog reaches
${\gtrsim}90\%$ purity for $\log M_h^{\rm AM} {\gtrsim} 12.5$.  More
important, this halo-based group finder is less biased towards red galaxy
colors than the red sequence-based methods \citep{Golden-Marx2023},
allowing a galaxy color-independent measurement of \Ns.  The Y21 catalog
derived halo mass estimates ($M_h^{\rm Y21}$) via abundance matching (AM)
based on total cluster luminosity. Since this AM was performed over a broad
redshift range, the resulting masses exhibit an unphysical trend with
redshift, which we correct for by applying an evolution correction: $\log
M_h^{\rm AM}{=}\log M_{h,0}^{\rm AM} + 13.74\log(1+z)-1.03$.  We emphasize
that this AM-based halo mass $M_h^{\rm AM}$ is used solely for binning,
while the average halo mass of each $M_h^{\rm AM}$ bin is directly measured
from weak lensing.

We select clusters with $\log M_h^{\rm AM} {\geq} 12.7$ and divide them
into five $\log M_h^{\rm AM}$ bins with [12.7, 13.0, 13.4, 13.8, 14.2,
15.0] as the bin edges. We set the positions of the brightest cluster
galaxies (BCGs) as the cluster centers and only keep systems that are
well-centered according to the X-ray study of miscentering by
\citet{Shao2025}. In particular, we remove systems whose BCG position is
offset from the luminosity-weighted center by more than 20\% of the cluster
radius from each bin.

\subsection{SDSS Shear Catalog}
\label{subsec:shear}

To measure galaxy--galaxy lensing, we use the public shear
catalog\footnote{\url{https://www.andrew.cmu.edu/user/rmandelb/data.html}}
presented in \citet{Reyes2012}.  This catalog is based on Sloan Digital Sky
Survey (SDSS) Data Release 7 \citep{SDSSDR7} and adopts the
re-Gaussianization method \citep{Hirata2003} to measure the shapes of
galaxies. The catalog was systematically tested in
\citet{Mandelbaum2012,Mandelbaum2013}.  The photometric redshifts of the
sources are estimated using the Zurich Extragalactic Bayesian Redshift Analyzer
\citep[ZEBRA;][]{Feldmann2006}.  We refer to \citet{Reyes2012} for
technical details on the shear catalog. We have also derived our
galaxy--galaxy lensing measurements using other public shear catalogs,
including HSC Y3 \citep{Li2022HSC} and DECADE \citep{Anbajagane2025}, and
the results are generally consistent with each other.  We adopt the SDSS
shear catalog for our fiducial galaxy--galaxy lensing measurements because
of its unique combination of optimal lensing kernel for dwarfs, homogeneity
of imaging data, and large overlap with DESI DR1 ($\sim5400~{\rm deg}^2$).

\section{Large-scale structure observables}
\label{sec:statistics}

We employ three types of LSS observables for constraining the HODs of each
sample, including the projected auto-correlation function $\wpgg$, the
galaxy--galaxy lensing $\ds$, and the satellite HOD measurement $\Ns$.
Following the practice in \citetalias{ZM2015}, we make a deliberate choice
to omit the observed galaxy number density $\ngobs$ from our fiducial HOD
analysis. The rationale is as follows.  Firstly, we hope to derive a clean
HSMR for dwarf central galaxies from the $\ds$ profiles (after effectively
subtracting the satellite contributions) instead of $\ngobs$, which does
not carry any direct halo mass information~(but only does so indirectly via
abundance matching). Secondly, the uncertainties in the dwarf galaxy
abundances are difficult to estimate due to shredding and other photometric
systematics \citep{Manwadkar2026,Moore2025}.  Nonetheless, we have verified
that including $\ngobs$ as an additional constraint does not change the
main results of this paper (see Appendix~\ref{appendix:ngobs}).

\subsection{Projected Auto-Correlation Function}
\label{subsec:clustering}

For each volume-limited sample, we measure its projected auto-correlation
function $\wpgg$ by integrating the corresponding 2D redshift-space
auto-correlation function $\xiggrs$
\begin{equation}
    \wpgg (r_p) = 2 \int_0^{r_\pi^{\rm max}} \xiggrs (r_p, r_\pi) \diff r_\pi,
    \label{eq:wpgg}
\end{equation}
where $r_p$ and $r_\pi$ are the projected and line-of-sight (LOS)
separations between galaxy pairs, respectively, and $r_\pi^{\rm max}$ is
the integration limit.  For computing $\xiggrs$, we adopt the Landy--Szalay
estimator \citep{Landy1993}
\begin{equation}
    \xiggrs (r_p, r_\pi) = \frac{DD - 2DR + RR}{RR},
\end{equation}
where $DD$, $DR$, and $RR$ are the data--data, data--random, and
random--random pair counts, respectively.  We employ two types of weights
when counting pairs, including one applied to pairs of galaxies and the
other to individual galaxies.  The pair weights include the so-called
``pairwise-inverse-probability'' \citep[PIP;][]{Bianchi2017} weight and the
angular up-weight \citep{Percival2017}. Both pair weights are designed to
correct for small-scale incompleteness due to DESI fiber assignment, and
are extensively validated in \citet{Bianchi2018,Bianchi2025}. We refer
readers to Section 5.2 of \citet{Bianchi2025} for technical details on
the PIP weights.  The individual weight is computed as $w_{\rm tot}'=w_{\rm
sys}\times w_{\rm zfail}$, for each galaxy. This is different from the LSS
weight as the $w_{\rm comp}$ term has been absorbed into the pair weights.

We use 11 logarithmic $r_p$ bins from $0.1\,\hmpc$ to $30\,\hmpc$ for all
the $\wpgg$ measurements, but adopt different values of $r_\pi^{\rm max}$
depending on the redshift range of the sample.  In particular, we set
$r_\pi^{\rm max} {=} 40\,\hmpc$ for MW-like and massive galaxy samples
(i.e., $\log M_\star {\geq} 9.6$), and $r_\pi^{\rm max}=20\,\hmpc$ for the
dwarf samples. We explicitly model the residual redshift-space distortion
effect \citep[RRSD;][]{vdbosch2013} when predicting $\wpgg$, so that our
result is insensitive to the choice of $r_\pi^{\rm max}$.

For the majority of our samples, we estimate the uncertainty matrices of
$\wpgg$ by resampling 100 spatially contiguous jackknife regions on the
sky \citep{Norberg2009}. However, the volumes of the two least massive
samples ($\log M_\star{\leq}8.8$) are too small to yield robust jackknife
covariances. To circumvent this issue, we estimate the error matrices using
mock galaxy samples constructed from the Uchuu N-body simulation
\citep{Ishiyama2021}.  For each of the two samples, we first find a
reasonable HOD model that roughly reproduces the observed $\wpgg$ and use
this model to populate halos in the simulation. Next, we divide the full
simulation box into $N$ (68400 and 12400) sub-boxes that have roughly the
volume occupied by that sample, and then compute the covariance matrices
using these mock $\wpgg$ measurements as an approximation for the DESI
result \citep{Norberg2009}.

\subsection{Galaxy--Galaxy Lensing $\ds$}
\label{subsec:ggl}

The observable quantity of galaxy--galaxy lensing is the so-called surface
density contrast profile ($\ds$).  We compute $\ds$ using the tangential
ellipticities $e_{\rm t}$ of background source galaxies via
\begin{equation}
    \ds(r_p) = \frac{\Sigma_{\rm ls}w_{\rm ls}e_{\rm t}^{\rm ls}
    \Sigma_{\rm crit}(z_{\rm l}, z_{\rm s})}
    {2\mathcal{R}\sum_{\rm ls} w_{\rm ls}}\times
    \frac{\sum_{\rm ls} w_{\rm ls}}{\sum_{\rm rs} w_{\rm rs}},
    \label{eq:deltasigma}
\end{equation}
where $z_{\rm l}$ and $z_{\rm s}$ are the redshifts of the lens and source
galaxies, respectively, $w_{\rm ls}$ ($w_{\rm rs}$) is the inverse-variance
weight of each lens (random)--source pair, $\Sigma_{\rm crit}$ is the critical
surface mass density, and $\mathcal{R}{\approx}0.87$ is the responsivity factor
that converts tangential ellipticity to tangential shear \citep{Singh2017}.  In
particular, the critical surface density is defined by
\begin{equation}
     \Sigma^{-1}_{\rm crit} (z_{\rm l}, z_{\rm s}) \equiv
     \frac{4\pi G}{c^2}\frac{D_{\rm ls}D_{\rm l}(1+z_{\rm l})^2}{D_{\rm s}},
\end{equation}
where $D_{\rm l}$, $D_{\rm s}$, and $D_{\rm ls}$ are the angular diameter
distances to the lens, to the source, and between the lens and source,
respectively. The factor of $(1+z_{\rm l})^2$ comes from our use of
comoving coordinates. The weight of each lens--source pair is given by
\begin{equation}
     w_{\rm ls} = \frac{w_{\rm tot}}{\Sigma_{\rm crit}^2 (z_{\rm l}, z_{\rm s})
     (\sigma_{s}^2 + \sigma_{\rm SN}^2)},
    \label{eq:w_ls}
\end{equation}
where $\sigma_s$ is the shape measurement error of each source galaxy and
$\sigma_{\rm SN}=0.365$ is the intrinsic shape noise \citep{Reyes2012}.
The weight of each random--source pair is computed in the same way, with
the random samples controlled to have the same redshift distribution as the
lens samples.

We perform the summation in Equation~\ref{eq:deltasigma} over all lens
(random)--source pairs at $r_p$ but with redshift differences $z_s-z_l>0.1$. The
second term $\sum_{\rm ls} w_{\rm ls}/\sum_{\rm rs} w_{\rm rs}$ in
Equation~\ref{eq:deltasigma} is to account for the dilution of lensing signals
by galaxies that are physically associated with the lens but incorrectly
included in the source catalog due to photo-z errors
\citep{Sheldon2004,Mandelbaum2005}.  Following \citet{Singh2017}, we repeat the
$\ds$ calculation for the corresponding random samples and subtract the results
around randoms to obtain the final $\ds$ profiles.  We measure $\ds$ in 10
logarithmic $r_p$ bins from $0.01\,\hmpc$ to $30\,\hmpc$.

As described in Section \ref{subsubsec:complete_sample}, we use the {\it
total} galaxy samples to boost the S/N of our galaxy--galaxy lensing
measurements.  However, galaxies below the completeness limit are
preferentially low-$\Upsilon_\star$ systems.  To correct for this
incompleteness, we apply an additional $1/\vmax$ weight to every lens
galaxy, so that our fiducial $\ds$ measurements from the total samples are
statistically consistent with them being volume-limited.  In particular,
for a galaxy with $r$-band K-corrected absolute magnitude $M_r$, we compute
the maximum comoving distance $D_{\rm max}$ up to which this galaxy can be
observed by BGS Bright, yielding the $1/\vmax$ weight $D^{-3}_{\rm
max}(M_r)$.  We demonstrate in Appendix~\ref{appendix:vmax} that after
applying the $1/\vmax$ weights, the effective $\Upsilon_\star$ distribution
of the total sample is the same as that of the corresponding volume-limited
sample. The uncertainty matrices of galaxy--galaxy lensing are estimated
using the jackknife resampling technique with 100 spatially contiguous
subregions within the overlapping area between SDSS and DESI BGS DR1
($\sim5400\,{\rm deg}^2$).

\begin{figure}[!t]
     \centering
     \includegraphics[width=.99\linewidth]{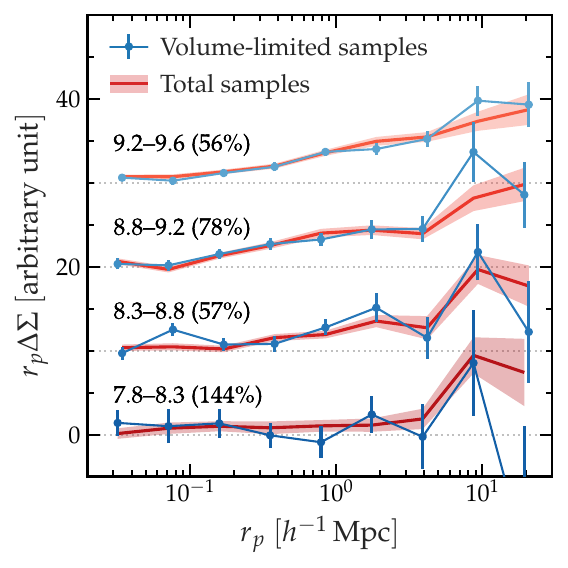}
     \caption{Comparison between $\ds$ signals measured from the four
     volume-limited dwarf samples~(blue solid circles with error bars) and
     from the total samples after applying proper $\vmax$ weights (red
     curves with shaded error bands). The numbers in the parentheses
     indicate the relative enhancement of S/N after using the total
     samples. We shift the $\ds$ measurements vertically by some arbitrary
     amount to avoid clutter.}
     \label{fig:ds_vmax}
\end{figure}

Figure~\ref{fig:ds_vmax} compares our fiducial galaxy--galaxy lensing
measurements (red curves with error bands) from the total samples to those
from the volume-limited samples (blue circles with error bars), for the
four lowest stellar mass bins. The two sets of $\ds$ measurements are in
excellent agreement, demonstrating the efficacy of our $1/\vmax$ weighting
scheme.  After switching to the total samples, the S/N of $\ds$ generally
increases by $50\%$--$140\%$ (numbers in the parentheses), effectively
increasing the number of available lenses by a factor of two to six.  We
list the S/N values of all measurements in Table~\ref{tab:sample}.

\subsection{Satellite HOD $\Ns$}
\label{subsec:Nsat}

Traditional HOD constraints primarily rely on $\wpgg(r_p)$ for separating
the central and satellite contributions to the HOD. While adding $\ds(r_p)$
can help break the degeneracy between the central and satellite HODs, the
constraining power remains dominated by clustering due to its higher S/N.
This standard approach works well for galaxy samples occupying a
cosmologically meaningful volume, enabling robust $\wpgg$ measurements out
to scales of $r_p=30\,\hmpc$. However, despite the unprecedented
depth of DESI, our volume-limited dwarf galaxy samples are limited to
$z\leq 0.08$, with the lowest-$M_\star$ sample below $z=0.02$. Consequently,
their $\wpgg$ measurements are strongly impacted by systematic
uncertainties associated with cosmic variance and integral constraints
\citep{Guo2017,Chen2019}.  Although our simulation-based covariance
matrices partially mitigate these effects, we seek to shift the
constraining power to a new observable that is inherently more robust to
small-volume systematics.

To this end, we follow the method developed by \citet{Shao2025} to directly
measure the satellite HOD $\Ns$, and then combine it with $\ds$ and $\wpgg$
to measure the HSMR of central galaxies. The $\Ns$ measurements are not
only insensitive to cosmic variance, but also enable a straightforward
subtraction of satellite contributions to the $\ds$ of dwarf galaxies.  The
measurement of $\Ns$ requires: 1) the halo mass estimates from cluster weak
lensing, and 2) the satellite occupations inferred from cluster--galaxy
cross-correlations.  We briefly describe the method for measuring $\Nsatmh$
below and refer interested readers to \citet{Shao2025} for the technical
details.

We start by measuring the average halo mass $M_h^{\rm WL}$ of the five
cluster samples binned by $M_h^{\rm AM}$ using cluster weak lensing (see
Appendix~\ref{appendix:sathod}).  Following \citet{Zu2021}, we assume an
isotropic Navarro--Frenk--White \citep[NFW;][]{NFW1997} profile for the
clusters and a shape-2 Gamma distribution to describe the miscentering
of BCGs \citep{Shao2025}.  Next, for each stellar mass sample, we measure
its projected cross-correlation function $\wphg(r_p)$ with each of the five
cluster samples (see Appendix~\ref{appendix:sathod}). We then infer the 3D
isotropic cluster--galaxy cross-correlation function $\xihg(r)$ by fitting
the observed $\wphg(r_p)$ with an analytic model of
\begin{equation}
    \wphg (r_p) = 2 \int_0^{100\,\hmpc} \xihg \left(r=\sqrt{r_p^2 + r_\pi^2}\right)\,\diff r_\pi.
\end{equation}
By integrating $\xihg(r)$, we obtain a ``raw'' measurement of the satellite
occupation as
\begin{equation}
    \langle N_{\rm sat}'\rangle = \ngobs \int_0^{r_h^{\rm WL}} 4\pi r^2\left[\xihg (r) + 1\right]\,\diff r,
    \label{eq:rawnsat}
\end{equation}
where $r_h^{\rm WL}$ is the average halo radius computed from $\mhwl$.

Finally, we obtain the satellite HOD measurements by applying a
cluster-dependent correction to all the raw measurements
\begin{equation}
     \Nsatmhwl = (1+q)\cdot\langle N_{\rm sat}'\rangle,
\end{equation}
where $q=[1.26, 1.24, 1.21, 1.19, 1.18]$ is the so-called ``galaxy occupation
bias'' for the five cluster samples. This bias is caused by the
miscentering and the scatter in the halo mass distribution of the cluster
sample, which will distribute some satellite galaxies outside the average halo
radius $r_h^{\rm WL}$ and lead to an underestimate of $\Nsat$. We calibrate the
$q$ values using a suite of simulation-based mock samples of clusters and
galaxies, following the prescription proposed in \citet{Shao2025}. We add an
additional 10\% uncertainty in quadrature to the error of $\Nsatmhwl$ to account
for any residual systematics in the calibration of $q$.

\begin{figure*}
     \includegraphics[width=.99\linewidth]{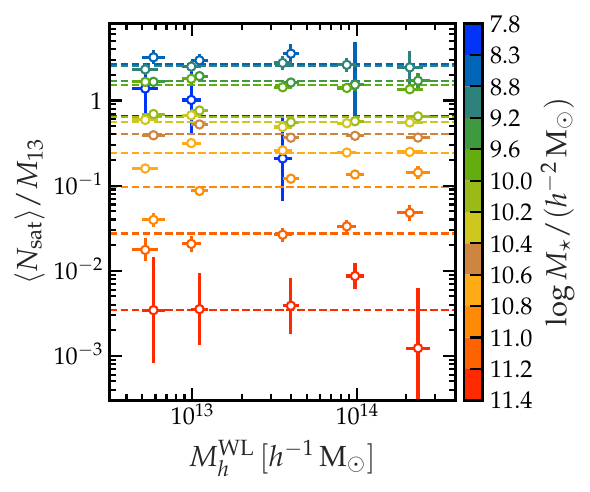}
     \caption{Satellite occupation measurements~($\Ns$). Open circles with
     error bars represent the satellite occupation number per
     $10^{13}\,\hmsun$ halo mass, $\Nsat/M_{13}$, as a function
     of the weak lensing halo mass $\mhwl$ for the 12 stellar mass bins,
     color-coded according to the colorbar on the right.  Circles in the same
     column share exactly the same weak lensing halo mass, but we add a small
     horizontal shift to avoid clutter.  The horizontal dashed lines are the
     average $\Nsat/M_{13}$ of individual stellar mass bins.}
     \label{fig:NsatMh}
\end{figure*}

Figure~\ref{fig:NsatMh} presents our \Ns\ measurement (circles with error bars)
for the 12 stellar mass bins, color-coded by $\log M_\star$ according to the
colorbar on the right. To highlight the linear relation between $\Nsat$ and
$\mhwl$, we plot the satellite occupation number per $10^{13}~\hmsun$ halo mass,
$\Nsat/M_{13}$, where $M_{13}\equiv M_h/(10^{13}\,\hmsun)$, as a function of
$\mhwl$. Note that we add slight horizontal offsets to the data points to avoid
clutter, while the weak lensing halo mass measurement for each cluster mass bin
is shared among all the stellar mass bins.  To help guide the eye, we show the
average $\Nsat/M_{13}$ of each stellar mass bin as a dashed horizontal line of a
matching color. The average $\Nsat/M_{13}$ follows a decreasing trend with
increasing stellar mass, except at the lowest stellar mass ($\log
M_\star=[7.8,8.3]$), where the occupation number per halo mass is a factor of
two below the second lowest stellar mass bin ($\log M_\star=[8.3,8.8]$).
Overall, the \Ns\ measurements of all 12 stellar mass bins are roughly
consistent with having a power-law slope of unity, i.e., a constant value of
$\Nsat/M_{13}$ \citep{Kravtsov2004,Zheng2005}.

We present the cluster lensing profiles and cluster--galaxy
cross-correlations in Appendix~\ref{appendix:sathod}, and the individual
$\mhwl$ and $\Nsatmhwl$ measurements are listed in Table~\ref{tab:sathod}.

\section{Method}
\label{sec:method}

\subsection{Model Parameterization}
\label{subsec:method_hod}

{
\begin{deluxetable}{lcc}
\tabletypesize{\footnotesize}
\tablecaption{Descriptions and priors of our model parameters for each sample.
\label{tab:hod_params}}
\tablehead{
\colhead{Parameter} & \colhead{Description} & \colhead{Prior}
}

\startdata
$\log M_0$             & Mean of the log-normal $\Ncen$            & [9.0, 15.0] \\
$\sigma_{\log M}$      & Width of the log-normal $\Ncen$           & [0.1, 1.0] \\
$N_0$                  & Amplitude of $\Ncen$                      & [0.1, 2.0] \\
$\log M_1$             & Amplitude of $\Nsat$                      & [10.0, 15.0] \\
$\log\kappa$           & Scale at which $\Nsat$ is truncated       & [-1.0, 1.0] \\
$\sigma_{\rm cut}$     & Width of the $\Nsat$ truncation           & [0.1, 2.0] \\
$\alphasat$            & Power-law slope of $\Nsat$                & $\mathcal{N}(1, 0.05)$ \\
$\mathcal{A}_{\rm c}$  & Satellite concentration w.r.t matter & [0.1, 3.0] \\
\enddata
\end{deluxetable}
}

As the first step of our SHMR constraint, we infer the mean HSMR, $\hsmr$,
from a joint modeling of $\wpgg$, $\ds$, and \Ns\ measurements.  We perform
separate HOD analyses for each of the 12 stellar mass bins, instead of
adopting a single global HOD model for the entire galaxy sample, as was
done in \citetalias{ZM2015}.  This discrete approach ensures that the HSMR
inferred in the dwarf galaxy regime is not simply an extrapolation from the
high-mass end, where the average S/N of the three observables is higher.

For each stellar mass bin, its HOD is the sum of a central and a satellite
component
\begin{equation}
    \langle N_{\rm gal}(M_h) \rangle = \Ncenmh + \Nsatmh.
\end{equation}
We model the central HOD $\Ncenmh$ as a flat-top log-normal distribution
with a maximum occupation of unity
\begin{equation}
    \Ncenmh = \min\left \{N_0 \exp{\left[-\frac{\log (M_h/M_0)}{2\sigma_{\log M}^2}\right]},\ 1.0\right \},
\end{equation}
where $M_0$, $\sigma_{\log M}$ and $N_0$ are the center, dispersion, and
amplitude of the log-normal.  While a log-normal HOD of centrals is roughly
expected from a log-normal SHMR, we regard the parameters that govern the
shape of the central HOD as nuisance parameters. Instead, we focus on the
derived quantity, i.e., the average halo mass of the central galaxies
\begin{equation}
    \avgmh = \frac{\int M_h\Ncenmh\phi(M_h)\,\diff M_h}{\int \Ncenmh\phi(M_h)\,\diff M_h},
\end{equation}
where $\phi(M_h)\equiv{\rm d}n_h/{\rm d}M_h$ is the halo mass function. We
have tested different functional forms of $\Ncenmh$ and the inferred
$\avgmh$ is insensitive to this choice; however, the galaxy number density
depends sensitively on the shape of its central HOD (e.g., the spread in
halo mass), which is why $M_0$, $\sigma_{\log M}$, and $N_0$ are nuisance
parameters in our HOD constraint.  For an extreme example,
\citet{Mandelbaum2006} inferred $\langle M_h\rangle$ from their HOD
modeling despite adopting a Dirac delta function for the $\Ncenmh$.

For modeling the observed satellite HOD, we use a power-law function of
halo mass with a flexible transition to zero at the low-mass end
\begin{equation}
    \Nsatmh =
    \frac{1}{2}\left[1+{\rm erf}\left(\frac{\log(M_h/(\kappa M_0))}{\sqrt{2}\sigma_{\rm cut}}\right)\right]
    \left(\frac{M_h}{M_1}\right)^{\alphasat}.
    \label{eq:Nsat}
\end{equation}
In addition, we set $\Nsatmh=\min\{\Nsatmh, \Ncenmh\}$ when $M_h{<}M_0$,
as a physical constraint that low-mass halos should not host more satellites
than centrals of the same stellar mass. The satellite fraction of the sample is
therefore
\begin{equation}
    f_{\rm sat} = \frac{1}{n_g}\int \Nsatmh\phi(M_h)\,\diff M_h,
    \label{eq:fsat}
\end{equation}
where
\begin{equation}
    n_g = \int \langle N_{\rm gal}(M_h) \rangle \phi(M_h)\,\diff M_h
\end{equation}
is the predicted number density of the sample.

In addition to the HOD parameters, we describe the spatial distributions of
dark matter and satellite galaxies within halos using the NFW profiles.
Following \citetalias{ZM2015}, we adopt the halo mass--concentration
relation of \citet{Zhao2009} for describing the dark matter concentration
$c_m$, and introduce $\mathcal{A}_c\equiv c_g/c_m$ as a free parameter to
describe the concentration ratio between galaxies ($c_g$) and dark matter.
In total, we have eight free parameters for each stellar mass bin: \{$M_0$,
$\sigma_{\log M}$, $N_0$, $\kappa$, $\sigma_{\rm cut}$, $M_1$, $\alphasat$,
$\mathcal{A}_c$\}.

\subsection{Prediction of Observables}
\label{subsec:method_prediction}

With a given HOD, we predict $\wpgg$ and $\ds$ using the analytical halo
model of \citetalias{ZM2015} and compare our measured \Ns\ directly with
Equation~\ref{eq:Nsat}. We briefly describe the analytic model below and
refer interested readers to \citetalias{ZM2015} for more details.

We first predict the 3D galaxy auto-correlation function $\xigg(r)$ and
galaxy--matter cross-correlation $\xi_{gm}(r)$, and then convert them into
$\wpgg(r_p)$ and $\ds(r_p)$, respectively.  The 3D correlation between
galaxies and species $x$ can be decomposed into the 1-halo and 2-halo terms
as
\begin{equation}
     \xi_{gx}(r) + 1 = \left[\xi_{gx}^{1h}(r)+1\right] + \left[\xi_{gx}^{2h}(r)+1\right],
\end{equation}
where $x$ could be either galaxy ($g$) or matter ($m$).

We compute the 1-halo term of $\xigg(r)$ as the sum of ``cen--sat'' (cs)
and ``sat--sat'' (ss) contributions
\begin{equation}
\begin{aligned}
\xi_{gg}^{1h}(r)   + 1 &=  \frac{1}{4\pi r^2\,\bar n_g^2}\\
   &\times \int\phi(M_h)
    \Big[\langle N_{\rm cen}(M_h) N_{\rm sat}(M_h)\rangle\, F'_{\rm cs}(r|c_g)  \\
   & + \frac{1}{2}\langle N_{\rm sat}(M_h)(N_{\rm sat}(M_h)-1)\rangle\, F'_{\rm ss}(r|c_g)
   \Big]\diff M_h ,
\end{aligned}
\end{equation}
where $F'_{\rm cs}(r|c_g)$ is the normalized NFW profile describing the
spatial distribution of central--satellite pairs, and $F'_{\rm ss}(r|c_g)$,
defined as the self-convolution of $F'_{\rm cs}(r|c_g)$, describes the
distribution of satellite--satellite pairs \citep{Zheng2007}.

The 1-halo term of $\xi_{gm}$ includes contributions from the dark matter
associated with the main halo, the subhalo associated with the satellites,
and the stellar mass in the galaxy. Following the prescription of
\citetalias{ZM2015}, we model the stellar contribution as a point mass, and
the subhalo lensing as a tidally truncated NFW profile, using their Equations 25
and 43, respectively.  The 1-halo contribution due to the main halo can be
written as the sum of ``cen'' and ``sat'' components
\begin{equation}
     \begin{aligned}
     \xi_{gm}^{1h}(r) + 1
     &= \frac{1}{4\pi r^2\,\bar n_g \bar\rho_m}\\
     &\times \int \phi(M_h)
     \Big[
          \Ncenmh M_h F'_{\rm cm}(r|c_m) \\
     &+ \Nsatmh M_h F'_{\rm sm}(r|c_m,c_g)
     \Big]\diff M_h,
\end{aligned}
\end{equation}
where $F'_{\rm cm}(r|c_m)$ is the normalized NFW profile of dark matter,
and $F'_{\rm sm}(r|c_m,c_g)$ is the convolution between $F'_{\rm
cs}(r|c_g)$ and $F'_{\rm cm}(r|c_m)$.

We predict the respective 2-halo terms as
\begin{equation}
     \xi_{gg}^{2h}(r) = b_g^2\zeta^2(r) \xi_{mm}(r),
\end{equation}
and
\begin{equation}
     \xi_{gm}^{2h}(r) = b_g\zeta(r) \xi_{mm}(r),
\end{equation}
where $b_g$ is the linear bias of the galaxy sample, $\zeta(r)$ describes
the scale-dependent galaxy bias \citep{Tinker2005}, and $\xi_{mm}(r)$ is
the auto-correlation of matter. We calculate $\xi_{mm}(r)$ as the Fourier
transform of the non-linear matter power spectrum using \texttt{Halofit}
\citep{Takahashi2012}.  The galaxy bias $b_g$ is computed as
\begin{equation}
    b_g = \frac{1}{n_g}\int b_h(M_h)\langle N_{\rm gal}(M_h)\rangle\phi(M_h)\,\diff M_h,
\end{equation}
where $b_h(M_h)$ is the halo bias function predicted from
\citet{Tinker2010}.  Following \citetalias{ZM2015}, we explicitly calculate
the halo exclusion effect for the 1-halo to 2-halo transitions in $\xigg$
and $\xi_{gm}$ \citep{Tinker2005,Yoo2006}.

Similar to Equation~\ref{eq:wpgg}, we integrate $\xigg(r)$ along the LOS to
predict the projected galaxy auto-correlation function
\begin{equation}
    \wpgg (r_p) = 2 \int_0^{r_\pi^{\rm max}} \xigg \left(r=\sqrt{r_p^2 + r_\pi^2}\right)\,\diff r_\pi,
    \label{eq:wpggmodel}
\end{equation}
except that $r_\pi$ is the LOS distance in real space.  We adopt the same
value of $r_\pi^{\rm max}$ as used in the measurements and correct for the RRSD
effect using the analytic model of \citet{vdbosch2013}.

Using $\xi_{gm}$, we can straightforwardly predict the surface density contrast
profile $\ds(r_p)$ as
\begin{equation}
    \ds(r_p) = \langle\Sigma(<r_p)\rangle - \Sigma(r_p),
\end{equation}
where
\begin{equation}
    \Sigma(r_p) = 2\bar\rho_m \int_0^{\infty}\left[1 + \xi_{gm}\left(r=\sqrt{r_p^2 + r_\pi^2}\right)\right]\,\diff r_\pi,
\end{equation}
and
\begin{equation}
    \langle\Sigma(<r_p)\rangle = \frac{2}{r_p^2} \int_0^{r_p} r'_p \Sigma(r'_p) {\rm d}r'_p.
\end{equation}

\section{Parameter constraints}
\label{sec:parameter}

We constrain the HOD parameters for each stellar mass bin using a standard
Bayesian inference model.  For each stellar mass bin, the combined data vector
is $\bm{D}{=}\{\wpgg,\,\ds,\,\Nsatmh\}$, which constrains eight model
parameters: $\bm{\theta}\equiv\{\log M_0$, $\sigma_{\log M}$, $N_0$, $\log M_1$,
$\log \kappa$, $\sigma_{\rm cut}$, $\alphasat$, $\mathcal{A}_{\rm c}\}$.  The
descriptions and the prior ranges of these parameters are summarized in Table
\ref{tab:hod_params}.

We write the likelihood function as the product of three individual
likelihoods
\begin{equation}
    \ln \mathcal{L} = \ln \mathcal{L}_{w_p} + \ln \mathcal{L}_{\ds} + \ln \mathcal{L}_{\Ns}.
\end{equation}
The error matrices of the three observables are largely independent when
using the SDSS shear catalog: the uncertainties in the weak lensing
measurements (including $\ds$ and $M_h^{\rm WL}$) remain dominated by the
shape noise of source galaxies \citep{Mandelbaum2013}, while the
measurement of $\Nsatmhwl$ is limited by the uncertainties in $M_h^{\rm
WL}$ \citep{Shao2025}.  We assume standard Gaussian likelihood models for
$\mathcal{L}_{w_p}$ and $\mathcal{L}_{\ds}$, using the covariance matrices
estimated in \S\ref{subsec:clustering} and \S\ref{subsec:ggl},
respectively.

\begin{figure*}[!t]
     \centering
     \includegraphics[width=\linewidth]{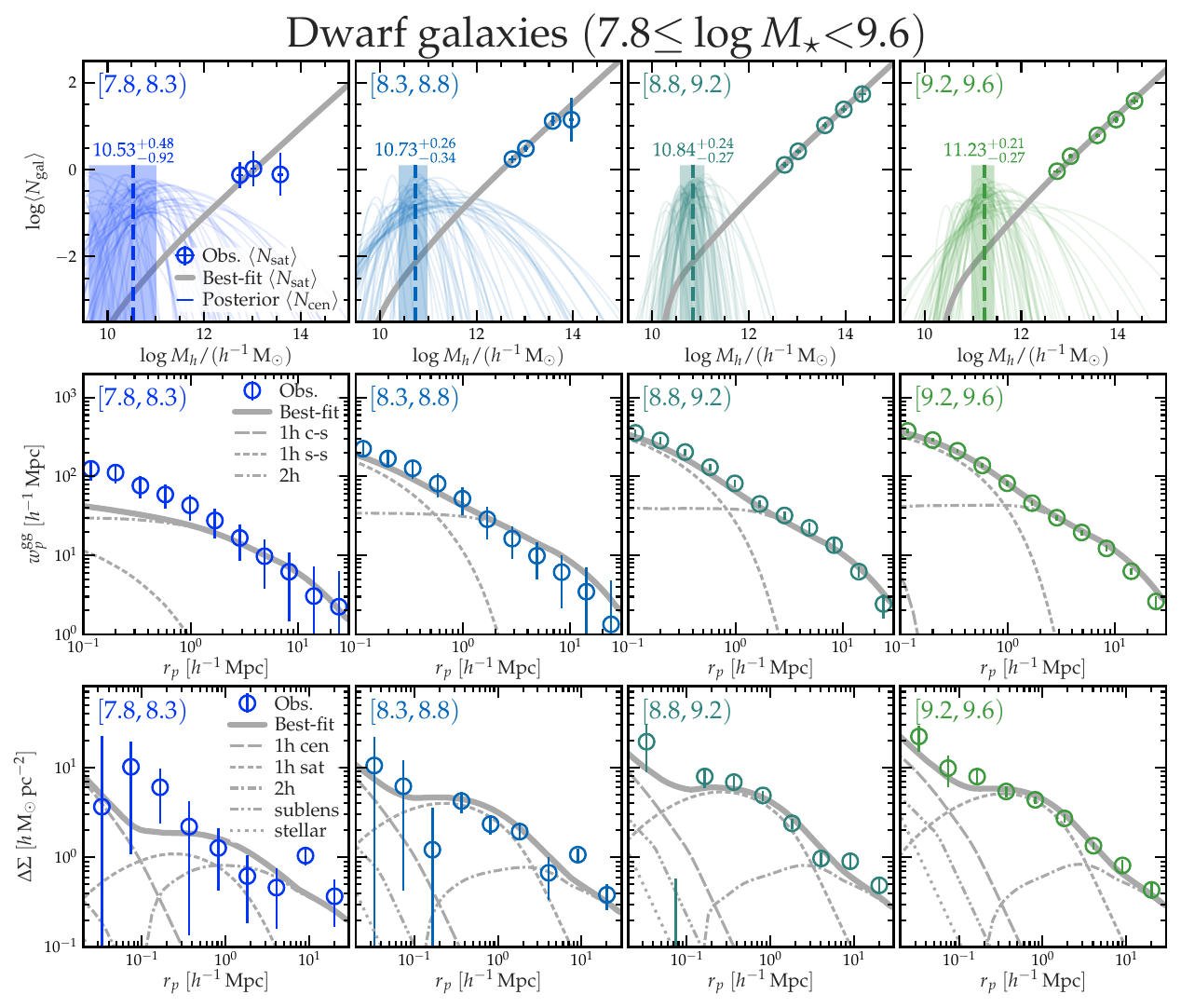}
     \caption{Comparison between the observed~(open circles with error bars)
     and predicted~(thick gray solid curves) observables for the four dwarf
     galaxy samples in different columns~(marked in the top left corner of
     each panel). Top, middle, and bottom rows compare the satellite
     occupation number as a function of halo mass~($\Ns$),  projected
     galaxy auto-correlation as a function of projected radius~($\wpgg$),
     and surface density contrast as a function of projected
     radius~($\ds$), respectively. In each panel of the top row, we also
     show the posterior median and $1\sigma$ constraint on the average halo
     mass of the central galaxies as the thick vertical dashed line with a
     shaded band~(with the confidence limit marked on top).  The bundle of
     light-colored thin curves indicates 50 random draws from the posterior
     distributions of the central HOD~($\Ncenmh$). In panels of the middle
     and bottom rows, thin gray curves of different line styles indicate
     the different components~(indicated by the legend in the left-most
     panels) that contribute to the overall 1-halo and 2-halo signals~(see
     text for details).} \label{fig:obsdwarf}
\end{figure*}

\begin{figure*}[!t]
     \centering
     \includegraphics[width=\linewidth]{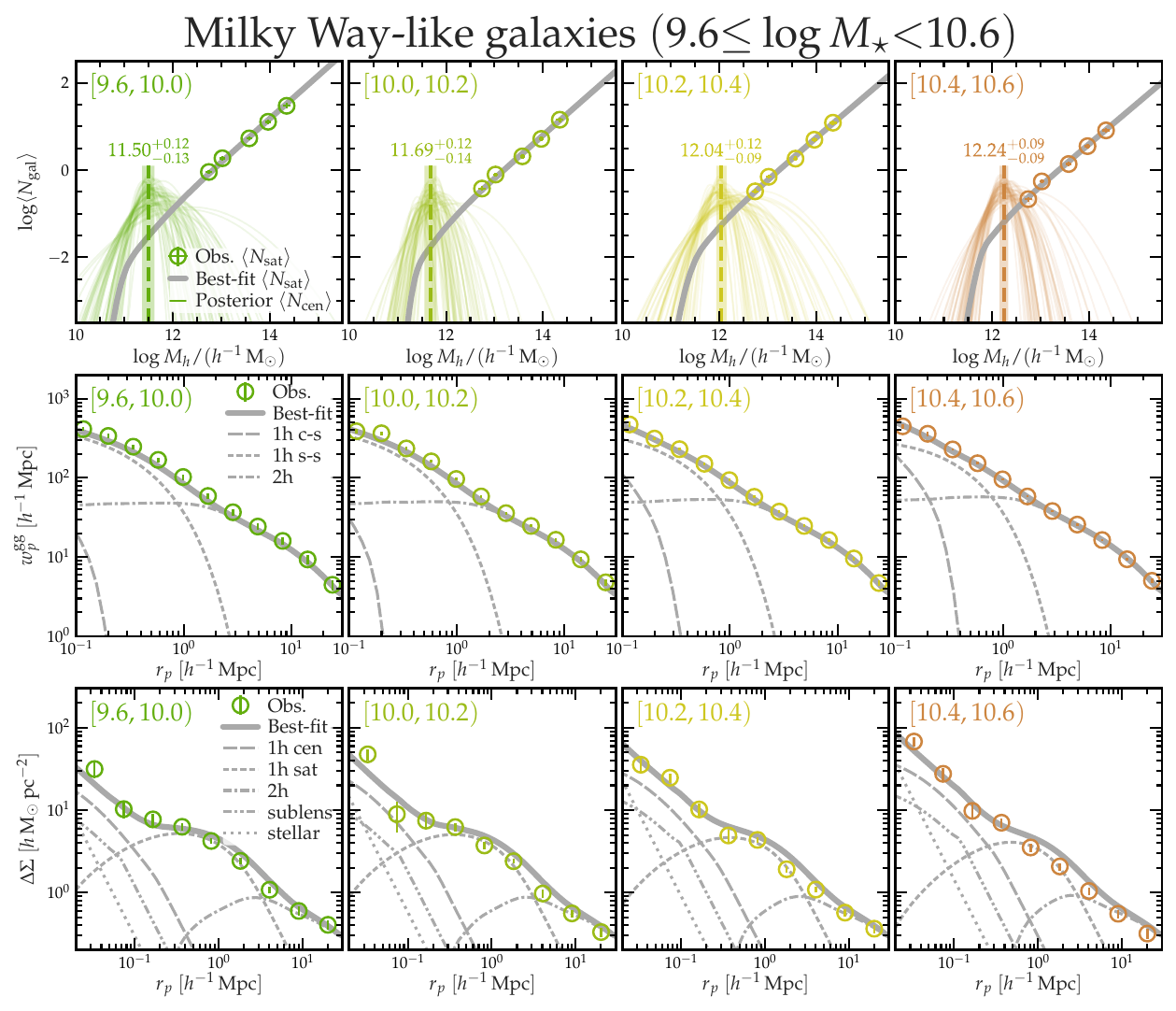}
     \caption{Similar to Fig.~\ref{fig:obsdwarf}, but for the MW-like
     galaxy samples ($9.6{\leq}\log M_\star{<}10.6$).}
     \label{fig:obsmw}
\end{figure*}

For \Ns, we need to account for the covariance between the measurement
uncertainties of $\mhwl$ and $\Nsat$ in the likelihood evaluation
\citep{Shao2025}. Let us consider the $i$-th cluster sample for which we
have measured the average weak lensing halo mass $M_{h,i}^{\rm WL}$.  For
any given stellar mass bin, we measure the average satellite occupation
$\Nsat_i$ and estimate the covariance matrix $\mathbf{C}_i$ between
$M_{h,i}^{\rm WL}$ and $\Nsat_i$. Assuming the two quantities follow a 2D
Gaussian distribution about the true values $(M_{h,i}^{\rm true}, \langle
N_{\rm sat}\rangle_i^{\rm true})$, we can write the log-likelihood function
as
\begin{align}
&\ln \mathcal{L}_{\Ns,i} = \nonumber\\
&-\frac{1}{2}
\begin{pmatrix}
M_{h,i} - M_{h,i}^{\rm true} \\
\Nsat_i - \langle N_{\rm sat}\rangle_i^{\rm true}
\end{pmatrix}^{T}
\mathbf{C}_i^{-1}
\begin{pmatrix}
M_{h,i} - M_{h,i}^{\rm true} \\
\Nsat_i - \langle N_{\rm sat}\rangle_i^{\rm true}
\end{pmatrix}.
\end{align}
In the above expression, $M_{h,i}^{\rm true}$ is a latent variable that
will be constrained by $M_{h,i}^{\rm WL}$ during the inference, and
$\langle N_{\rm sat}\rangle_i^{\rm true}{=}\langle N_{\rm sat}(M_{h,i}^{\rm
true})\rangle$ is directly predicted from $M_{h,i}^{\rm true}$ via
Equation~\ref{eq:Nsat}.  Finally, the combined log-likelihood of $\Nsat$
from all five cluster samples is given by
\begin{equation}
          \ln \mathcal{L}_{\Ns} =
          \sum_{i=1}^{5} \ln \mathcal{L}_{\Ns,i}.
\end{equation}

We make use of the Python package
\texttt{NAUTILUS}\footnote{\url{https://nautilus-sampler.readthedocs.io/en/latest/index.html}}
\citep{Nautilus2023} to sample the posterior distributions of our model
parameters.  To expedite the sampling, \texttt{NAUTILUS} combines
importance nested sampling with neural networks to efficiently explore the
8D parameter space.  We adopt 1500 live points and set the stopping
criterion to be $f_{\rm live}<10^{-2}$ when running the sampler.

\begin{figure*}[!t]
     \centering
     \includegraphics[width=\linewidth]{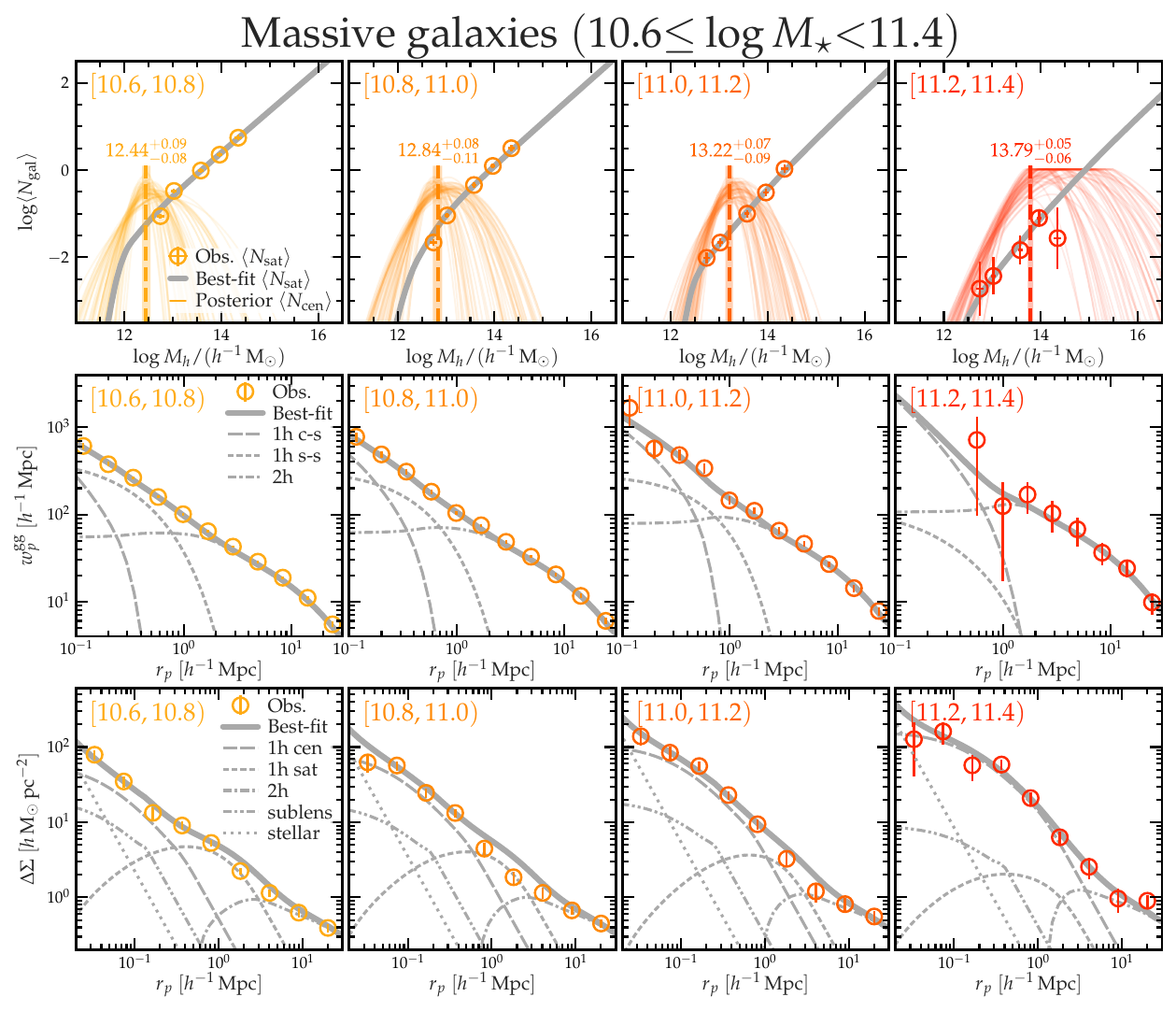}
     \caption{Similar to Fig.~\ref{fig:obsdwarf}, but for the massive
     galaxy samples ($10.6{\leq}\log M_\star{<}11.4$).}
     \label{fig:obsbcg}
\end{figure*}

\section{Results}
\label{sec:results}

\subsection{HODs of Dwarf, MW-like, and Massive galaxies}
\label{subsec:results_hod}

We present the best-fitting HOD models and compare their predictions to the
observations in Figures~\ref{fig:obsdwarf}, \ref{fig:obsmw}, and
\ref{fig:obsbcg}, for the dwarf, MW-like, and massive galaxies,
respectively.  The 1$\sigma$ constraints on model parameters, $\avgmh$,
$\fsat$, and $b_g$ for all 12 stellar mass bins are summarized in
Table~\ref{tab:bestfit_params}.

Figure~\ref{fig:obsdwarf} presents the HOD constraints for our dwarf galaxy
samples, with each column displaying the result for one of the four stellar
mass bins (ranging from $\log M_\star{=}7.8$ to $\log M_\star{=}9.6$ from
left to right). Within each column, colored circles with error bars in the
top, middle, and bottom panels indicate the observed $\Nsatmhwl$,
$\wpgg(r_p)$, and $\ds(r_p)$, respectively, for that stellar mass bin. Gray
thick solid curves in each panel are the corresponding predictions from the
posterior mean model. Overall, our model predictions provide excellent
descriptions of the observed data for all the dwarf samples, yielding
reasonable values for the goodness-of-fit ($\chi^2/{\rm dof}{=}$1--2).

In each panel of the top row, thin colored curves are the central HODs
predicted by 50 random draws from the posterior samples. Despite the large
variation in the individual central HODs, the average halo mass of the
central galaxies is tightly constrained, indicated by the vertical dashed
line with a 1$\sigma$ shaded error band. Notably, $\hsmr$ exhibits a
relatively shallow trend with stellar mass in the dwarf regime, increasing
by only $\sim0.6\,{\rm dex}$ over a $\sim1.4\,{\rm dex}$ range in
$M_\star$.

In each middle panel, we decompose the $\wpgg$ prediction into
contributions from the 1-halo ``cen--sat'' (dashed), 1-halo ``sat--sat''
(short dashed), and 2-halo (dot-dashed) terms.  Due to the lack of overlap
in $M_h$ between the central and satellite HODs, the 1-halo cen--sat term
is almost absent in the four dwarf samples.  Consequently, the 1-halo term
of each observed $\wpgg$ is dominated by the ``sat--sat'' component, which
agrees well with the satellite HOD measurement shown in the corresponding
top panel.  The only exception is in the lowest stellar mass bin, where the
1-halo ``sat--sat'' component is subdominant compared to the 2-halo term,
causing the small-scale clustering to be underpredicted by the model. This
discrepancy is driven by the relatively small amplitude of \Ns, as seen in
Figure~\ref{fig:NsatMh}. Meanwhile, the small satellite occupation number
is consistent with the low amplitude of galaxy--galaxy lensing at $\sim
0.5\,\hmpc$, where the 1-halo satellite term dominates. Therefore, we
suspect that this $w_p$ discrepancy could be a manifestation of a more complex
galaxy--halo connection beyond our standard HOD model, e.g., galaxy
assembly bias or satellite conformity.  Regardless, the $\wpgg$ of dwarf
galaxies carries little information on the HSMR, as the bias dependence on
halo mass is flat in the dwarf regime ($b_h\propto M_h^{0.04}$).

{
\setlength{\tabcolsep}{4pt}
\begin{deluxetable*}{lccccrccccccc}[!t]
\tabletypesize{\footnotesize} \tablecaption{Posterior ranges of model
parameters and derived HOD properties for different samples.  The first
eight parameters are our model parameters described in Table
\ref{tab:hod_params}; $\log \avgmh$, $\fsat$, and $b_g$ are the average
halo mass of the central galaxies, the satellite fraction, and the average
bias of each sample, respectively.  \label{tab:bestfit_params}}
\tablehead{
\colhead{$\log M_\star$} & \colhead{$\log M_0$} & \colhead{$\sigma_{\log M}$} &
\colhead{$n_{\rm cen}$} & \colhead{$\log M_1$} & \colhead{$\log\kappa$} &
\colhead{$\sigma_{\rm cut}$} & \colhead{$\alphasat$} &
\colhead{$\mathcal{A}_{\rm c}$} & \colhead{$\log\langle M_h\rangle$} &
\colhead{$f_{\rm sat}$} & \colhead{$b_g$}
}
\startdata
\multicolumn{12}{l}{Dwarf} \\
7.8--8.3 & $10.69_{-0.82}^{+0.59}$ & $0.42_{-0.23}^{+0.32}$ & $0.31_{-0.15}^{+0.25}$ & $13.02_{-0.22}^{+0.21}$ & $-0.03_{-0.66}^{+0.68}$ & $0.97_{-0.67}^{+0.70}$ & $0.99_{-0.05}^{+0.05}$ & $1.79_{-1.02}^{+0.83}$ & $10.53_{-0.92}^{+0.48}$ & $0.08_{-0.07}^{+0.07}$ & $0.73_{-0.10}^{+0.07}$ \\
8.3--8.8 & $10.86_{-0.41}^{+0.47}$ & $0.33_{-0.17}^{+0.34}$ & $0.27_{-0.13}^{+0.24}$ & $12.49_{-0.07}^{+0.07}$ & $0.03_{-0.69}^{+0.67}$ & $1.06_{-0.72}^{+0.65}$ & $1.01_{-0.05}^{+0.05}$ & $1.81_{-0.62}^{+0.70}$ & $10.73_{-0.34}^{+0.26}$ & $0.26_{-0.05}^{+0.06}$ & $0.89_{-0.03}^{+0.04}$ \\
8.8--9.2 & $10.87_{-0.28}^{+0.27}$ & $0.16_{-0.04}^{+0.12}$ & $0.33_{-0.15}^{+0.20}$ & $12.58_{-0.07}^{+0.05}$ & $0.17_{-0.76}^{+0.59}$ & $1.23_{-0.76}^{+0.54}$ & $1.02_{-0.04}^{+0.04}$ & $1.38_{-0.22}^{+0.26}$ & $10.84_{-0.27}^{+0.24}$ & $0.38_{-0.05}^{+0.06}$ & $1.02_{-0.02}^{+0.02}$ \\
9.2--9.6 & $11.31_{-0.30}^{+0.31}$ & $0.27_{-0.12}^{+0.19}$ & $0.23_{-0.09}^{+0.15}$ & $12.75_{-0.07}^{+0.05}$ & $0.16_{-0.74}^{+0.59}$ & $1.14_{-0.74}^{+0.59}$ & $1.02_{-0.04}^{+0.04}$ & $0.90_{-0.15}^{+0.17}$ & $11.23_{-0.27}^{+0.21}$ & $0.41_{-0.05}^{+0.06}$ & $1.09_{-0.02}^{+0.02}$ \\
\hline
\multicolumn{12}{l}{MW-like} \\
9.6--10.0 & $11.56_{-0.15}^{+0.19}$ & $0.25_{-0.10}^{+0.16}$ & $0.31_{-0.13}^{+0.21}$ & $12.79_{-0.06}^{+0.05}$ & $-0.24_{-0.52}^{+0.58}$ & $0.78_{-0.54}^{+0.77}$ & $0.98_{-0.04}^{+0.04}$ & $0.77_{-0.16}^{+0.19}$ & $11.50_{-0.13}^{+0.12}$ & $0.48_{-0.07}^{+0.07}$ & $1.14_{-0.02}^{+0.02}$ \\
10.0--10.2 & $11.72_{-0.14}^{+0.14}$ & $0.15_{-0.04}^{+0.10}$ & $0.30_{-0.12}^{+0.17}$ & $13.18_{-0.05}^{+0.04}$ & $-0.02_{-0.61}^{+0.48}$ & $0.55_{-0.40}^{+0.79}$ & $0.97_{-0.04}^{+0.04}$ & $0.76_{-0.16}^{+0.18}$ & $11.69_{-0.14}^{+0.12}$ & $0.45_{-0.06}^{+0.06}$ & $1.17_{-0.02}^{+0.02}$ \\
10.2--10.4 & $12.12_{-0.14}^{+0.25}$ & $0.31_{-0.14}^{+0.19}$ & $0.26_{-0.08}^{+0.17}$ & $13.22_{-0.05}^{+0.04}$ & $-0.54_{-0.32}^{+0.38}$ & $0.68_{-0.45}^{+0.69}$ & $0.95_{-0.04}^{+0.04}$ & $0.56_{-0.15}^{+0.19}$ & $12.04_{-0.09}^{+0.12}$ & $0.48_{-0.05}^{+0.05}$ & $1.21_{-0.01}^{+0.01}$ \\
10.4--10.6 & $12.29_{-0.11}^{+0.15}$ & $0.25_{-0.10}^{+0.13}$ & $0.37_{-0.11}^{+0.19}$ & $13.36_{-0.05}^{+0.04}$ & $-0.45_{-0.36}^{+0.32}$ & $0.59_{-0.36}^{+0.59}$ & $0.94_{-0.04}^{+0.04}$ & $0.35_{-0.13}^{+0.17}$ & $12.24_{-0.09}^{+0.09}$ & $0.46_{-0.04}^{+0.05}$ & $1.26_{-0.01}^{+0.01}$ \\
\hline
\multicolumn{12}{l}{Massive} \\
10.6--10.8 & $12.51_{-0.12}^{+0.20}$ & $0.27_{-0.12}^{+0.17}$ & $0.35_{-0.10}^{+0.22}$ & $13.57_{-0.05}^{+0.04}$ & $-0.42_{-0.36}^{+0.32}$ & $0.64_{-0.35}^{+0.68}$ & $0.95_{-0.04}^{+0.04}$ & $0.65_{-0.23}^{+0.29}$ & $12.44_{-0.08}^{+0.09}$ & $0.41_{-0.04}^{+0.04}$ & $1.30_{-0.01}^{+0.01}$ \\
10.8--11.0 & $12.99_{-0.20}^{+0.17}$ & $0.36_{-0.16}^{+0.11}$ & $0.41_{-0.10}^{+0.17}$ & $13.88_{-0.07}^{+0.05}$ & $-0.21_{-0.26}^{+0.31}$ & $0.53_{-0.25}^{+0.62}$ & $0.98_{-0.04}^{+0.04}$ & $0.65_{-0.37}^{+0.78}$ & $12.84_{-0.11}^{+0.08}$ & $0.29_{-0.02}^{+0.03}$ & $1.39_{-0.02}^{+0.02}$ \\
11.0--11.2 & $13.32_{-0.14}^{+0.11}$ & $0.28_{-0.09}^{+0.09}$ & $0.36_{-0.12}^{+0.17}$ & $14.23_{-0.12}^{+0.12}$ & $0.26_{-0.59}^{+0.50}$ & $1.45_{-0.39}^{+0.35}$ & $1.02_{-0.05}^{+0.05}$ & $0.43_{-0.27}^{+0.82}$ & $13.22_{-0.09}^{+0.07}$ & $0.27_{-0.03}^{+0.04}$ & $1.61_{-0.03}^{+0.03}$ \\
11.2--11.4 & $14.31_{-0.30}^{+0.41}$ & $0.50_{-0.14}^{+0.15}$ & $1.14_{-0.59}^{+0.57}$ & $14.74_{-0.23}^{+0.17}$ & $-0.05_{-0.63}^{+0.67}$ & $1.50_{-0.43}^{+0.34}$ & $1.00_{-0.05}^{+0.05}$ & $1.65_{-0.99}^{+0.92}$ & $13.79_{-0.06}^{+0.05}$ & $0.10_{-0.03}^{+0.07}$ & $1.96_{-0.07}^{+0.07}$ \\
\enddata
\end{deluxetable*}
}

Therefore, the main leverage for our HSMR constraint (top row of
Figure~\ref{fig:obsdwarf}) comes from the weak lensing measurements in the
bottom row.  In each bottom panel, we show the decomposition of our
best-fitting $\ds(r_p)$ into 1-halo central (dashed), 1-halo satellite
(short dashed), 2-halo (dot-dashed), subhalo lensing (dot-dot-dashed), and
stellar component (dotted) terms.  The galaxy--galaxy lensing profile is
dominated by the 1-halo satellite component on scales $\sim 1\,\hmpc$
where the lensing S/N is the highest, while the contribution by the central
halo mass (i.e., 1-halo central) is largely overwhelmed by the satellite
component except on scales below $\sim 0.1\,\hmpc$.  Thanks to our
direct measurements of the satellite HOD, however, we can robustly remove
the 1-halo satellite contribution from the overall $\ds$ profile, thereby
revealing the dark matter halos of the smallest central galaxies.

For the MW-like galaxies shown in Figure~\ref{fig:obsmw}, the central halo
mass constraints are significantly tighter compared to those for the dwarf
galaxies, primarily due to the improved S/N in the overall measurements. In
particular, the satellite HODs are pinned down by the observations at $\log
M_h>12.5$, thereby removing the satellite contribution to the 2-halo term
of $\wpgg$. This effectively measures the linear bias of the centrals,
which provides an indirect constraint on the central halo mass through the
halo bias function --- compared to the dwarf galaxies, the bias dependence
on halo mass becomes more sensitive in the MW-like regime ($b(M_h)\propto
M_h^{0.1}$).  Meanwhile, the 1-halo satellite terms in the $\ds(r_p)$ of
MW-like galaxies are less dominant than for the dwarfs, enabling a more
direct constraint on the $\hsmr$ using weak lensing. Likewise for the
massive galaxies in Figure~\ref{fig:obsbcg}, the combination of lower
satellite fractions and higher S/N yields even more stringent constraints
on the mean HSMR in the massive regime.

\subsection{Satellite Fractions}\label{subsec:results_fsat}

\begin{figure}[!t]
     \centering \includegraphics[width=.99\linewidth]{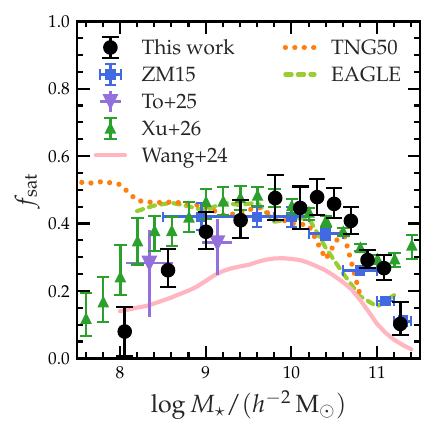}
     \caption{Comparison of the satellite fractions $\fsat$ derived in this
     work~(black circles with error bars) to previous observational constraints
     and simulation predictions.
     Blue squares with error bars are the \texttt{iHOD} constraints from
     SDSS~\citepalias{ZM2015}, whereas purple triangles, green triangles, and
     the pink solid curve are constraints from
     DESI~\citep{To2025,Xu2026,Wang2024CLF}.
     Orange dotted and green dashed curves are the
     predictions from the TNG50 \citep{Pillepich2019} and EAGLE
     \citep{Schaye2015}, respectively.}
     \label{fig:fsat}
\end{figure}

The small-scale $\wpgg$ is dominated by the 1-halo ``sat--sat'' term, which
depends on the satellite HOD \Ns\ as well as the satellite fraction $\fsat$.
With \Ns\ directly measured, our HOD analyses are able to place stringent
constraints on $\fsat$.  From the best-fitting HOD models, we predict the
satellite fractions ($f_{\rm sat}$) for the 12 stellar mass bins using
Equation~\ref{eq:fsat}. In Figure~\ref{fig:fsat}, we compare our inferred
$f_{\rm sat}$--$M_\star$ relation (black solid circles with error bars) with
various results from the literature: blue squares are inferred from the HOD
modeling of SDSS galaxies by \citetalias{ZM2015}. Purple triangles, green
triangles, and the pink solid curve are inferred from the weak lensing of DESI
dwarf galaxies~\citep{To2025}, the SHAM modeling of DESI
galaxies~\citep{Xu2026}, and counting galaxy membership in the DESI galaxy group
catalog~\citep{Wang2024CLF}, respectively. The dotted and dashed curves are
predictions from TNG50 \citep{Pillepich2019} and EAGLE \citep{Schaye2015}
hydrodynamical simulations, respectively.

At the MW-like galaxy scale and above, our inferred $\fsat$--$M_\star$ relation
exhibits a plateau between $\log M_\star = 9.6$ and 10.6, and declines towards the
massive end. This trend is consistent with the SDSS constraint from
\citetalias{ZM2015}. Although the amplitude of our plateau is higher than that
of \citetalias{ZM2015} by 20\%, the two results are consistent with each other
within 1$\sigma$.  The central versus satellite dichotomy is also reasonably
reproduced in hydro simulations at $\log M_\star>9.6$, as the predictions from
TNG50 and EAGLE are roughly consistent with our constraint. However, the $\fsat$
measurements from group catalogs are significantly lower than our HOD
constraint.  This discrepancy likely originates from the low-mass halos in the
group catalog. For example, for the $\log M_\star=10.0$--$10.2$ sample, our
best-fitting HOD predicts that $\sim 30\%$ of the satellites live in halos with
$M_h<10^{12.5}\,\hmsun$, whereas only $\sim 10\%$ of satellites
belong to such halos in the DESI group catalog.

Into the dwarf galaxy regime, we find a declining trend of $\fsat$ with
decreasing stellar mass, dropping rapidly from $\fsat=0.4$ at $\log M_\star=9.4$
to $\fsat=0.1$ at $\log M_\star=8.0$.  A similar trend at the dwarf mass scale
is also seen independently by \citet{Wang2024CLF}, \citet{To2025}, and
\citet{Xu2026}. This declining $\fsat$ into the low-mass end also echoes the
non-monotonic trend of \Ns\ with stellar mass shown in Figure~\ref{fig:NsatMh}.
In contrast, this non-monotonic trend is not reproduced by TNG50 and EAGLE, both
of which predict a roughly constant $\fsat$ below $\log M_\star{=}9.6$. 
It is still unclear whether this discrepancy is due to observational
systematics or the more fundamental differences in our understanding of galaxy
formation physics, which we will explore in future work.

\subsection{Isolating the Weak Lensing of Central Galaxies}
\label{subsec:results_dscen}

\begin{figure*}[!t]
     \centering \includegraphics[width=.99\linewidth]{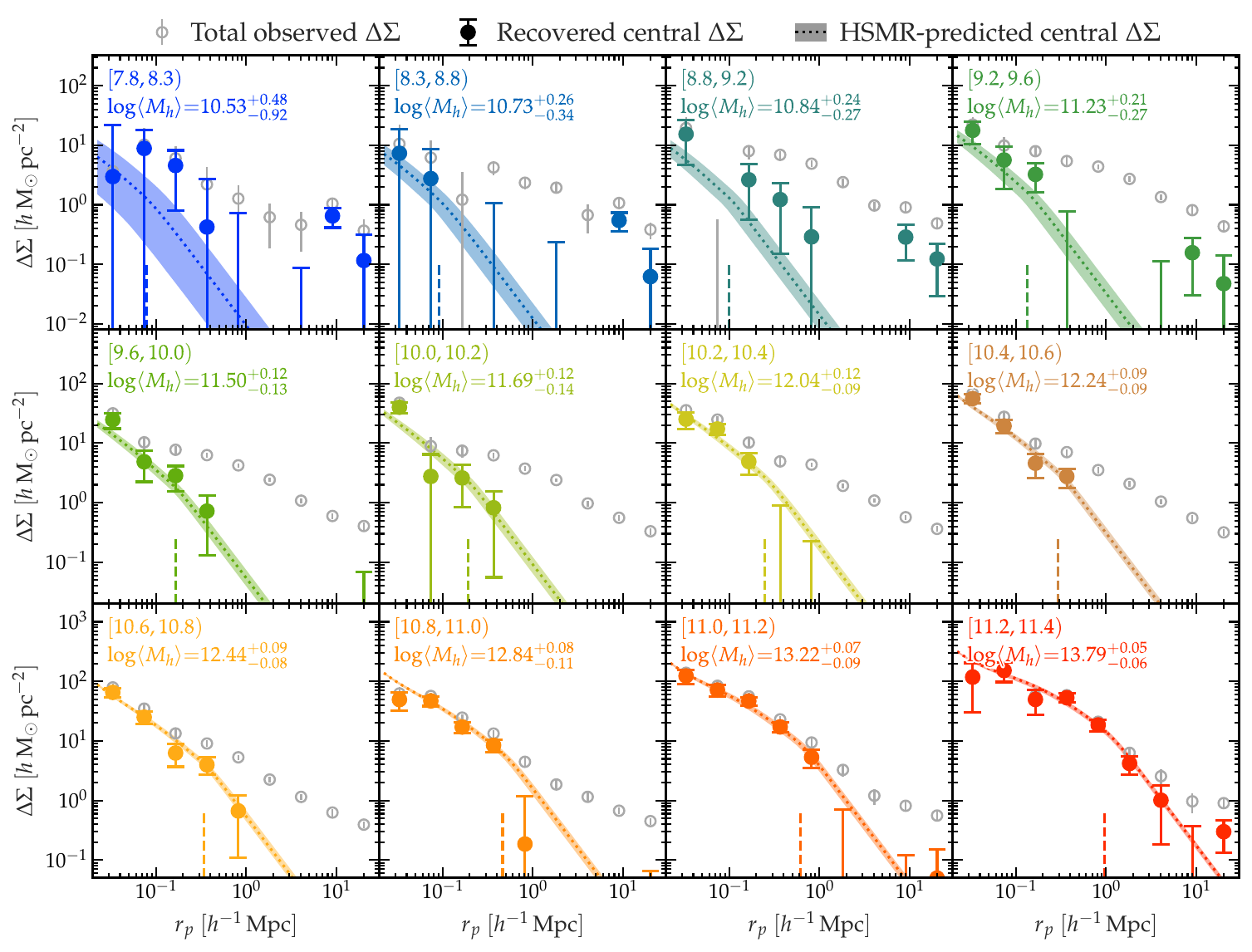}
     \caption{ Weak lensing signals of the central galaxies in the 12
     stellar mass bins~(marked in the top left corner of each panel).  In
     each panel, colored filled circles with error bars are the weak lensing
     profile of the central galaxies, recovered from subtracting the
     satellite contribution to the total observed profile~(gray open
     circles with error bars). The colored dotted curve with an uncertainty band
     is the $\ds$ profile predicted from the posterior median constraint of
     our fiducial HOD analysis~(listed in the legend). The vertical dashed
     line denotes the average halo radius of the central galaxies.}
     \label{fig:dscen}
\end{figure*}

Several recent studies have employed the weak lensing of dwarf galaxies to infer
the mean HSMR at the low-mass end. Using photometrically-selected dwarf
galaxies, \citet{Thornton2024} inferred the mean halo masses of low-mass
galaxies with $\log M_\star\sim8$--$9$, using the weak lensing shear catalog
from the Dark Energy Survey \citep{DES2018,DES2021}. \citet{To2025} performed a
similar analysis using spectroscopic lenses from DESI DR1 BGS Bright and the
DECADE shear catalog \citep{Anbajagane2025}. In order to achieve a higher S/N in
weak lensing, \citet{Treiber2025} combined DES photometric lenses with the DESI
spectroscopic ones at the same dwarf mass scale.  Among them,
\citet{Thornton2024} and \citet{To2025} employed a simulation-based inference
method to model the satellite contribution using subhalos (downsampled according
to a given satellite fraction), and \citet{Treiber2025} only fitted to
small-scale ($r_p<1\,\hmpc$) $\ds$ to derive halo mass.

Unlike prior approaches, our method uniquely enables the direct measurement
of the satellite HODs. In particular, while the \Ns\ measurement directly
constrains the satellite HOD at the high-mass end, its combination with
$\wpgg$ on small scales pins down $\fsat$. With both $\Nsatmh$ and $\fsat$
tightly constrained, our Bayesian method effectively ``subtracts'' the
satellite contribution from the overall galaxy--galaxy lensing signals to
isolate the weak lensing profile of centrals.

We demonstrate the efficacy of our method by performing an explicit
satellite subtraction from the observed $\ds$ profiles across 12 stellar
mass bins in Figure~\ref{fig:dscen}. In each panel, gray open circles with
error bars indicate the original overall $\ds$ measurements for the total
sample of the stellar mass bin, while the colored solid circles with error
bars are the recovered $\ds$ profiles for the central galaxies within that
bin. When computing the error bars associated with the isolated central
$\ds$ after subtraction, we take into account the full posterior
distribution of our HOD predictions for the 1-halo satellite contributions.

\begin{figure*}[!t]
     \centering
     \includegraphics[width=.9\linewidth]{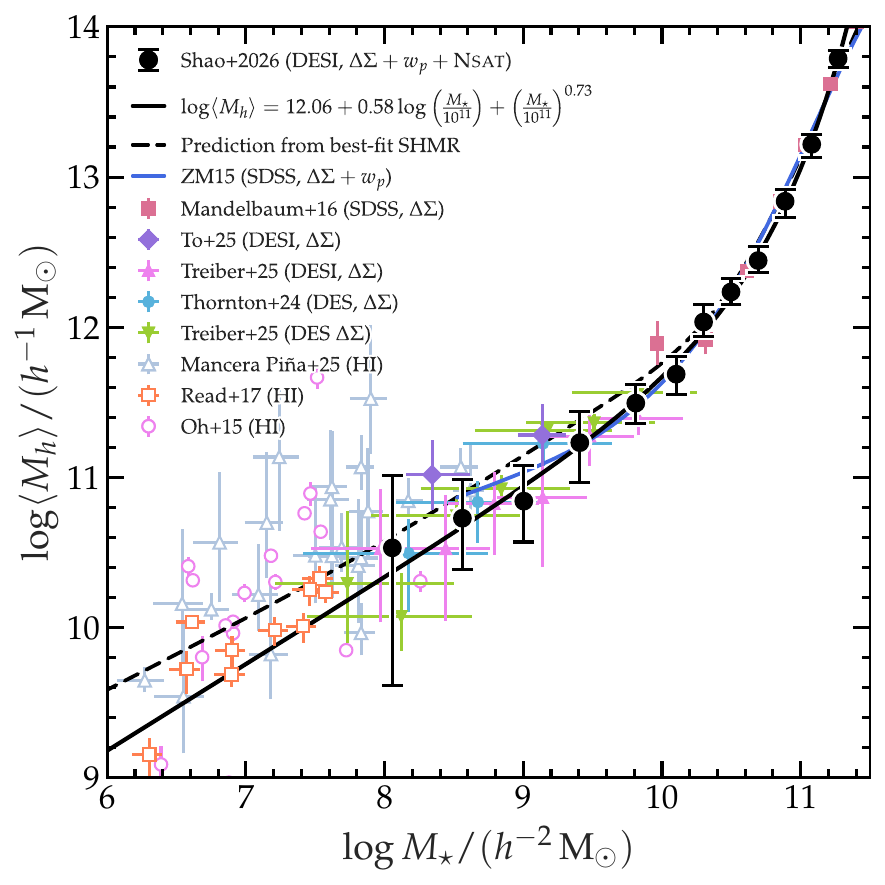}
     \caption{The halo-to-stellar mass relation~(HSMR) derived in this work
     using DESI DR1. Black circles with error bars show our average halo mass
     measurements for the 12 individual stellar mass-binned samples, and the
     black solid curve is an accurate fitting formula to the HSMR. The black
     dashed curve shows the prediction from our best-fitting SHMR presented in
     \S\ref{sec:shmr}, while the blue solid curve is the HSMR prediction from
     \citetalias{ZM2015}. Various filled symbols with error bars include the
     weak lensing constraints from photometric~\citep{Thornton2024,Treiber2025}
     and spectroscopic~\citep{Mandelbaum2016,To2025,Treiber2025} lenses.
     Different open symbols with error bars are the individual halo mass
     measurements for the nearby dwarf galaxies using \hi\ rotation
     curves~\citep{ManceraPina2025,Read2017,Oh2015}.}
     \label{fig:MhMs}
\end{figure*}

For comparison, each colored dotted curve with an uncertainty band is the
$\ds$ profile predicted from our posterior constraint on the average halo
mass $\avgmh$, listed in the top left corner of each panel. The
corresponding halo radius is marked by the vertical dashed line in each
panel. The excellent agreement between the predicted and reconstructed
lensing profiles further validates the internal consistency between the
three observables in our HOD analysis. Focusing on the dwarf galaxies (top
row), we find the large discrepancy between the total (gray) and central (colored)
lensing profiles highlights the challenges of inferring dwarf HSMR using
galaxy--galaxy lensing without the knowledge of satellite HOD.

\subsection{HSMR: Average Halo Masses of Central Galaxies}
\label{subsec:hsmr}

Figure~\ref{fig:MhMs} presents our fiducial mean HSMR, i.e., the average
halo mass measurements for the central galaxies within 12 stellar mass
bins, shown as the black solid circles with error bars.  The mean HSMR
increases monotonically with stellar mass but with a steepening slope at
the high-mass end. We find an accurate fitting formula to describe the
black circles as
\begin{equation}
    \log \hsmr = 12.06+0.58\log\left(\frac{M_\star}{10^{11}}\right)+
    \left(\frac{M_\star}{10^{11}}\right)^{0.73},
    \label{eqn:bestfithsmr}
\end{equation}
which is shown as the solid black curve. In addition to the
phenomenological fit, we also plot a black dashed curve representing the
prediction from our final best-fitting SHMR (to be presented in
\S\ref{sec:shmr}). This prediction incorporates additional information from
the observed stellar mass function down to $\log M_\star=8$, causing the
0.2-dex offset between the two curves below $\log M_\star=10$.
Nevertheless, the SHMR prediction remains consistent with our fiducial HSMR
measurement (black solid circles) within 1$\sigma$.

For the MW-like and massive galaxies, our mean HSMR is consistent with the
results from SDSS, including the HOD constraint of \citetalias{ZM2015}
(blue solid curve) and the weak lensing halo mass\footnote{The original
measurements by \citet{Mandelbaum2016} were for red and blue galaxies
separately, and we use the number-weighted average of the two to derive the
mean HSMR for the overall sample.} measurements by \citet{Mandelbaum2016}
(pink squares with error bars).  This agreement underscores the robustness
of the mean HSMR for galaxies with intermediate-to-high stellar masses.

Entering the dwarf mass range covered by the DESI BGS ($\log
M_\star=7.8$--$9.6$), our mean HSMRs are broadly consistent with the weak
lensing measurements from \citet{To2025} (DESI BGS lenses),
\citet{Thornton2024} (DES photometric lenses), and \citet{Treiber2025}
(DESI spectroscopic + DES photometric lenses). Compared with the results
from photometric lenses \citep{Thornton2024,Treiber2025}, our measurements
have much smaller uncertainties in the mean stellar mass thanks to the
large and homogeneous spectroscopic samples of DESI lenses.  \citet{To2025}
used the same DESI BGS DR1 as in our work, but adopted the newly released
DECADE shear catalog \citep{Anbajagane2025} that has a higher source density
($4.59\,{\rm arcmin^{-2}}$) than the SDSS ($1.2\,{\rm arcmin^{-2}}$).  In
addition, \citet{Treiber2025} combined all available DESI spectroscopic
dwarfs (without specific completeness cuts) and shear catalogs (including
DES, KiDS, and SDSS), aiming at extracting the maximum S/N from the current
dataset.  However, the halo mass uncertainties of our measurements are
similar to \citet{To2025} and \citet{Treiber2025}, thanks to the joint
constraint from $\ds$, $\wpgg$, and \Ns. Our fiducial measurements (black
circles) are consistent with the \citet{Treiber2025} results, but are
slightly lower than \citet{To2025}, likely due to the different treatments
of the satellite contribution to $\ds$. Nevertheless, our black solid and
dashed curves roughly encapsulate the spread between the three sets of weak
lensing measurements.

Extrapolating our HSMR measurement into even lower mass ($\log
M_\star<7.8$), we find that both black curves agree reasonably well with
the individual halo mass measurements (various open symbols) from three
rotation curve studies, including \citet{ManceraPina2025} (triangles),
\citet{Oh2015} (circles), and \citet{Read2017} (squares).  However, despite
covering the same range of stellar mass, the three sets of data points are
segregated in halo mass. In particular, the measurements from
\citet{ManceraPina2025} and \citet{Oh2015} appear to be generally higher
than our extrapolated curves. This can be at least partially explained by
selection effects: the observed galaxies are selected to be \hi-rich
systems which may preferentially reside in more massive halos at fixed
$M_\star$ \citep[e.g.,][]{Lagos2011,Guo2023}.  In contrast,
\citet{Read2017} measured the halo masses of isolated dwarf galaxies, and
their results (open squares) are roughly sandwiched by the extrapolations
of our two black curves.  Therefore, it is unclear whether the
$\avgmh$--$M_\star$ relation between $\log M_\star=6$--$9$ is more elevated
than the extrapolation from DESI dwarfs.

\section{Constraints on the Stellar-to-Halo Mass Relation}
\label{sec:shmr}

In this second part of the paper, we present our constraints on the SHMR
from the fiducial measurements of the mean HSMR, $\hsmr$. The rationale of
our methodology is as follows. The mean HSMR is in essence a measurement of
\begin{equation}
     \hsmr = \frac{\int M_hP(M_h|M_\star)\,\diff M_h}{\int P(M_h|M_\star)\,\diff M_h},
    \label{eq:hsmr}
\end{equation}
where the HSMR is tied to the SHMR via Bayes' theorem,
\begin{equation}
     P(M_h|M_\star) = \frac{\phi(M_h)}{\phi_{\rm cen}(M_\star)}\times P(M_\star|M_h),
\end{equation}
where
\begin{equation}
     \phi_{\rm cen}(M_\star) = \int P(M_\star|M_h)\phi(M_h)\,\diff M_h
    \label{eq:phicen_shmr}
\end{equation}
is the SMF of central galaxies (without any
observational incompleteness). Therefore, by combining our $\hsmr$
measurement with an independent central SMF, we can place strong
constraints on both the mean SHMR ($f_{\rm SHMR}$) and its logarithmic
scatter ($\sigmalogm$).

To model the SHMR, we adopt a flexible functional form of $f_{\rm SHMR}$
proposed by \citet{Behroozi2019},
\begin{equation}
\log (M_\star/M_p)=\epsilon - \log\left(10^{-\alpha x} + 10^{-\beta x}\right) +
          \gamma e^{-\frac{x^2}{2\delta^2}},
\end{equation}
where $x\equiv\log(M_h/M_p)$. This $f_{\rm SHMR}$ is a double power-law
plus a Gaussian at the pivot halo mass $M_p$.

For the scatter about the mean SHMR, \citetalias{ZM2015} found that $\sigmalogm$
is constant for massive halos but starts increasing at $M_p$ towards the
low-mass end (also see \citealt{Moster2018,Behroozi2019,Munshi2021,Tinker2021}).
To fully explore the variation of scatter at the dwarf scale, we parameterize
$\sigmalogm$ as a piecewise-linear function,
\begin{equation}
     \sigmalogm(M_h) =
     \begin{cases}
          \sigma_p, & M_h{\geq}M_p \\
          \sigma_p + \left(\sigma_l-\sigma_p\right)\frac{\log M_h - \log M_p}{\log M_l - \log M_p}, & M_h{<}M_p
     \end{cases},
    \label{eq:siglogm}
\end{equation}
where we set $M_l{=}10^{10.5}\,\hmsun$, and
$\sigma_l$ and $\sigma_p$ are free parameters representing the scatter
at halo masses $M_l$ and $M_p$, respectively.

To estimate the central SMF $\phi_{\rm cen}(M_\star)$ in observations, we adopt
the recent measurement of the total SMF $\phi_{\rm tot}(M_\star)$ from
\citet{Xu2025}. Leveraging the cross-correlation between DESI BGS galaxies and a
much fainter DESI imaging sample, \citet{Xu2025} used the Photometric objects
Around Cosmic webs (PAC) method to measure the total SMF in a much larger volume
than that covered by current spectroscopic surveys. In the stellar mass range of our
interest ($\log M_\star{\geq}7.8$), their measurements cover a volume of
$200^3~h^{-3}{\rm Mpc}^3$ (for comparison, BGS Bright only covers
$50^3~h^{-3}{\rm Mpc}^3$), thereby significantly reducing the systematics from
cosmic variance \citep{Chen2019}.  We then combine $\phi_{\rm tot}$ with our
measurements of satellite fraction $\fsat(M_\star)$ to estimate
\begin{equation}
     \phi_{\rm cen}(M_\star) = \phi_{\rm tot}(M_\star)\times\left[1-\fsat(M_\star)\right],
    \label{eq:phicen}
\end{equation}
while properly taking into account the measurement uncertainties in both
$\phi_{\rm tot}$ and $\fsat$. In particular, we take the uncertainty of $\fsat$
derived from the posterior distribution of our HOD parameters, and add an
additional 30\% fractional error in quadrature to account for the systematic
uncertainties in the measurements of SMF and $\fsat$.

For each of the 12 stellar mass bins ($M_\star^1{\leq} M_\star{<}M_\star^2$), we
predict its $\phi_{\rm cen}$ and $\avgmh$
from the SHMR by extending Equations~\ref{eq:hsmr} and~\ref{eq:phicen_shmr},
\begin{equation}
     \phi_{\rm cen} =
     \int_{M_\star^1}^{M_\star^2}\phi_{\rm cen}(M_\star)\,\diff M_\star
\end{equation}
and
\begin{equation}
     \avgmh = \frac{\int M_hP(M_h|M_\star^1{\leq} M_\star{<}M_\star^2)\,\diff M_h}
     {\int P(M_h|M_\star^1{\leq} M_\star{<}M_\star^2)\,\diff M_h},
\end{equation}
where
\begin{equation}
     P(M_h|M_\star^1{\leq} M_\star{<}M_\star^2) =
     \frac{\phi(M_h)\int_{M_\star^1}^{M_\star^2}P(M_\star|M_h)\,\diff M_\star}
     {\phi_{\rm cen}}.
\end{equation}
The predicted $\phi_{\rm cen}$ and $\avgmh$ are then compared to our fiducial
mean HSMR and central SMF measurements. Similarly, we adopt a Bayesian inference
framework to constrain the model parameters from this comparison. We adopt a
Gaussian likelihood model and the uncertainties of the $\avgmh$ measurements are
estimated from the 1$\sigma$ posterior constraints from the HOD analyses.
Unlike the first part of the paper in which we treat each stellar mass bin
separately, for the SHMR constraint we fit the mean HSMR and central SMF jointly
over the 12 stellar mass bins. We have nine free parameters in total, including
six for the mean SHMR \{$M_p$, $\epsilon$, $\alpha$, $\beta$, $\gamma$,
$\delta$\}, two for the scatter \{$\sigma_l$, $\sigma_p$\}, and one nuisance
parameter $\delta\log M_\star$ to account for any systematic offset in the
stellar mass estimates between \citet{Xu2025} and ours.  During the inference,
we adopt Gaussian priors $\sigma_p\sim\mathcal{N}(0.2, 0.02)$ and $\delta\log
M_\star\sim\mathcal{N}(0, 0.05)$ based on the results from \citetalias{ZM2015},
and uniform priors for the other parameters.

\begin{figure*}[!t]
     \centering
     \includegraphics[width=.8\linewidth]{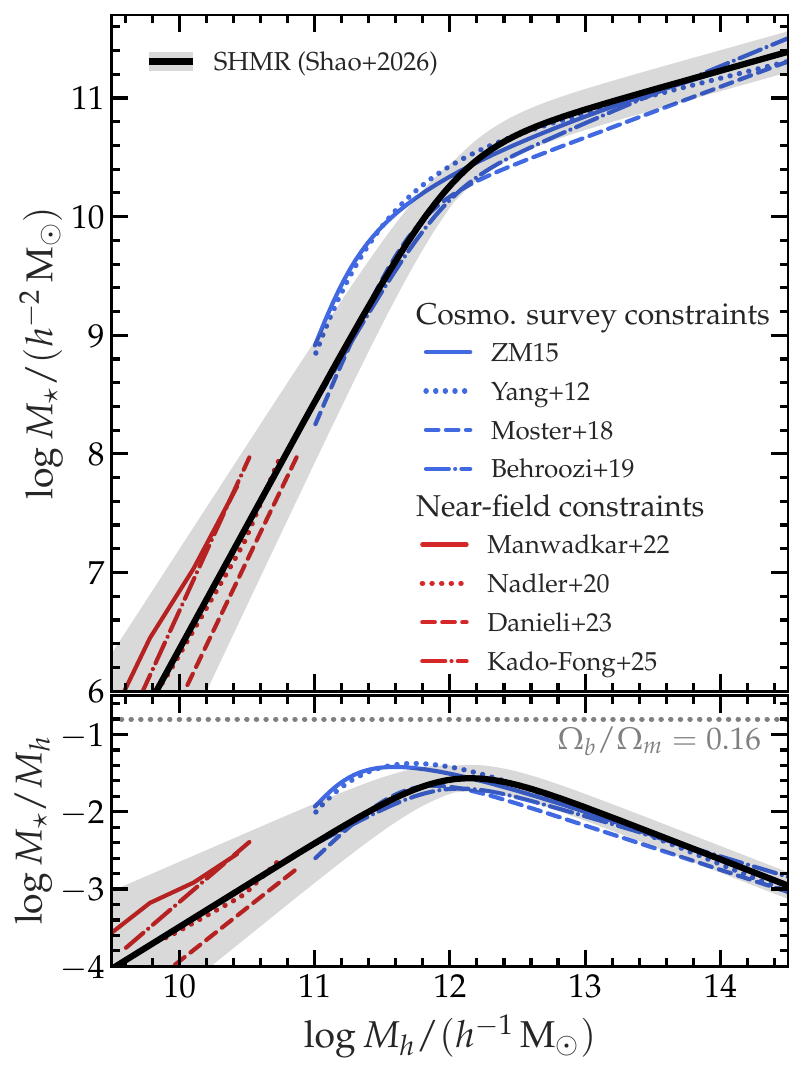}
     \caption{The stellar-to-halo mass relation (SHMR; top) and the
     stellar-to-halo mass ratio (bottom) derived in this work using DESI DR1. In
     each panel, the black solid curve with gray shaded band indicates the
     best-fitting mean SHMR and the scatter.
     For comparison, the four blue curves at
     $M_h>10^{11}\,\hmsun$ indicate four previous SHMR constraints using
     cosmological surveys, including the SDSS \texttt{iHOD} analysis~(blue
     solid; \citetalias{ZM2015}), conditional luminosity function~\citep[blue
     dotted;][]{Yang2012}, UniverseMachine~\citep[blue
     dot-dashed;][]{Behroozi2019}, and EMERGE~\citep[blue
     dashed;][]{Moster2018}. The four red curves at $M_\star{<}10^8\,\hsqmsun$
     are recent near-field constraints from \citet[red solid;][]{Manwadkar2022},
     \citet[red dotted;][]{Nadler2020}, \citet[red dashed;][]{Danieli2023}, and
     \citet[red dot-dashed;][]{Kado-Fong2025}.
     The horizontal dotted line in the bottom panel indicates the cosmic mean
     baryon fraction from \citet{Planck18}.}
     \label{fig:SHMR}
\end{figure*}

\begin{figure*}[!t]
     \centering \includegraphics[width=.99\linewidth]{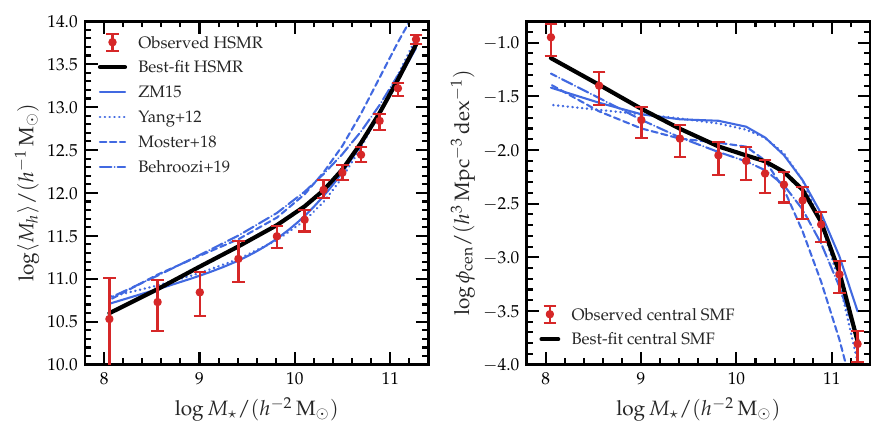}
     \caption{Comparison between the observed and best-fitting HSMR~(left
     panel) and central galaxy stellar mass functions~(right panel). In
     each panel, red circles with error bars are the observations and the black
     solid curve is the prediction from the best-fitting SHMR in
     Figure~\ref{fig:SHMR}. Predictions from other cosmological survey
     constraints are also shown in the same styles as Figure~\ref{fig:SHMR}
     for comparison.}
     \label{fig:HSMR_SMF}
\end{figure*}

\begin{deluxetable*}{ccccccccc}[!t]
\tablewidth{0pt}
\tablecaption{Posterior ranges of model parameters in the stellar-to-halo mass relation.}
\label{tab:shmr_params}
\tablehead{
\colhead{$\log M_p$} &
\colhead{$\epsilon$} &
\colhead{$\alpha$} &
\colhead{$\beta$} &
\colhead{$\log\gamma$} &
\colhead{$\delta$} &
\colhead{$\sigma_l$} &
\colhead{$\sigma_p$} &
\colhead{$\delta \log M_\star$}
}
\startdata
$12.03_{-0.13}^{+0.13}$ &
$-1.59_{-0.06}^{+0.05}$ &
$2.08_{-0.19}^{+0.22}$ &
$0.32_{-0.06}^{+0.06}$ &
$-2.80_{-1.47}^{+1.51}$ &
$1.03_{-0.63}^{+0.66}$ &
$0.68_{-0.33}^{+0.21}$ &
$0.17_{-0.02}^{+0.02}$ &
$0.13_{-0.03}^{+0.03}$ \\
\enddata
\end{deluxetable*}

We present our final constraint on the SHMR and the stellar-to-halo mass ratio
in the top and bottom panels of Figure~\ref{fig:SHMR}, respectively. The median
and 1$\sigma$ constraints of the SHMR parameters are presented in
Table~\ref{tab:shmr_params}. Exploiting the maximum constraining power from the
DESI BGS, our posterior median predictions for the mean SHMR (thick black solid
curve) and logarithmic scatter (gray shaded band) cover almost five orders of
magnitude in both the stellar and halo masses, extending a robust galaxy--halo
relation into the dwarf galaxy regime.  Our posterior mean SHMR is characterized
by a power-law slope change from $0.32{\pm}0.06$ above the MW mass to
$2.08{\pm}0.21$ in the dwarf regime. This suggests that as halo mass decreases
below the MW mass, the average star formation efficiency continues to decline at
a steady rate, consistent with the rapid increase of the mass-loading factor of
galactic outflows~\citep{Lin2023}.

In addition, we compare our mean SHMR with various constraints derived from
other studies using cosmological survey data, including the HOD constraint from
SDSS \citepalias[blue solid curve;][]{ZM2015}, the conditional luminosity
function (CLF) constraint from SDSS \citep[blue dotted curve;][]{Yang2012}, and
two empirical galaxy formation models based on SHAM, including EMERGE
\citep[blue dashed curve;][]{Moster2018} and UniverseMachine \citep[blue
dot-dashed curve;][]{Behroozi2019}. These mean SHMRs are roughly consistent with
one another for galaxies above MW halo masses, but they differ significantly in
the range $\log M_h=11$--$12$, leading to a divergent sequence~(not shown) when
extrapolated to $\log M_h<11$. The two panels of Figure~\ref{fig:HSMR_SMF} show
the comparison between posterior median model predictions (black solid curve)
and observations (red solid circles with error bars) for the mean HSMR (left
panel) and central SMF (right panel), respectively. The various other curves are
predicted from the four SHMRs derived from cosmological surveys, with their line
styles matched to those shown in the Figure~\ref{fig:SHMR}.

At $\log M_\star<8$, we compare our SHMR constraint with the various mean
SHMRs derived from near-field studies, including the constraints from MW
satellites by \citet{Manwadkar2026} (red solid curve) and
\citet{Nadler2020} (red dotted curve), as well as from galaxies in the
local volume by \citet[ELVES;][]{Danieli2023} and
\citet[SAGAbg;][]{Kado-Fong2025} (red dashed and red dot-dashed curves,
respectively).

Although the four near-field constraints differ in the amplitudes of SHMR,
they exhibit very similar slopes $M_\star\propto M_h^2$.  Interestingly,
the slope of our mean SHMR is in excellent agreement with the four
near-field constraints, demonstrating a strong coherence in the galaxy
formation efficiency from MW-like systems down to dwarfs.  More important,
while the systematic offsets among the near-field constraints are likely
dominated by cosmic variance (including halo-to-halo variance), the
amplitude of our mean SHMR from DESI falls within the median range of the
four curves. The strong overall consistency between our DESI constraints
and the near-field SHMRs is highly encouraging, demonstrating a successful
bridging of the methodological gap between large-scale surveys and
near-field studies.

Interestingly, while the mean SHMR maintains a roughly constant slope at the
low-mass end, the scatter about the SHMR grows from $0.17{\pm}0.02$ dex at
MW-like scales to $0.68_{-0.33}^{+0.21}$ dex for LMC-like dwarfs. This suggests
that the pathways of galaxy formation become increasingly diverse among
lower-mass dark matter halos, consistent with the predictions from
semi-empirical models~\citep{Kim2026} and hydro simulations~\citep{Cruz2025} of
dwarf galaxies.  However, one caveat is that our constraining power on the
scatter is still relatively weak, and the increase of scatter could be
degenerate with a second slope change of the mean SHMR for LMC-like systems. For
example, \citet{Xu2026} recently found the mean SHMR favors an upturn at the LMC
scale with the scatter being roughly constant. We will examine this particular
degeneracy with the upcoming final data release of DESI.

\section{Conclusion}
\label{sec:conclusion}

In this work, we derive stringent constraints on the galaxy-halo connection
using DESI DR1 BGS galaxies with stellar mass
$M_\star>10^{7.8}\,\hsqmsun$ at $0.01<z<0.2$. After dividing these
galaxies into 12 $M_\star$-binned samples, we measure three types of
observables for each sample: two conventional two-point
statistics~(projected clustering $w_p$ and galaxy--galaxy lensing $\ds$)
and a novel observable, the satellite HOD~(\Ns).  We then perform a
comprehensive suite of HOD analyses over these observables and measure the
average halo mass $\langle M_h \rangle$ of central galaxies for each of the
12 galaxy samples.

To enhance our constraining power on dwarf galaxy scales, we include all
observed galaxies using a proper $1/V_{\max}$ weighting scheme, thereby
improving the S/N of dwarf lensing by 50\%--140\% compared to the
traditional volume-limited samples. More important, we apply the method of
\citet{Shao2025} to directly measure the \Ns\ of DESI BGS galaxies using
their cross-correlation with the Y21 DESI group catalog. The addition of
\Ns\ enables robust subtraction of satellite contributions to both $w_p$
and $\ds$ during the individual HOD analyses, yielding an average
halo-to-stellar mass relation~(HSMR) of
\begin{equation}
    \log \langle M_h \rangle = 12.06+0.58\log\left(\frac{M_\star}{10^{11}}\right)+
    \left(\frac{M_\star}{10^{11}}\right)^{0.73}.
\end{equation}

Combining this HSMR with the stellar mass function measured by
\citet{Xu2025}, we then place tight constraints on the stellar-to-halo
mass relation~(SHMR) down to the LMC scale,
bridging the methodological gap between large-scale and near-field
studies. Our best-fitting SHMR is characterized by a power-law slope change
from $0.32{\pm}0.06$ above the MW mass to $2.08{\pm}0.21$ in the
dwarf regime. Meanwhile, the scatter about the SHMR grows from
$0.17{\pm}0.02$ dex at MW-like scales to $0.68_{-0.33}^{+0.21}$ dex
for systems comparable to the LMC. This increasing scatter towards the
low-mass end likely signifies the diverse evolutionary paths that dwarf
galaxies have taken since reionization~\citep{Kim2026}, which may further
require a lower level of stellar feedback under
$\Lambda$CDM~\citep{Cruz2025} or a more exotic type of dark
matter~\citep[e.g., self-interacting;][]{Zeng2025}.

Our work demonstrates the efficacy of DESI for constraining the
galaxy--halo connection in the dwarf galaxy regime, owing to its
unprecedented survey depth and volume. In the near future, our method can
be directly applied to the upcoming final data release of DESI, which will
include a five-times-larger sample of dwarf galaxies. This will further pin
down the slope and scatter of the SHMR at the LMC scale and, more
importantly, uncover the dependence of the dwarf--halo connection on the
environment~\citep{Bhattacharya2025} and galaxy properties~(e.g., size and
morphology). Meanwhile, ongoing and future imaging surveys by
ground-based telescopes and space missions, including the Vera C.  Rubin
Observatory~\citep[LSST;][]{LSST2019}, Nancy Grace Roman Space
Telescope~\citep[{\it Roman};][]{Roman2019}, {\it
Euclid}~\citep{Euclid2025}, and China Space Station Telescope~\citep[{\it
CSST};][]{CSST2025}, will reduce the uncertainties of the weak lensing halo
mass for dwarf galaxies by at least a factor of several. The improved weak
lensing signals of galaxy groups will also significantly increase the
accuracy of the \Ns\ measurements, further tightening the SHMR. In the longer
term, DESI~Run~2~\citep{DESI2}, Spec-S5~\citep{SpecS5}, MUltiplexed Survey
Telescope~\citep[MUST;][]{MUST2024}, and Jiaotong University Spectroscopic
Telescope~\citep[JUST;][]{JUST2024} will provide a richer sample of dwarf
galaxies down to $M_\star\sim10^6\,\msun$~\citep{Nadler2024}. With
next-generation spectroscopic and imaging surveys, our method provides a
sharp probe of the threshold of galaxy formation physics~\citep{Simon2019,
Fitts2017, Sales2022}, which is modulated by the very nature of dark
matter~\citep{Governato2015, Tulin2018, Adhikari2025}.

\begin{acknowledgments}
We thank Zheng Zheng and Yifei Luo for the DESI internal review.
Z.S. thanks Rachel Mandelbaum and Kun Xu for helpful
comments and discussions. This work is
supported by the National Key Basic Research and Development Program of
China (No. 2023YFA1607800, 2023YFA1607804), the National Natural Science
Foundation of China (12595313, 12173024), and the China Manned Space
Program (No. CMS-CSST-2025-A04).  Z.S. is supported by the T.D. Lee
scholarship. Y.Z. acknowledges the generous sponsorship from Yangyang
Development Fund.  Y.Z. thanks Cathy Huang for her hospitality at the
Zhangjiang High-tech Park.

This material is based upon work supported by the U.S. Department of Energy
(DOE), Office of Science, Office of High-Energy Physics, under Contract No.
DE-AC02-05CH11231, and by the National Energy Research Scientific Computing
Center, a DOE Office of Science User Facility under the same contract.
Additional support for DESI was provided by the U.S. National Science Foundation
(NSF), Division of Astronomical Sciences under Contract No. AST-0950945 to the
NSF's National Optical-Infrared Astronomy Research Laboratory; the Science and
Technology Facilities Council of the United Kingdom; the Gordon and Betty Moore
Foundation; the Heising-Simons Foundation; the French Alternative Energies and
Atomic Energy Commission (CEA); the Secretariat of Science, Humanities,
Technology and Innovation (SECIHTI) of Mexico; the Ministry of Science,
Innovation and Universities of Spain (MICIU/AEI/10.13039/501100011033), and by
the DESI Member Institutions:
\url{https://www.desi.lbl.gov/collaborating-institutions}. Any opinions,
findings, and conclusions or recommendations expressed in this material are
those of the author(s) and do not necessarily reflect the views of the U. S.
National Science Foundation, the U. S. Department of Energy, or any of the
listed funding agencies.

The authors are honored to be permitted to conduct scientific research on
I'oligam Du'ag (Kitt Peak), a mountain with particular significance to the
Tohono O'odham Nation.

This work made use of the Gravity Supercomputer at the Department of
Astronomy, Shanghai Jiao Tong University.

The data points corresponding to the most relevant figures in this paper
are available on Zenodo at DOI: \url{https://doi.org/10.5281/zenodo.22234808}.
\end{acknowledgments}

%



\software{NumPy \citep{numpy2020},
          SciPy \citep{scipy2020},
          Astropy \citep{astropy2013,astropy2018,astropy2022},
          dsigma \citep{dsigma2022},
          nautilus \citep{Nautilus2023}.
          }



\appendix
\twocolumngrid

\section{Validation of $\vmax$ weighting}
\label{appendix:vmax}

\begin{figure*}[!t]
     \centering
     \includegraphics[width=.9\linewidth]{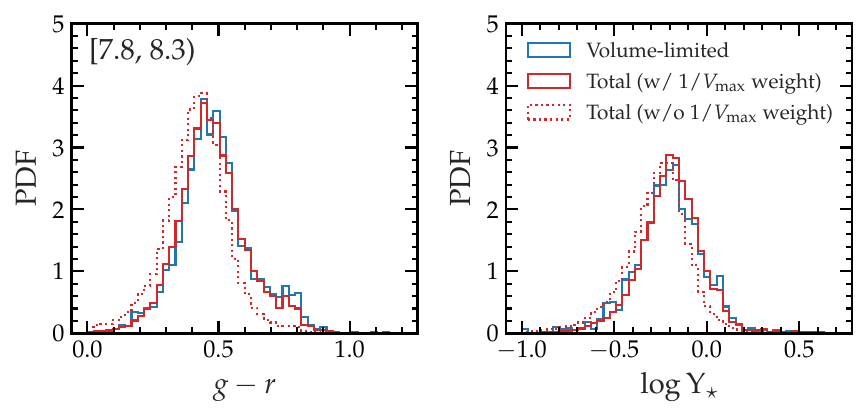}
     \caption{Comparison of the $g-r$ color distribution (left) and the
     mass-to-light ratio distribution (right) between the volume-limited sample
     (blue solid curve) and the total sample with $1/\vmax$ weights (red solid
     curve), for the $\log M_\star{=}7.8-8.3$ sample. The distributions from the
     total sample without $1/\vmax$ weights are also shown as the red dotted
     curves.}
     \label{fig:pdf_color_upsilon}
\end{figure*}

As shown in Figure~\ref{fig:pdf_color_upsilon}, when the $1/\vmax$ weights
are applied, both the $g-r$ color distribution (left) and the mass-to-light
ratio distribution (right) of the total sample (red solid curves) are
identical to those of the volume-limited sample (blue solid curve).  In
contrast, without the $1/\vmax$ weights (red dotted curves), the total
sample is biased towards bluer and lower mass-to-light ratio galaxies,
which is expected for a flux-limited sample.  This validation confirms that
our $1/\vmax$ weighting scheme effectively corrects for the selection bias
in the flux-limited sample, allowing us to recover the intrinsic
distributions of galaxy properties as if we were working with a
volume-limited sample.

\section{Measurement of \Ns}
\label{appendix:sathod}

\begin{figure*}
     \centering
     \includegraphics[width=.99\linewidth]{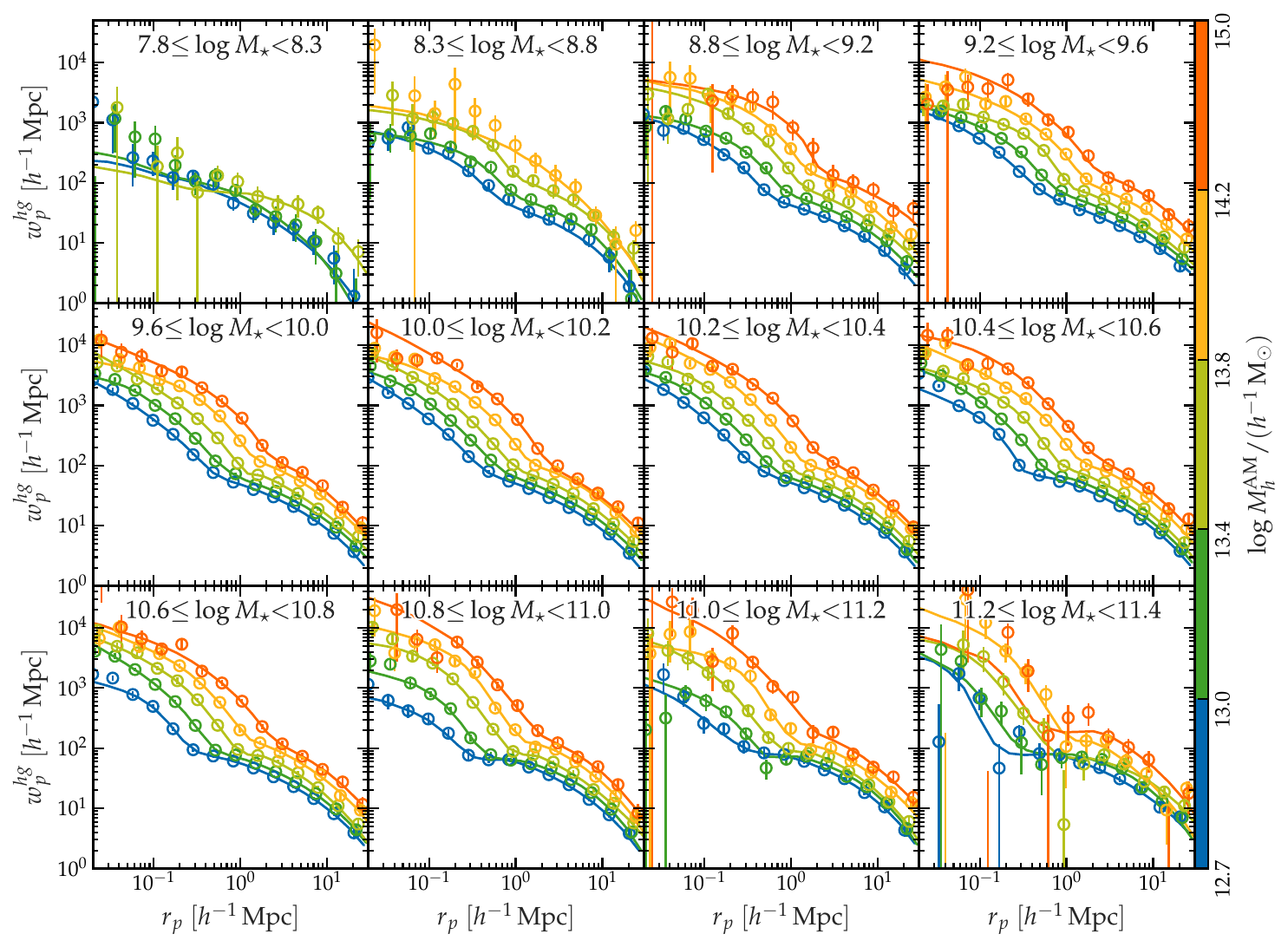}
     \caption{Projected halo-galaxy cross-correlation functions $\wphg$
     between the five Y21 DESI group samples~(different colors) and 12 stellar
     mass-binned galaxy samples~(different panels).  In each panel, circles
     with error bars are the measurements and solid curves are the
     best-fitting predictions. The AM halo mass ranges of the five
     Y21 group samples are color-coded according to the colorbar on the
     right.}
     \label{fig:crosswp}
\end{figure*}

\begin{figure*}
     \centering
     \includegraphics[width=.99\linewidth]{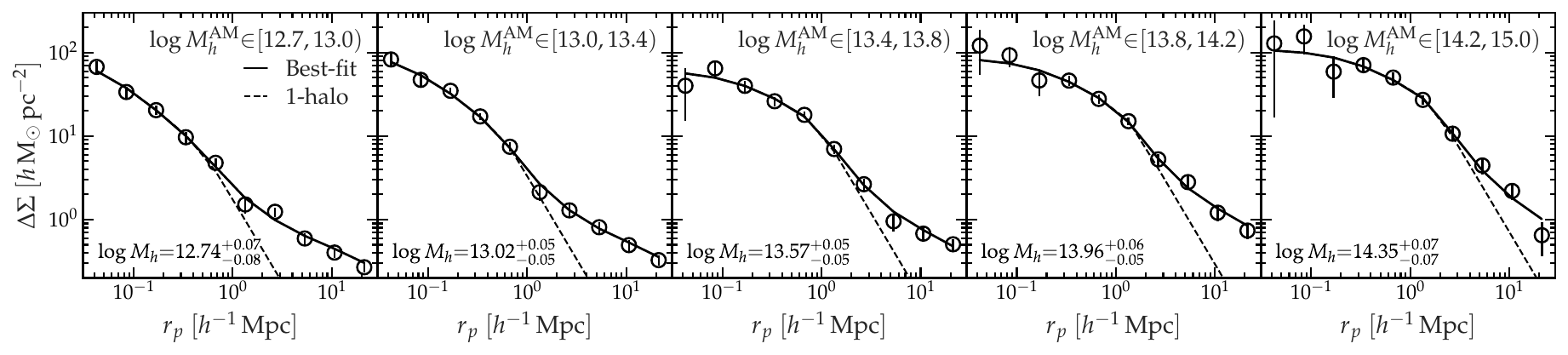}
     \caption{Comparison between the observed weak lensing
     profiles~(circles with error bars) and the best-fitting model
     predictions (solid curves) for the five Y21 DESI group samples~(different
     panels).  We list the weak lensing halo mass constraint in the bottom
     left corner of each panel, with the dashed curve showing the contribution
     from the 1-halo term.}
     \label{fig:groupmass}
\end{figure*}

Figures~\ref{fig:crosswp} and \ref{fig:groupmass} present our measurements
of the cluster--galaxy cross-correlations and the cluster weak lensing
profiles, whereas the values of $\log \Nsat$ and $\mhwl$ are summarized in
Table~\ref{tab:sathod}.

\begin{deluxetable}{lccccc}
\tablecaption{Measurements of $\log \Nsat$ for different galaxy samples and
halo mass bins.  The second row of the header lists the average halo mass
$\mhwl$ of each group sample measured from weak lensing.}\label{tab:sathod}
\tablehead{
\colhead{\shortstack{$\log M_\star$ \\ $ $}} &
\colhead{\shortstack{$M_h^{\rm AM}\in[12.7, 13.0)$ \\ $\log M_h{=}12.74{\pm}0.07$}} &
\colhead{\shortstack{$M_h^{\rm AM}\in[13.0, 13.4)$ \\ $\log M_h{=}13.02{\pm}0.05$}} &
\colhead{\shortstack{$M_h^{\rm AM}\in[13.4, 13.8)$ \\ $\log M_h{=}13.57{\pm}0.05$}} &
\colhead{\shortstack{$M_h^{\rm AM}\in[13.8, 14.2)$ \\ $\log M_h{=}13.96{\pm}0.06$}} &
\colhead{\shortstack{$M_h^{\rm AM}\in[14.2, 15.0)$ \\ $\log M_h{=}14.35{\pm}0.07$}}
}
\startdata
$7.8$--$8.3$   & $-0.13\pm0.30$ & $0.02\pm0.41$  & $-0.11\pm0.50$ & \nodata          & \nodata \\
$8.3$--$8.8$   & $0.24\pm0.08$  & $0.48\pm0.08$  & $1.12\pm0.12$  & $1.15\pm0.50$  & \nodata \\
$8.8$--$9.2$   & $0.11\pm0.06$  & $0.42\pm0.06$  & $1.02\pm0.08$  & $1.38\pm0.09$  & $1.74\pm0.20$ \\
$9.2$--$9.6$   & $-0.04\pm0.05$ & $0.31\pm0.05$  & $0.79\pm0.06$  & $1.15\pm0.07$  & $1.58\pm0.09$ \\
$9.6$--$10.0$  & $-0.04\pm0.05$ & $0.27\pm0.05$  & $0.73\pm0.05$  & $1.11\pm0.05$  & $1.47\pm0.06$ \\
$10.0$--$10.2$ & $-0.42\pm0.05$ & $-0.10\pm0.05$ & $0.31\pm0.05$  & $0.72\pm0.05$  & $1.15\pm0.05$ \\
$10.2$--$10.4$ & $-0.49\pm0.05$ & $-0.16\pm0.05$ & $0.26\pm0.05$  & $0.70\pm0.05$  & $1.09\pm0.06$ \\
$10.4$--$10.6$ & $-0.67\pm0.05$ & $-0.26\pm0.05$ & $0.14\pm0.05$  & $0.55\pm0.05$  & $0.91\pm0.06$ \\
$10.6$--$10.8$ & $-1.06\pm0.06$ & $-0.48\pm0.05$ & $-0.01\pm0.05$ & $0.35\pm0.05$  & $0.74\pm0.06$ \\
$10.8$--$11.0$ & $-1.66\pm0.09$ & $-1.04\pm0.06$ & $-0.34\pm0.05$ & $0.09\pm0.05$  & $0.50\pm0.08$ \\
$11.0$--$11.2$ & $-2.02\pm0.14$ & $-1.66\pm0.10$ & $-1.00\pm0.08$ & $-0.52\pm0.08$ & $0.03\pm0.09$ \\
$11.2$--$11.4$ & $-2.72\pm0.62$ & $-2.43\pm0.43$ & $-1.84\pm0.33$ & $-1.10\pm0.16$ & $-1.56\pm0.71$
\enddata
\end{deluxetable}

\section{Influence of $\ngobs$}
\label{appendix:ngobs}

\begin{figure*}
     \centering
     \includegraphics[width=.95\linewidth]{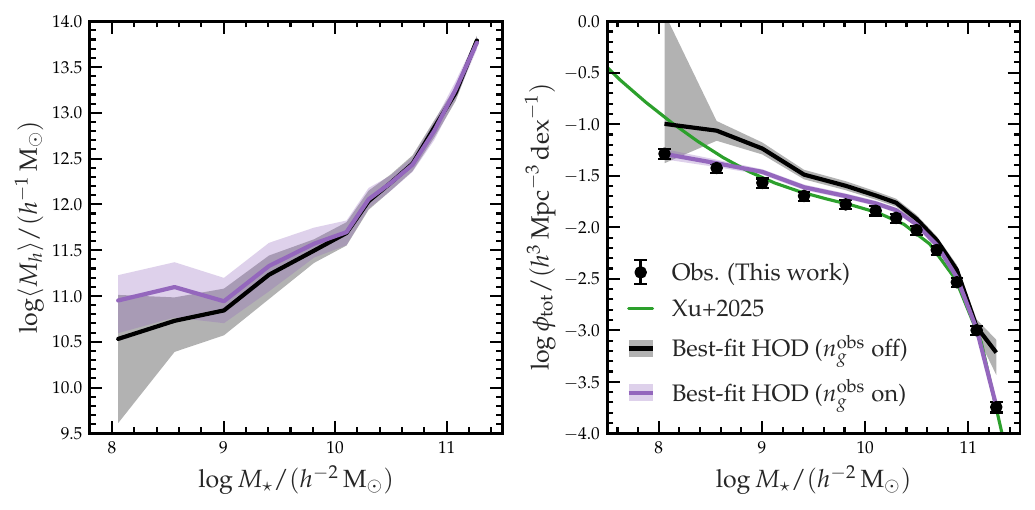}
     \caption{Constraints on the $\hsmr$~(left) and the SMF~(right), where the
     galaxy number density $\ngobs$ is included (purple) or excluded (black)
     during the HOD inference, respectively. The 1$\sigma$ uncertainties are
     also shown in shaded regions of corresponding colors. The green solid curve
     in the right panel is the best-fitting Schechter function to the
     photometric stellar mass function measured by \citet{Xu2025}.}
     \label{fig:constraint_with_ngobs}
\end{figure*}

This appendix presents how the inclusion of observed galaxy abundance
($\ngobs$) will influence our HOD constraints.
Figure~\ref{fig:constraint_with_ngobs} compares the predicted $\hsmr$ (left) and
SMF (right) from the best-fitting HOD model when excluding (black circles with
error bars) and including (purple circles with error bars) $\ngobs$ during the
inference.  Despite the fact that predictions from including $\ngobs$ match the
observed SMF (black solid curve with gray shaded region) better, the predicted
$\fsat$ and $\hsmr$ are generally consistent with each other, indicating the
robustness of our main results against the inclusion of $\ngobs$ in the HOD
constraint.

\clearpage
\bibliography{shmr}{}

@ARTICLE{Lin2023,
       author = {{Lin}, Yuanye and {Zu}, Ying},
        title = "{Constraints on galactic outflows from the metallicity-stellar mass-SFR relation of EAGLE simulation and SDSS galaxies}",
      journal = {\mnras},
         year = 2023,
        month = may,
       volume = {521},
       number = {1},
        pages = {411-432},
          doi = {10.1093/mnras/stad502},
archivePrefix = {arXiv},
       eprint = {2212.01402},
 primaryClass = {astro-ph.GA},
       adsurl = {https://ui.adsabs.harvard.edu/abs/2023MNRAS.521..411L}
}

@ARTICLE{DiValentino2026,
       author = {{Di Valentino}, Eleonora and {Said}, Jackson Levi and {Saridakis}, Emmanuel N.},
        title = "{Tensions in Cosmology 2025}",
      journal = {Nature Astronomy},
         year = 2026,
        month = feb,
       volume = {10},
        pages = {180-182},
          doi = {10.1038/s41550-026-02781-1},
archivePrefix = {arXiv},
       eprint = {2509.25288},
 primaryClass = {astro-ph.CO},
       adsurl = {https://ui.adsabs.harvard.edu/abs/2026NatAs..10..180D}
}

@ARTICLE{Mandelbaum2018,
       author = {{Mandelbaum}, Rachel},
        title = "{Weak Lensing for Precision Cosmology}",
      journal = {\araa},
         year = 2018,
        month = sep,
       volume = {56},
        pages = {393-433},
          doi = {10.1146/annurev-astro-081817-051928},
archivePrefix = {arXiv},
       eprint = {1710.03235},
 primaryClass = {astro-ph.CO},
       adsurl = {https://ui.adsabs.harvard.edu/abs/2018ARA&A..56..393M}
}

@ARTICLE{Sales2022,
       author = {{Sales}, Laura V. and {Wetzel}, Andrew and {Fattahi}, Azadeh},
        title = "{Baryonic solutions and challenges for cosmological models of dwarf galaxies}",
      journal = {Nature Astronomy},
         year = 2022,
        month = jun,
       volume = {6},
        pages = {897-910},
          doi = {10.1038/s41550-022-01689-w},
archivePrefix = {arXiv},
       eprint = {2206.05295},
 primaryClass = {astro-ph.GA},
       adsurl = {https://ui.adsabs.harvard.edu/abs/2022NatAs...6..897S}
}

@ARTICLE{Zeng2025,
       author = {{Zeng}, Zhichao Carton and {Peter}, Annika H.~G. and {Du}, Xiaolong and {Benson}, Andrew and {Li}, Jiaxuan and {Mace}, Charlie and {Yang}, Shengqi},
        title = "{Diversity and universality: Evolution of dwarf galaxies with self-interacting dark matter}",
      journal = {\prd},
         year = 2025,
        month = sep,
       volume = {112},
       number = {6},
          eid = {063008},
        pages = {063008},
          doi = {10.1103/x9t4-3zy7},
archivePrefix = {arXiv},
       eprint = {2412.14621},
 primaryClass = {astro-ph.GA},
       adsurl = {https://ui.adsabs.harvard.edu/abs/2025PhRvD.112f3008Z}
}

@ARTICLE{Bhattacharya2025,
       author = {{Bhattacharyya}, Joy and {Peter}, Annika H.~G. and {Leauthaud}, Alexie},
        title = "{Dwarf Galaxies in the TNG50 Field: connecting their Star-formation Rates with their Environments}",
      journal = {The Open Journal of Astrophysics},
         year = 2025,
        month = apr,
       volume = {8},
          eid = {43},
        pages = {43},
          doi = {10.33232/001c.137130},
archivePrefix = {arXiv},
       eprint = {2501.01946},
 primaryClass = {astro-ph.GA},
       adsurl = {https://ui.adsabs.harvard.edu/abs/2025OJAp....8E..43B}
}

@ARTICLE{Cruz2025,
       author = {{Cruz}, Akaxia and {Brooks}, Alyson and {Lisanti}, Mariangela and {Peter}, Annika H.~G. and {Geda}, Robel and {Quinn}, Thomas and {Tremmel}, Michael and {Munshi}, Ferah and {Keller}, Ben and {Wadsley}, James},
        title = "{Dwarf diversity in $Λ$CDM with baryons}",
      journal = {arXiv e-prints},
         year = 2025,
        month = oct,
          eid = {arXiv:2510.11800},
        pages = {arXiv:2510.11800},
          doi = {10.48550/arXiv.2510.11800},
archivePrefix = {arXiv},
       eprint = {2510.11800},
 primaryClass = {astro-ph.GA},
       adsurl = {https://ui.adsabs.harvard.edu/abs/2025arXiv251011800C}
}

@ARTICLE{Governato2015,
       author = {{Governato}, F. and {Weisz}, D. and {Pontzen}, A. and {Loebman}, S. and {Reed}, D. and {Brooks}, A.~M. and {Behroozi}, P. and {Christensen}, C. and {Madau}, P. and {Mayer}, L. and {Shen}, S. and {Walker}, M. and {Quinn}, T. and {Keller}, B.~W. and {Wadsley}, J.},
        title = "{Faint dwarfs as a test of DM models: WDM versus CDM}",
      journal = {\mnras},
         year = 2015,
        month = mar,
       volume = {448},
       number = {1},
        pages = {792-803},
          doi = {10.1093/mnras/stu2720},
archivePrefix = {arXiv},
       eprint = {1407.0022},
 primaryClass = {astro-ph.GA},
       adsurl = {https://ui.adsabs.harvard.edu/abs/2015MNRAS.448..792G}
}

@ARTICLE{Adhikari2025,
       author = {{Adhikari}, Susmita and {Banerjee}, Arka and {Boddy}, Kimberly K. and {Cyr-Racine}, Francis-Yan and {Desmond}, Harry and {Dvorkin}, Cora and {Jain}, Bhuvnesh and {Kahlhoefer}, Felix and {Kaplinghat}, Manoj and {Nierenberg}, Anna and {Peter}, Annika H.~G. and {Robertson}, Andrew and {Sakstein}, Jeremy and {Zavala}, Jes{\'u}s},
        title = "{Astrophysical tests of dark matter self-interactions}",
      journal = {Reviews of Modern Physics},
         year = 2025,
        month = oct,
       volume = {97},
       number = {4},
          eid = {045004},
        pages = {045004},
          doi = {10.1103/m2vm-59y3},
archivePrefix = {arXiv},
       eprint = {2207.10638},
 primaryClass = {astro-ph.CO},
       adsurl = {https://ui.adsabs.harvard.edu/abs/2025RvMP...97d5004A}
}

@ARTICLE{Tulin2018,
       author = {{Tulin}, Sean and {Yu}, Hai-Bo},
        title = "{Dark matter self-interactions and small scale structure}",
      journal = {\physrep},
         year = 2018,
        month = feb,
       volume = {730},
        pages = {1-57},
          doi = {10.1016/j.physrep.2017.11.004},
archivePrefix = {arXiv},
       eprint = {1705.02358},
 primaryClass = {hep-ph},
       adsurl = {https://ui.adsabs.harvard.edu/abs/2018PhR...730....1T}
}

@ARTICLE{Fitts2017,
       author = {{Fitts}, Alex and {Boylan-Kolchin}, Michael and {Elbert}, Oliver D. and {Bullock}, James S. and {Hopkins}, Philip F. and {O{\~n}orbe}, Jose and {Wetzel}, Andrew and {Wheeler}, Coral and {Faucher-Gigu{\`e}re}, Claude-Andr{\'e} and {Kere{\v{s}}}, Du{\v{s}}an and {Skillman}, Evan D. and {Weisz}, Daniel R.},
        title = "{fire in the field: simulating the threshold of galaxy formation}",
      journal = {\mnras},
         year = 2017,
        month = nov,
       volume = {471},
       number = {3},
        pages = {3547-3562},
          doi = {10.1093/mnras/stx1757},
archivePrefix = {arXiv},
       eprint = {1611.02281},
 primaryClass = {astro-ph.GA},
       adsurl = {https://ui.adsabs.harvard.edu/abs/2017MNRAS.471.3547F}
}

@ARTICLE{Simon2019,
       author = {{Simon}, Joshua D.},
        title = "{The Faintest Dwarf Galaxies}",
      journal = {\araa},
         year = 2019,
        month = aug,
       volume = {57},
        pages = {375-415},
          doi = {10.1146/annurev-astro-091918-104453},
archivePrefix = {arXiv},
       eprint = {1901.05465},
 primaryClass = {astro-ph.GA},
       adsurl = {https://ui.adsabs.harvard.edu/abs/2019ARA&A..57..375S}
}

@ARTICLE{Nadler2024,
       author = {{Nadler}, Ethan O. and {Gluscevic}, Vera and {Driskell}, Trey and {Wechsler}, Risa H. and {Moustakas}, Leonidas A. and {Benson}, Andrew and {Mao}, Yao-Yuan},
        title = "{Forecasts for Galaxy Formation and Dark Matter Constraints from Dwarf Galaxy Surveys}",
      journal = {\apj},
         year = 2024,
        month = may,
       volume = {967},
       number = {1},
          eid = {61},
        pages = {61},
          doi = {10.3847/1538-4357/ad3bb1},
archivePrefix = {arXiv},
       eprint = {2401.10318},
 primaryClass = {astro-ph.GA},
       adsurl = {https://ui.adsabs.harvard.edu/abs/2024ApJ...967...61N}
}

@ARTICLE{MUST2024,
       author = {{Zhao}, Cheng and {Huang}, Song and {He}, Mengfan and {Montero-Camacho}, Paulo and {Liu}, Yu and {Renard}, Pablo and {Tang}, Yunyi and {Verdier}, Aurelien and {Xu}, Wenshuo and {Yang}, Xiaorui and {Yu}, Jiaxi and {Zhang}, Yao and {Zhao}, Siyi and {Zhou}, Xingchen and {He}, Shengyu and {Kneib}, Jean-Paul and {Li}, Jiayi and {Li}, Zhuoyang and {Wang}, Wen-Ting and {Xianyu}, Zhong-Zhi and {Zhang}, Yidian and {Gsponer}, Rafaela and {Li}, Xiao-Dong and {Rocher}, Antoine and {Zou}, Siwei and {Tan}, Ting and {Huang}, Zhiqi and {Wang}, Zhuoxiao and {Li}, Pei and {Rombach}, Maxime and {Dong}, Chenxing and {Forero-Sanchez}, Daniel and {Ning}, Yuanhang and {Shan}, Huanyuan and {Wang}, Tao and {Li}, Yin and {Zhai}, Zhongxu and {Wang}, Yuting and {Zhao}, Gong-Bo and {Shi}, Yong and {Mao}, Shude and {Huang}, Lei and {Guo}, Liquan and {Cai}, Zheng},
        title = "{MUltiplexed Survey Telescope (MUST) Science White Paper I: Overview of Large-Scale Structure Cosmology in the Era of Stage-V Spectroscopic Surveys}",
      journal = {arXiv e-prints},
         year = 2024,
        month = nov,
          eid = {arXiv:2411.07970},
        pages = {arXiv:2411.07970},
          doi = {10.48550/arXiv.2411.07970},
archivePrefix = {arXiv},
       eprint = {2411.07970},
 primaryClass = {astro-ph.CO},
       adsurl = {https://ui.adsabs.harvard.edu/abs/2024arXiv241107970Z}
}

@ARTICLE{JUST2024,
       author = {{JUST Team} and {Liu}, Chengze and {Zu}, Ying and {Feng}, Fabo and {Li}, Zhaoyu and {Yu}, Yu and {Bai}, Hua and {Cui}, Xiangqun and {Gu}, Bozhong and {Gu}, Yizhou and {Han}, Jiaxin and {Hou}, Yonghui and {Hu}, Zhongwen and {Ji}, Hangxin and {Jing}, Yipeng and {Li}, Wei and {Qi}, Zhaoxiang and {Tan}, Xianyu and {Tian}, Cairang and {Yang}, Dehua and {Yuan}, Xiangyan and {Zhai}, Chao and {Zhang}, Congcong and {Zhang}, Jun and {Zhang}, Haotong and {Zhang}, Pengjie and {Zhang}, Yong and {Zhao}, Yi and {Zheng}, Xianzhong and {Zhu}, Qingfeng and {Yang}, Xiaohu},
        title = "{The Jiao Tong University Spectroscopic Telescope (JUST) Project}",
      journal = {Astronomical Techniques and Instruments},
         year = 2024,
        month = jan,
       volume = {1},
       number = {1},
        pages = {16-30},
          doi = {10.61977/ati2024008},
archivePrefix = {arXiv},
       eprint = {2402.14312},
 primaryClass = {astro-ph.IM},
       adsurl = {https://ui.adsabs.harvard.edu/abs/2024AstTI...1...16J}
}

@ARTICLE{astropy2018,
       author = {{Astropy Collaboration} and {Price-Whelan}, A.~M. and {Sip{\H{o}}cz}, B.~M. and {G{\"u}nther}, H.~M. and {Lim}, P.~L. and {Crawford}, S.~M. and {Conseil}, S. and {Shupe}, D.~L. and {Craig}, M.~W. and {Dencheva}, N. and {Ginsburg}, A. and {VanderPlas}, J.~T. and {Bradley}, L.~D. and {P{\'e}rez-Su{\'a}rez}, D. and {de Val-Borro}, M. and {Aldcroft}, T.~L. and {Cruz}, K.~L. and {Robitaille}, T.~P. and {Tollerud}, E.~J. and {Ardelean}, C. and {Babej}, T. and {Bach}, Y.~P. and {Bachetti}, M. and {Bakanov}, A.~V. and {Bamford}, S.~P. and {Barentsen}, G. and {Barmby}, P. and {Baumbach}, A. and {Berry}, K.~L. and {Biscani}, F. and {Boquien}, M. and {Bostroem}, K.~A. and {Bouma}, L.~G. and {Brammer}, G.~B. and {Bray}, E.~M. and {Breytenbach}, H. and {Buddelmeijer}, H. and {Burke}, D.~J. and {Calderone}, G. and {Cano Rodr{\'\i}guez}, J.~L. and {Cara}, M. and {Cardoso}, J.~V.~M. and {Cheedella}, S. and {Copin}, Y. and {Corrales}, L. and {Crichton}, D. and {D'Avella}, D. and {Deil}, C. and {Depagne}, {\'E}. and {Dietrich}, J.~P. and {Donath}, A. and {Droettboom}, M. and {Earl}, N. and {Erben}, T. and {Fabbro}, S. and {Ferreira}, L.~A. and {Finethy}, T. and {Fox}, R.~T. and {Garrison}, L.~H. and {Gibbons}, S.~L.~J. and {Goldstein}, D.~A. and {Gommers}, R. and {Greco}, J.~P. and {Greenfield}, P. and {Groener}, A.~M. and {Grollier}, F. and {Hagen}, A. and {Hirst}, P. and {Homeier}, D. and {Horton}, A.~J. and {Hosseinzadeh}, G. and {Hu}, L. and {Hunkeler}, J.~S. and {Ivezi{\'c}}, {\v{Z}}. and {Jain}, A. and {Jenness}, T. and {Kanarek}, G. and {Kendrew}, S. and {Kern}, N.~S. and {Kerzendorf}, W.~E. and {Khvalko}, A. and {King}, J. and {Kirkby}, D. and {Kulkarni}, A.~M. and {Kumar}, A. and {Lee}, A. and {Lenz}, D. and {Littlefair}, S.~P. and {Ma}, Z. and {Macleod}, D.~M. and {Mastropietro}, M. and {McCully}, C. and {Montagnac}, S. and {Morris}, B.~M. and {Mueller}, M. and {Mumford}, S.~J. and {Muna}, D. and {Murphy}, N.~A. and {Nelson}, S. and {Nguyen}, G.~H. and {Ninan}, J.~P. and {N{\"o}the}, M. and {Ogaz}, S. and {Oh}, S. and {Parejko}, J.~K. and {Parley}, N. and {Pascual}, S. and {Patil}, R. and {Patil}, A.~A. and {Plunkett}, A.~L. and {Prochaska}, J.~X. and {Rastogi}, T. and {Reddy Janga}, V. and {Sabater}, J. and {Sakurikar}, P. and {Seifert}, M. and {Sherbert}, L.~E. and {Sherwood-Taylor}, H. and {Shih}, A.~Y. and {Sick}, J. and {Silbiger}, M.~T. and {Singanamalla}, S. and {Singer}, L.~P. and {Sladen}, P.~H. and {Sooley}, K.~A. and {Sornarajah}, S. and {Streicher}, O. and {Teuben}, P. and {Thomas}, S.~W. and {Tremblay}, G.~R. and {Turner}, J.~E.~H. and {Terr{\'o}n}, V. and {van Kerkwijk}, M.~H. and {de la Vega}, A. and {Watkins}, L.~L. and {Weaver}, B.~A. and {Whitmore}, J.~B. and {Woillez}, J. and {Zabalza}, V. and {Astropy Contributors}},
        title = "{The Astropy Project: Building an Open-science Project and Status of the v2.0 Core Package}",
      journal = {\aj},
         year = 2018,
        month = sep,
       volume = {156},
       number = {3},
          eid = {123},
        pages = {123},
          doi = {10.3847/1538-3881/aabc4f},
archivePrefix = {arXiv},
       eprint = {1801.02634},
 primaryClass = {astro-ph.IM},
       adsurl = {https://ui.adsabs.harvard.edu/abs/2018AJ....156..123A}
}

@ARTICLE{astropy2013,
       author = {{Astropy Collaboration} and {Robitaille}, Thomas P. and
         {Tollerud}, Erik J. and {Greenfield}, Perry and {Droettboom}, Michael and
         {Bray}, Erik and {Aldcroft}, Tom and {Davis}, Matt and
         {Ginsburg}, Adam and {Price-Whelan}, Adrian M. and
         {Kerzendorf}, Wolfgang E. and {Conley}, Alexander and {Crighton}, Neil and
         {Barbary}, Kyle and {Muna}, Demitri and {Ferguson}, Henry and
         {Grollier}, Fr{\'e}d{\'e}ric and {Parikh}, Madhura M. and
         {Nair}, Prasanth H. and {Unther}, Hans M. and {Deil}, Christoph and
         {Woillez}, Julien and {Conseil}, Simon and {Kramer}, Roban and
         {Turner}, James E.~H. and {Singer}, Leo and {Fox}, Ryan and
         {Weaver}, Benjamin A. and {Zabalza}, Victor and {Edwards}, Zachary I. and
         {Azalee Bostroem}, K. and {Burke}, D.~J. and {Casey}, Andrew R. and
         {Crawford}, Steven M. and {Dencheva}, Nadia and {Ely}, Justin and
         {Jenness}, Tim and {Labrie}, Kathleen and {Lim}, Pey Lian and
         {Pierfederici}, Francesco and {Pontzen}, Andrew and {Ptak}, Andy and
         {Refsdal}, Brian and {Servillat}, Mathieu and {Streicher}, Ole},
        title = "{Astropy: A community Python package for astronomy}",
      journal = {\aap},
         year = "2013",
        month = "Oct",
       volume = {558},
          eid = {A33},
        pages = {A33},
          doi = {10.1051/0004-6361/201322068},
archivePrefix = {arXiv},
       eprint = {1307.6212},
 primaryClass = {astro-ph.IM},
       adsurl = {https://ui.adsabs.harvard.edu/abs/2013A&A...558A..33A}
}

@ARTICLE{astropy2022,
       author = {{Astropy Collaboration} and {Price-Whelan}, Adrian M. and {Lim}, Pey Lian and {Earl}, Nicholas and {Starkman}, Nathaniel and {Bradley}, Larry and {Shupe}, David L. and {Patil}, Aarya A. and {Corrales}, Lia and {Brasseur}, C.~E. and {N{\"o}the}, Maximilian and {Donath}, Axel and {Tollerud}, Erik and {Morris}, Brett M. and {Ginsburg}, Adam and {Vaher}, Eero and {Weaver}, Benjamin A. and {Tocknell}, James and {Jamieson}, William and {van Kerkwijk}, Marten H. and {Robitaille}, Thomas P. and {Merry}, Bruce and {Bachetti}, Matteo and {G{\"u}nther}, H. Moritz and {Aldcroft}, Thomas L. and {Alvarado-Montes}, Jaime A. and {Archibald}, Anne M. and {B{\'o}di}, Attila and {Bapat}, Shreyas and {Barentsen}, Geert and {Baz{\'a}n}, Juanjo and {Biswas}, Manish and {Boquien}, M{\'e}d{\'e}ric and {Burke}, D.~J. and {Cara}, Daria and {Cara}, Mihai and {Conroy}, Kyle E. and {Conseil}, Simon and {Craig}, Matthew W. and {Cross}, Robert M. and {Cruz}, Kelle L. and {D'Eugenio}, Francesco and {Dencheva}, Nadia and {Devillepoix}, Hadrien A.~R. and {Dietrich}, J{\"o}rg P. and {Eigenbrot}, Arthur Davis and {Erben}, Thomas and {Ferreira}, Leonardo and {Foreman-Mackey}, Daniel and {Fox}, Ryan and {Freij}, Nabil and {Garg}, Suyog and {Geda}, Robel and {Glattly}, Lauren and {Gondhalekar}, Yash and {Gordon}, Karl D. and {Grant}, David and {Greenfield}, Perry and {Groener}, Austen M. and {Guest}, Steve and {Gurovich}, Sebastian and {Handberg}, Rasmus and {Hart}, Akeem and {Hatfield-Dodds}, Zac and {Homeier}, Derek and {Hosseinzadeh}, Griffin and {Jenness}, Tim and {Jones}, Craig K. and {Joseph}, Prajwel and {Kalmbach}, J. Bryce and {Karamehmetoglu}, Emir and {Ka{\l}uszy{\'n}ski}, Miko{\l}aj and {Kelley}, Michael S.~P. and {Kern}, Nicholas and {Kerzendorf}, Wolfgang E. and {Koch}, Eric W. and {Kulumani}, Shankar and {Lee}, Antony and {Ly}, Chun and {Ma}, Zhiyuan and {MacBride}, Conor and {Maljaars}, Jakob M. and {Muna}, Demitri and {Murphy}, N.~A. and {Norman}, Henrik and {O'Steen}, Richard and {Oman}, Kyle A. and {Pacifici}, Camilla and {Pascual}, Sergio and {Pascual-Granado}, J. and {Patil}, Rohit R. and {Perren}, Gabriel I. and {Pickering}, Timothy E. and {Rastogi}, Tanuj and {Roulston}, Benjamin R. and {Ryan}, Daniel F. and {Rykoff}, Eli S. and {Sabater}, Jose and {Sakurikar}, Parikshit and {Salgado}, Jes{\'u}s and {Sanghi}, Aniket and {Saunders}, Nicholas and {Savchenko}, Volodymyr and {Schwardt}, Ludwig and {Seifert-Eckert}, Michael and {Shih}, Albert Y. and {Jain}, Anany Shrey and {Shukla}, Gyanendra and {Sick}, Jonathan and {Simpson}, Chris and {Singanamalla}, Sudheesh and {Singer}, Leo P. and {Singhal}, Jaladh and {Sinha}, Manodeep and {Sip{\H{o}}cz}, Brigitta M. and {Spitler}, Lee R. and {Stansby}, David and {Streicher}, Ole and {{\v{S}}umak}, Jani and {Swinbank}, John D. and {Taranu}, Dan S. and {Tewary}, Nikita and {Tremblay}, Grant R. and {de Val-Borro}, Miguel and {Van Kooten}, Samuel J. and {Vasovi{\'c}}, Zlatan and {Verma}, Shresth and {de Miranda Cardoso}, Jos{\'e} Vin{\'\i}cius and {Williams}, Peter K.~G. and {Wilson}, Tom J. and {Winkel}, Benjamin and {Wood-Vasey}, W.~M. and {Xue}, Rui and {Yoachim}, Peter and {Zhang}, Chen and {Zonca}, Andrea and {Astropy Project Contributors}},
        title = "{The Astropy Project: Sustaining and Growing a Community-oriented Open-source Project and the Latest Major Release (v5.0) of the Core Package}",
      journal = {\apj},
         year = 2022,
        month = aug,
       volume = {935},
       number = {2},
          eid = {167},
        pages = {167},
          doi = {10.3847/1538-4357/ac7c74},
archivePrefix = {arXiv},
       eprint = {2206.14220},
 primaryClass = {astro-ph.IM},
       adsurl = {https://ui.adsabs.harvard.edu/abs/2022ApJ...935..167A}
}

@ARTICLE{numpy2020,
       author = {{Harris}, Charles R. and {Millman}, K. Jarrod and {van der Walt}, St{\'e}fan J. and {Gommers}, Ralf and {Virtanen}, Pauli and {Cournapeau}, David and {Wieser}, Eric and {Taylor}, Julian and {Berg}, Sebastian and {Smith}, Nathaniel J. and {Kern}, Robert and {Picus}, Matti and {Hoyer}, Stephan and {van Kerkwijk}, Marten H. and {Brett}, Matthew and {Haldane}, Allan and {del R{\'\i}o}, Jaime Fern{\'a}ndez and {Wiebe}, Mark and {Peterson}, Pearu and {G{\'e}rard-Marchant}, Pierre and {Sheppard}, Kevin and {Reddy}, Tyler and {Weckesser}, Warren and {Abbasi}, Hameer and {Gohlke}, Christoph and {Oliphant}, Travis E.},
        title = "{Array programming with NumPy}",
      journal = {\nat},
         year = 2020,
        month = sep,
       volume = {585},
       number = {7825},
        pages = {357-362},
          doi = {10.1038/s41586-020-2649-2},
archivePrefix = {arXiv},
       eprint = {2006.10256},
 primaryClass = {cs.MS},
       adsurl = {https://ui.adsabs.harvard.edu/abs/2020Natur.585..357H}
}

@ARTICLE{scipy2020,
       author = {{Virtanen}, Pauli and {Gommers}, Ralf and {Oliphant}, Travis E. and {Haberland}, Matt and {Reddy}, Tyler and {Cournapeau}, David and {Burovski}, Evgeni and {Peterson}, Pearu and {Weckesser}, Warren and {Bright}, Jonathan and {van der Walt}, St{\'e}fan J. and {Brett}, Matthew and {Wilson}, Joshua and {Millman}, K. Jarrod and {Mayorov}, Nikolay and {Nelson}, Andrew R.~J. and {Jones}, Eric and {Kern}, Robert and {Larson}, Eric and {Carey}, C.~J. and {Polat}, {\.I}lhan and {Feng}, Yu and {Moore}, Eric W. and {VanderPlas}, Jake and {Laxalde}, Denis and {Perktold}, Josef and {Cimrman}, Robert and {Henriksen}, Ian and {Quintero}, E.~A. and {Harris}, Charles R. and {Archibald}, Anne M. and {Ribeiro}, Ant{\^o}nio H. and {Pedregosa}, Fabian and {van Mulbregt}, Paul and {SciPy 1. 0 Contributors}},
        title = "{SciPy 1.0: fundamental algorithms for scientific computing in Python}",
      journal = {Nature Methods},
         year = 2020,
        month = feb,
       volume = {17},
        pages = {261-272},
          doi = {10.1038/s41592-019-0686-2},
archivePrefix = {arXiv},
       eprint = {1907.10121},
 primaryClass = {cs.MS},
       adsurl = {https://ui.adsabs.harvard.edu/abs/2020NatMe..17..261V}
}

@software{dsigma2022,
       author = {{Lange}, Johannes and {Huang}, Song},
        title = "{dsigma: Galaxy-galaxy lensing Python package}",
 howpublished = {Astrophysics Source Code Library, record ascl:2204.006},
         year = 2022,
        month = apr,
          eid = {ascl:2204.006},
       adsurl = {https://ui.adsabs.harvard.edu/abs/2022ascl.soft04006L}
}

@ARTICLE{DESIDR2cosmology,
       author = {{Abdul Karim}, M. and {Aguilar}, J. and {Ahlen}, S. and {Alam}, S. and {Allen}, L. and {Prieto}, C. Allende and {Alves}, O. and {Anand}, A. and {Andrade}, U. and {Armengaud}, E. and et al.},
        title = "{DESI DR2 results. II. Measurements of baryon acoustic oscillations and cosmological constraints}",
      journal = {\prd},
         year = 2025,
        month = oct,
       volume = {112},
       number = {8},
          eid = {083515},
        pages = {083515},
          doi = {10.1103/tr6y-kpc6},
archivePrefix = {arXiv},
       eprint = {2503.14738},
 primaryClass = {astro-ph.CO},
       adsurl = {https://ui.adsabs.harvard.edu/abs/2025PhRvD.112h3515A}
}

@ARTICLE{DESIDR1cosmology,
       author = {{Adame}, A.~G. and {Aguilar}, J. and {Ahlen}, S. and {Alam}, S. and {Alexander}, D.~M. and {Allende Prieto}, C. and {Alvarez}, M. and {Alves}, O. and {Anand}, A. and {Andrade}, U. and et al.},
        title = "{DESI 2024 VII: cosmological constraints from the full-shape modeling of clustering measurements}",
      journal = {\jcap},
         year = 2025,
        month = jul,
       volume = {2025},
       number = {7},
          eid = {028},
        pages = {028},
          doi = {10.1088/1475-7516/2025/07/028},
archivePrefix = {arXiv},
       eprint = {2411.12022},
 primaryClass = {astro-ph.CO},
       adsurl = {https://ui.adsabs.harvard.edu/abs/2025JCAP...07..028A}
}

@ARTICLE{DESIDR1,
       author = {{DESI Collaboration} and {Abdul-Karim}, M. and {Adame}, A.~G. and {Aguado}, D. and {Aguilar}, J. and {Ahlen}, S. and {Alam}, S. and {Aldering}, G. and {Alexander}, D.~M. and {Alfarsy}, R. and et al.},
        title = "{Data Release 1 of the Dark Energy Spectroscopic Instrument}",
      journal = {arXiv e-prints},
         year = 2025,
        month = mar,
          eid = {arXiv:2503.14745},
        pages = {arXiv:2503.14745},
          doi = {10.48550/arXiv.2503.14745},
archivePrefix = {arXiv},
       eprint = {2503.14745},
 primaryClass = {astro-ph.CO},
       adsurl = {https://ui.adsabs.harvard.edu/abs/2025arXiv250314745D}
}

@ARTICLE{DESIDR1Sample,
       author = {{Adame}, A.~G. and {Aguilar}, J. and {Ahlen}, S. and {Alam}, S. and {Alexander}, D.~M. and {Alvarez}, M. and {Alves}, O. and {Anand}, A. and {Andrade}, U. and {Armengaud}, E. and et al.},
        title = "{DESI 2024 II: sample definitions, characteristics, and two-point clustering statistics}",
      journal = {\jcap},
         year = 2025,
        month = jul,
       volume = {2025},
       number = {7},
          eid = {017},
        pages = {017},
          doi = {10.1088/1475-7516/2025/07/017},
archivePrefix = {arXiv},
       eprint = {2411.12020},
 primaryClass = {astro-ph.CO},
       adsurl = {https://ui.adsabs.harvard.edu/abs/2025JCAP...07..017A}
}

@ARTICLE{Poppett2024DESI,
       author = {{Poppett}, Claire and {Tyas}, Luke and {Aguilar}, J. and {Bebek}, Christopher and {Bramall}, D. and {Claybaugh}, T. and {Edelstein}, J. and {Fagrelius}, P. and {Heetderks}, H. and {Jelinsky}, P. and et al.},
        title = "{Overview of the Fiber System for the Dark Energy Spectroscopic Instrument}",
      journal = {\aj},
         year = 2024,
        month = dec,
       volume = {168},
       number = {6},
          eid = {245},
        pages = {245},
          doi = {10.3847/1538-3881/ad76a4},
       adsurl = {https://ui.adsabs.harvard.edu/abs/2024AJ....168..245P}
}

@ARTICLE{Miller2024DESI,
       author = {{Miller}, Timothy N. and {Doel}, Peter and {Gutierrez}, Gaston and {Besuner}, Robert and {Brooks}, David and {Gallo}, Giuseppe and {Heetderks}, Henry and {Jelinsky}, Patrick and {Kent}, Stephen M. and {Lampton}, Michael and et al.},
        title = "{The Optical Corrector for the Dark Energy Spectroscopic Instrument}",
      journal = {\aj},
         year = 2024,
        month = aug,
       volume = {168},
       number = {2},
          eid = {95},
        pages = {95},
          doi = {10.3847/1538-3881/ad45fe},
archivePrefix = {arXiv},
       eprint = {2306.06310},
 primaryClass = {astro-ph.IM},
       adsurl = {https://ui.adsabs.harvard.edu/abs/2024AJ....168...95M}
}

@ARTICLE{Schlafly2023DESI,
       author = {{Schlafly}, Edward F. and {Kirkby}, David and {Schlegel}, David J. and {Myers}, Adam D. and {Raichoor}, Anand and {Dawson}, Kyle and {Aguilar}, Jessica and {Allende Prieto}, Carlos and {Bailey}, Stephen and {BenZvi}, Segev and et al.},
        title = "{Survey Operations for the Dark Energy Spectroscopic Instrument}",
      journal = {\aj},
         year = 2023,
        month = dec,
       volume = {166},
       number = {6},
          eid = {259},
        pages = {259},
          doi = {10.3847/1538-3881/ad0832},
archivePrefix = {arXiv},
       eprint = {2306.06309},
 primaryClass = {astro-ph.CO},
       adsurl = {https://ui.adsabs.harvard.edu/abs/2023AJ....166..259S}
}

@ARTICLE{Guy2023DESI,
       author = {{Guy}, J. and {Bailey}, S. and {Kremin}, A. and {Alam}, Shadab and {Alexander}, D.~M. and {Allende Prieto}, C. and {BenZvi}, S. and {Bolton}, A.~S. and {Brooks}, D. and {Chaussidon}, E. and et al.},
        title = "{The Spectroscopic Data Processing Pipeline for the Dark Energy Spectroscopic Instrument}",
      journal = {\aj},
         year = 2023,
        month = apr,
       volume = {165},
       number = {4},
          eid = {144},
        pages = {144},
          doi = {10.3847/1538-3881/acb212},
archivePrefix = {arXiv},
       eprint = {2209.14482},
 primaryClass = {astro-ph.IM},
       adsurl = {https://ui.adsabs.harvard.edu/abs/2023AJ....165..144G}
}

@ARTICLE{DESI2022Overview,
       author = {{DESI Collaboration} and {Abareshi}, B. and {Aguilar}, J. and {Ahlen}, S. and {Alam}, Shadab and {Alexander}, David M. and {Alfarsy}, R. and {Allen}, L. and {Allende Prieto}, C. and {Alves}, O. and et al.},
        title = "{Overview of the Instrumentation for the Dark Energy Spectroscopic Instrument}",
      journal = {\aj},
         year = 2022,
        month = nov,
       volume = {164},
       number = {5},
          eid = {207},
        pages = {207},
          doi = {10.3847/1538-3881/ac882b},
archivePrefix = {arXiv},
       eprint = {2205.10939},
 primaryClass = {astro-ph.IM},
       adsurl = {https://ui.adsabs.harvard.edu/abs/2022AJ....164..207D}
}

@ARTICLE{DESI2016a,
       author = {{DESI Collaboration} and {Aghamousa}, Amir and {Aguilar}, Jessica and {Ahlen}, Steve and {Alam}, Shadab and {Allen}, Lori E. and {Allende Prieto}, Carlos and {Annis}, James and {Bailey}, Stephen and {Balland}, Christophe and et al.},
        title = "{The DESI Experiment Part I: Science,Targeting, and Survey Design}",
      journal = {arXiv e-prints},
         year = 2016,
        month = oct,
          eid = {arXiv:1611.00036},
        pages = {arXiv:1611.00036},
          doi = {10.48550/arXiv.1611.00036},
archivePrefix = {arXiv},
       eprint = {1611.00036},
 primaryClass = {astro-ph.IM},
       adsurl = {https://ui.adsabs.harvard.edu/abs/2016arXiv161100036D}
}

@ARTICLE{DESI2016b,
       author = {{DESI Collaboration} and {Aghamousa}, Amir and {Aguilar}, Jessica and {Ahlen}, Steve and {Alam}, Shadab and {Allen}, Lori E. and {Allende Prieto}, Carlos and {Annis}, James and {Bailey}, Stephen and {Balland}, Christophe and et al.},
        title = "{The DESI Experiment Part II: Instrument Design}",
      journal = {arXiv e-prints},
         year = 2016,
        month = oct,
          eid = {arXiv:1611.00037},
        pages = {arXiv:1611.00037},
          doi = {10.48550/arXiv.1611.00037},
archivePrefix = {arXiv},
       eprint = {1611.00037},
 primaryClass = {astro-ph.IM},
       adsurl = {https://ui.adsabs.harvard.edu/abs/2016arXiv161100037D}
}

@ARTICLE{Hahn2023,
       author = {{Hahn}, ChangHoon and {Wilson}, Michael J. and {Ruiz-Macias}, Omar and {Cole}, Shaun and {Weinberg}, David H. and {Moustakas}, John and {Kremin}, Anthony and {Tinker}, Jeremy L. and {Smith}, Alex and {Wechsler}, Risa H. and et al.},
        title = "{The DESI Bright Galaxy Survey: Final Target Selection, Design, and Validation}",
      journal = {\aj},
         year = 2023,
        month = jun,
       volume = {165},
       number = {6},
          eid = {253},
        pages = {253},
          doi = {10.3847/1538-3881/accff8},
archivePrefix = {arXiv},
       eprint = {2208.08512},
 primaryClass = {astro-ph.CO},
       adsurl = {https://ui.adsabs.harvard.edu/abs/2023AJ....165..253H}
}

@ARTICLE{Ross2025,
       author = {{Ross}, A.~J. and {Aguilar}, J. and {Ahlen}, S. and {Alam}, S. and {Anand}, A. and {Bailey}, S. and {Bianchi}, D. and {Brieden}, S. and {Brooks}, D. and {Burtin}, E. and et al.},
        title = "{The construction of large-scale structure catalogs for the Dark Energy Spectroscopic Instrument}",
      journal = {\jcap},
         year = 2025,
        month = jan,
       volume = {2025},
       number = {1},
          eid = {125},
        pages = {125},
          doi = {10.1088/1475-7516/2025/01/125},
archivePrefix = {arXiv},
       eprint = {2405.16593},
 primaryClass = {astro-ph.CO},
       adsurl = {https://ui.adsabs.harvard.edu/abs/2025JCAP...01..125R}
}

@ARTICLE{Zou2024,
       author = {{Zou}, Hu and {Sui}, Jipeng and {Saintonge}, Am{\'e}lie and {Scholte}, Dirk and {Moustakas}, John and {Siudek}, Malgorzata and {Dey}, Arjun and {Juneau}, Stephanie and {Guo}, Weijian and {Canning}, Rebecca and et al.},
        title = "{A Large Sample of Extremely Metal-poor Galaxies at z < 1 Identified from the DESI Early Data}",
      journal = {\apj},
         year = 2024,
        month = feb,
       volume = {961},
       number = {2},
          eid = {173},
        pages = {173},
          doi = {10.3847/1538-4357/ad1409},
archivePrefix = {arXiv},
       eprint = {2312.00300},
 primaryClass = {astro-ph.GA},
       adsurl = {https://ui.adsabs.harvard.edu/abs/2024ApJ...961..173Z}
}

@ARTICLE{Boquien2019cigale,
       author = {{Boquien}, M. and {Burgarella}, D. and {Roehlly}, Y. and {Buat}, V. and {Ciesla}, L. and {Corre}, D. and {Inoue}, A.~K. and {Salas}, H.},
        title = "{CIGALE: a python Code Investigating GALaxy Emission}",
      journal = {\aap},
         year = 2019,
        month = feb,
       volume = {622},
          eid = {A103},
        pages = {A103},
          doi = {10.1051/0004-6361/201834156},
archivePrefix = {arXiv},
       eprint = {1811.03094},
 primaryClass = {astro-ph.GA},
       adsurl = {https://ui.adsabs.harvard.edu/abs/2019A&A...622A.103B}
}

@ARTICLE{Yang2022cigale,
       author = {{Yang}, Guang and {Boquien}, M{\'e}d{\'e}ric and {Brandt}, W.~N. and {Buat}, V{\'e}ronique and {Burgarella}, Denis and {Ciesla}, Laure and {Lehmer}, Bret D. and {Ma{\l}ek}, Katarzyna and {Mountrichas}, George and {Papovich}, Casey and et al.},
        title = "{Fitting AGN/Galaxy X-Ray-to-radio SEDs with CIGALE and Improvement of the Code}",
      journal = {\apj},
         year = 2022,
        month = mar,
       volume = {927},
       number = {2},
          eid = {192},
        pages = {192},
          doi = {10.3847/1538-4357/ac4971},
archivePrefix = {arXiv},
       eprint = {2201.03718},
 primaryClass = {astro-ph.GA},
       adsurl = {https://ui.adsabs.harvard.edu/abs/2022ApJ...927..192Y}
}

@ARTICLE{Yang2020cigale,
       author = {{Yang}, G. and {Boquien}, M. and {Buat}, V. and {Burgarella}, D. and {Ciesla}, L. and {Duras}, F. and {Stalevski}, M. and {Brandt}, W.~N. and {Papovich}, C.},
        title = "{X-CIGALE: Fitting AGN/galaxy SEDs from X-ray to infrared}",
      journal = {\mnras},
         year = 2020,
        month = jan,
       volume = {491},
       number = {1},
        pages = {740-757},
          doi = {10.1093/mnras/stz3001},
archivePrefix = {arXiv},
       eprint = {2001.08263},
 primaryClass = {astro-ph.GA},
       adsurl = {https://ui.adsabs.harvard.edu/abs/2020MNRAS.491..740Y}
}

@ARTICLE{Dey2019,
       author = {{Dey}, Arjun and {Schlegel}, David J. and {Lang}, Dustin and {Blum}, Robert and {Burleigh}, Kaylan and {Fan}, Xiaohui and {Findlay}, Joseph R. and {Finkbeiner}, Doug and {Herrera}, David and {Juneau}, St{\'e}phanie and et al.},
        title = "{Overview of the DESI Legacy Imaging Surveys}",
      journal = {\aj},
         year = 2019,
        month = may,
       volume = {157},
       number = {5},
          eid = {168},
        pages = {168},
          doi = {10.3847/1538-3881/ab089d},
archivePrefix = {arXiv},
       eprint = {1804.08657},
 primaryClass = {astro-ph.IM},
       adsurl = {https://ui.adsabs.harvard.edu/abs/2019AJ....157..168D}
}

@ARTICLE{Kauffmann2003,
       author = {{Kauffmann}, Guinevere and {Heckman}, Timothy M. and {White}, Simon D.~M. and {Charlot}, St{\'e}phane and {Tremonti}, Christy and {Brinchmann}, Jarle and {Bruzual}, Gustavo and {Peng}, Eric W. and {Seibert}, Mark and {Bernardi}, Mariangela and et al.},
        title = "{Stellar masses and star formation histories for {}10$^{5}$ galaxies from the Sloan Digital Sky Survey}",
      journal = {\mnras},
         year = 2003,
        month = may,
       volume = {341},
       number = {1},
        pages = {33-53},
          doi = {10.1046/j.1365-8711.2003.06291.x},
archivePrefix = {arXiv},
       eprint = {astro-ph/0204055},
 primaryClass = {astro-ph},
       adsurl = {https://ui.adsabs.harvard.edu/abs/2003MNRAS.341...33K}
}

@ARTICLE{Salim2007,
       author = {{Salim}, Samir and {Rich}, R. Michael and {Charlot}, St{\'e}phane and {Brinchmann}, Jarle and {Johnson}, Benjamin D. and {Schiminovich}, David and {Seibert}, Mark and {Mallery}, Ryan and {Heckman}, Timothy M. and {Forster}, Karl and et al.},
        title = "{UV Star Formation Rates in the Local Universe}",
      journal = {\apjs},
         year = 2007,
        month = dec,
       volume = {173},
       number = {2},
        pages = {267-292},
          doi = {10.1086/519218},
archivePrefix = {arXiv},
       eprint = {0704.3611},
 primaryClass = {astro-ph},
       adsurl = {https://ui.adsabs.harvard.edu/abs/2007ApJS..173..267S}
}

@ARTICLE{To2025,
       author = {{To}, Chun-Hao and {Chang}, Chihway and {Anbajagane}, Dhayaa and {Wechsler}, Risa H. and {Drlica-Wagner}, Alex and {Adam{\'o}w}, M. and {Alarcon}, A. and {Becker}, M.~R. and {Carballo-Bello}, J.~A. and {Cawthon}, R. and et al.},
        title = "{A DECADE of dwarfs: first detection of weak lensing around spectroscopically confirmed low-mass galaxies}",
      journal = {arXiv e-prints},
         year = 2025,
        month = sep,
          eid = {arXiv:2509.20458},
        pages = {arXiv:2509.20458},
          doi = {10.48550/arXiv.2509.20458},
archivePrefix = {arXiv},
       eprint = {2509.20458},
 primaryClass = {astro-ph.GA},
       adsurl = {https://ui.adsabs.harvard.edu/abs/2025arXiv250920458T}
}

@ARTICLE{ZM2015,
       author = {{Zu}, Ying and {Mandelbaum}, Rachel},
        title = "{Mapping stellar content to dark matter haloes using galaxy clustering and galaxy-galaxy lensing in the SDSS DR7}",
      journal = {\mnras},
         year = 2015,
        month = dec,
       volume = {454},
       number = {2},
        pages = {1161-1191},
          doi = {10.1093/mnras/stv2062},
archivePrefix = {arXiv},
       eprint = {1505.02781},
 primaryClass = {astro-ph.CO},
       adsurl = {https://ui.adsabs.harvard.edu/abs/2015MNRAS.454.1161Z}
}

@ARTICLE{Yang2021,
       author = {{Yang}, Xiaohu and {Xu}, Haojie and {He}, Min and {Gu}, Yizhou and {Katsianis}, Antonios and {Meng}, Jiacheng and {Shi}, Feng and {Zou}, Hu and {Zhang}, Youcai and {Liu}, Chengze and et al.},
        title = "{An Extended Halo-based Group/Cluster Finder: Application to the DESI Legacy Imaging Surveys DR8}",
      journal = {\apj},
         year = 2021,
        month = mar,
       volume = {909},
       number = {2},
          eid = {143},
        pages = {143},
          doi = {10.3847/1538-4357/abddb2},
archivePrefix = {arXiv},
       eprint = {2012.14998},
 primaryClass = {astro-ph.GA},
       adsurl = {https://ui.adsabs.harvard.edu/abs/2021ApJ...909..143Y}
}

@ARTICLE{Schmidt1968,
       author = {{Schmidt}, Maarten},
        title = "{Space Distribution and Luminosity Functions of Quasi-Stellar Radio Sources}",
      journal = {\apj},
         year = 1968,
        month = feb,
       volume = {151},
        pages = {393},
          doi = {10.1086/149446},
       adsurl = {https://ui.adsabs.harvard.edu/abs/1968ApJ...151..393S}
}

@ARTICLE{Shao2025,
       author = {{Shao}, Zhiwei and {Zu}, Ying and {Salcedo}, Andr{\'e}s N. and {Wang}, Jiaqi and {Yang}, Xiaohu and {Weinberg}, David H. and {Xu}, Xiaoju and {Zhai}, Zhongxu and {Zhang}, Zhuowen and {Aguilar}, J. and et al.},
        title = "{Direct Measurement of Galaxy Assembly Bias using DESI DR1 Data}",
      journal = {arXiv e-prints},
         year = 2025,
        month = oct,
          eid = {arXiv:2510.20896},
        pages = {arXiv:2510.20896},
          doi = {10.48550/arXiv.2510.20896},
archivePrefix = {arXiv},
       eprint = {2510.20896},
 primaryClass = {astro-ph.GA},
       adsurl = {https://ui.adsabs.harvard.edu/abs/2025arXiv251020896S}
}

@ARTICLE{Landy1993,
       author = {{Landy}, Stephen D. and {Szalay}, Alexander S.},
        title = "{Bias and Variance of Angular Correlation Functions}",
      journal = {\apj},
         year = 1993,
        month = jul,
       volume = {412},
        pages = {64},
          doi = {10.1086/172900},
       adsurl = {https://ui.adsabs.harvard.edu/abs/1993ApJ...412...64L}
}

@ARTICLE{Bianchi2018,
       author = {{Bianchi}, Davide and {Burden}, Angela and {Percival}, Will J. and {Brooks}, David and {Cahn}, Robert N. and {Forero-Romero}, Jaime E. and {Levi}, Michael and {Ross}, Ashley J. and {Tarle}, Gregory},
        title = "{Unbiased clustering estimates with the DESI fibre assignment}",
      journal = {\mnras},
         year = 2018,
        month = dec,
       volume = {481},
       number = {2},
        pages = {2338-2348},
          doi = {10.1093/mnras/sty2377},
archivePrefix = {arXiv},
       eprint = {1805.00951},
 primaryClass = {astro-ph.CO},
       adsurl = {https://ui.adsabs.harvard.edu/abs/2018MNRAS.481.2338B}
}

@ARTICLE{Bianchi2017,
       author = {{Bianchi}, Davide and {Percival}, Will J.},
        title = "{Unbiased clustering estimation in the presence of missing observations}",
      journal = {\mnras},
         year = 2017,
        month = nov,
       volume = {472},
       number = {1},
        pages = {1106-1118},
          doi = {10.1093/mnras/stx2053},
archivePrefix = {arXiv},
       eprint = {1703.02070},
 primaryClass = {astro-ph.CO},
       adsurl = {https://ui.adsabs.harvard.edu/abs/2017MNRAS.472.1106B}
}

@ARTICLE{Percival2017,
       author = {{Percival}, Will J. and {Bianchi}, Davide},
        title = "{Using angular pair upweighting to improve 3D clustering measurements}",
      journal = {\mnras},
         year = 2017,
        month = nov,
       volume = {472},
       number = {1},
        pages = {L40-L44},
          doi = {10.1093/mnrasl/slx135},
archivePrefix = {arXiv},
       eprint = {1703.02071},
 primaryClass = {astro-ph.CO},
       adsurl = {https://ui.adsabs.harvard.edu/abs/2017MNRAS.472L..40P}
}

@ARTICLE{Bianchi2025,
       author = {{Bianchi}, D. and {Hanif}, M.~M.~S. and {Carnero Rosell}, A. and {Lasker}, J. and {Ross}, A.~J. and {Pinon}, M. and {de Mattia}, A. and {White}, M. and {Ahlen}, S. and {Bailey}, S. and et al.},
        title = "{Characterization of DESI fiber assignment incompleteness effect on 2-point clustering and mitigation methods for DR1 analysis}",
      journal = {\jcap},
         year = 2025,
        month = apr,
       volume = {2025},
       number = {4},
          eid = {074},
        pages = {074},
          doi = {10.1088/1475-7516/2025/04/074},
archivePrefix = {arXiv},
       eprint = {2411.12025},
 primaryClass = {astro-ph.CO},
       adsurl = {https://ui.adsabs.harvard.edu/abs/2025JCAP...04..074B}
}

@ARTICLE{Norberg2009,
       author = {{Norberg}, P. and {Baugh}, C.~M. and {Gazta{\~n}aga}, E. and {Croton}, D.~J.},
        title = "{Statistical analysis of galaxy surveys - I. Robust error estimation for two-point clustering statistics}",
      journal = {\mnras},
         year = 2009,
        month = jun,
       volume = {396},
       number = {1},
        pages = {19-38},
          doi = {10.1111/j.1365-2966.2009.14389.x},
archivePrefix = {arXiv},
       eprint = {0810.1885},
 primaryClass = {astro-ph},
       adsurl = {https://ui.adsabs.harvard.edu/abs/2009MNRAS.396...19N}
}

@ARTICLE{Ishiyama2021,
       author = {{Ishiyama}, Tomoaki and {Prada}, Francisco and {Klypin}, Anatoly A. and {Sinha}, Manodeep and {Metcalf}, R. Benton and {Jullo}, Eric and {Altieri}, Bruno and {Cora}, Sof{\'\i}a A. and {Croton}, Darren and {de la Torre}, Sylvain and et al.},
        title = "{The Uchuu simulations: Data Release 1 and dark matter halo concentrations}",
      journal = {\mnras},
         year = 2021,
        month = sep,
       volume = {506},
       number = {3},
        pages = {4210-4231},
          doi = {10.1093/mnras/stab1755},
archivePrefix = {arXiv},
       eprint = {2007.14720},
 primaryClass = {astro-ph.CO},
       adsurl = {https://ui.adsabs.harvard.edu/abs/2021MNRAS.506.4210I}
}

@ARTICLE{Reyes2012,
       author = {{Reyes}, R. and {Mandelbaum}, R. and {Gunn}, J.~E. and {Nakajima}, R. and {Seljak}, U. and {Hirata}, C.~M.},
        title = "{Optical-to-virial velocity ratios of local disc galaxies from combined kinematics and galaxy-galaxy lensing}",
      journal = {\mnras},
         year = 2012,
        month = oct,
       volume = {425},
       number = {4},
        pages = {2610-2640},
          doi = {10.1111/j.1365-2966.2012.21472.x},
archivePrefix = {arXiv},
       eprint = {1110.4107},
 primaryClass = {astro-ph.CO},
       adsurl = {https://ui.adsabs.harvard.edu/abs/2012MNRAS.425.2610R}
}

@ARTICLE{Mandelbaum2012,
       author = {{Mandelbaum}, Rachel and {Hirata}, Christopher M. and {Leauthaud}, Alexie and {Massey}, Richard J. and {Rhodes}, Jason},
        title = "{Precision simulation of ground-based lensing data using observations from space}",
      journal = {\mnras},
         year = 2012,
        month = feb,
       volume = {420},
       number = {2},
        pages = {1518-1540},
          doi = {10.1111/j.1365-2966.2011.20138.x},
archivePrefix = {arXiv},
       eprint = {1107.4629},
 primaryClass = {astro-ph.CO},
       adsurl = {https://ui.adsabs.harvard.edu/abs/2012MNRAS.420.1518M}
}

@ARTICLE{Mandelbaum2013,
       author = {{Mandelbaum}, Rachel and {Slosar}, An{\v{z}}e and {Baldauf}, Tobias and {Seljak}, Uro{\v{s}} and {Hirata}, Christopher M. and {Nakajima}, Reiko and {Reyes}, Reinabelle and {Smith}, Robert E.},
        title = "{Cosmological parameter constraints from galaxy-galaxy lensing and galaxy clustering with the SDSS DR7}",
      journal = {\mnras},
         year = 2013,
        month = jun,
       volume = {432},
       number = {2},
        pages = {1544-1575},
          doi = {10.1093/mnras/stt572},
archivePrefix = {arXiv},
       eprint = {1207.1120},
 primaryClass = {astro-ph.CO},
       adsurl = {https://ui.adsabs.harvard.edu/abs/2013MNRAS.432.1544M}
}

@ARTICLE{Hirata2003,
       author = {{Hirata}, Christopher and {Seljak}, Uro{\v{s}}},
        title = "{Shear calibration biases in weak-lensing surveys}",
      journal = {\mnras},
         year = 2003,
        month = aug,
       volume = {343},
       number = {2},
        pages = {459-480},
          doi = {10.1046/j.1365-8711.2003.06683.x},
archivePrefix = {arXiv},
       eprint = {astro-ph/0301054},
 primaryClass = {astro-ph},
       adsurl = {https://ui.adsabs.harvard.edu/abs/2003MNRAS.343..459H}
}

@ARTICLE{SDSSDR7,
       author = {{Abazajian}, Kevork N. and {Adelman-McCarthy}, Jennifer K. and {Ag{\"u}eros}, Marcel A. and {Allam}, Sahar S. and {Allende Prieto}, Carlos and {An}, Deokkeun and {Anderson}, Kurt S.~J. and {Anderson}, Scott F. and {Annis}, James and {Bahcall}, Neta A. and {Bailer-Jones}, C.~A.~L. and {Barentine}, J.~C. and {Bassett}, Bruce A. and {Becker}, Andrew C. and {Beers}, Timothy C. and {Bell}, Eric F. and {Belokurov}, Vasily and {Berlind}, Andreas A. and {Berman}, Eileen F. and {Bernardi}, Mariangela and {Bickerton}, Steven J. and {Bizyaev}, Dmitry and {Blakeslee}, John P. and {Blanton}, Michael R. and {Bochanski}, John J. and {Boroski}, William N. and {Brewington}, Howard J. and {Brinchmann}, Jarle and {Brinkmann}, J. and {Brunner}, Robert J. and {Budav{\'a}ri}, Tam{\'a}s and {Carey}, Larry N. and {Carliles}, Samuel and {Carr}, Michael A. and {Castander}, Francisco J. and {Cinabro}, David and {Connolly}, A.~J. and {Csabai}, Istv{\'a}n and {Cunha}, Carlos E. and {Czarapata}, Paul C. and {Davenport}, James R.~A. and {de Haas}, Ernst and {Dilday}, Ben and {Doi}, Mamoru and {Eisenstein}, Daniel J. and {Evans}, Michael L. and {Evans}, N.~W. and {Fan}, Xiaohui and {Friedman}, Scott D. and {Frieman}, Joshua A. and {Fukugita}, Masataka and {G{\"a}nsicke}, Boris T. and {Gates}, Evalyn and {Gillespie}, Bruce and {Gilmore}, G. and {Gonzalez}, Belinda and {Gonzalez}, Carlos F. and {Grebel}, Eva K. and {Gunn}, James E. and {Gy{\"o}ry}, Zsuzsanna and {Hall}, Patrick B. and {Harding}, Paul and {Harris}, Frederick H. and {Harvanek}, Michael and {Hawley}, Suzanne L. and {Hayes}, Jeffrey J.~E. and {Heckman}, Timothy M. and {Hendry}, John S. and {Hennessy}, Gregory S. and {Hindsley}, Robert B. and {Hoblitt}, J. and {Hogan}, Craig J. and {Hogg}, David W. and {Holtzman}, Jon A. and {Hyde}, Joseph B. and {Ichikawa}, Shin-ichi and {Ichikawa}, Takashi and {Im}, Myungshin and {Ivezi{\'c}}, {\v{Z}}eljko and {Jester}, Sebastian and {Jiang}, Linhua and {Johnson}, Jennifer A. and {Jorgensen}, Anders M. and {Juri{\'c}}, Mario and {Kent}, Stephen M. and {Kessler}, R. and {Kleinman}, S.~J. and {Knapp}, G.~R. and {Konishi}, Kohki and {Kron}, Richard G. and {Krzesinski}, Jurek and {Kuropatkin}, Nikolay and {Lampeitl}, Hubert and {Lebedeva}, Svetlana and {Lee}, Myung Gyoon and {Lee}, Young Sun and {French Leger}, R. and {L{\'e}pine}, S{\'e}bastien and {Li}, Nolan and {Lima}, Marcos and {Lin}, Huan and {Long}, Daniel C. and {Loomis}, Craig P. and {Loveday}, Jon and {Lupton}, Robert H. and {Magnier}, Eugene and {Malanushenko}, Olena and {Malanushenko}, Viktor and {Mandelbaum}, Rachel and {Margon}, Bruce and {Marriner}, John P. and {Mart{\'\i}nez-Delgado}, David and {Matsubara}, Takahiko and {McGehee}, Peregrine M. and {McKay}, Timothy A. and {Meiksin}, Avery and {Morrison}, Heather L. and {Mullally}, Fergal and {Munn}, Jeffrey A. and {Murphy}, Tara and {Nash}, Thomas and {Nebot}, Ada and {Neilsen}, Jr., Eric H. and {Newberg}, Heidi Jo and {Newman}, Peter R. and {Nichol}, Robert C. and {Nicinski}, Tom and {Nieto-Santisteban}, Maria and {Nitta}, Atsuko and {Okamura}, Sadanori and {Oravetz}, Daniel J. and {Ostriker}, Jeremiah P. and {Owen}, Russell and {Padmanabhan}, Nikhil and {Pan}, Kaike and {Park}, Changbom and {Pauls}, George and {Peoples}, Jr., John and {Percival}, Will J. and {Pier}, Jeffrey R. and {Pope}, Adrian C. and {Pourbaix}, Dimitri and {Price}, Paul A. and {Purger}, Norbert and {Quinn}, Thomas and {Raddick}, M. Jordan and {Re Fiorentin}, Paola and {Richards}, Gordon T. and {Richmond}, Michael W. and {Riess}, Adam G. and {Rix}, Hans-Walter and {Rockosi}, Constance M. and {Sako}, Masao and {Schlegel}, David J. and {Schneider}, Donald P. and {Scholz}, Ralf-Dieter and {Schreiber}, Matthias R. and {Schwope}, Axel D. and {Seljak}, Uro{\v{s}} and {Sesar}, Branimir and {Sheldon}, Erin and {Shimasaku}, Kazu and {Sibley}, Valena C. and {Simmons}, A.~E. and {Sivarani}, Thirupathi and {Allyn Smith}, J. and {Smith}, Martin C. and {Smol{\v{c}}i{\'c}}, Vernesa and {Snedden}, Stephanie A. and {Stebbins}, Albert and {Steinmetz}, Matthias and {Stoughton}, Chris and {Strauss}, Michael A. and {SubbaRao}, Mark and {Suto}, Yasushi and {Szalay}, Alexander S. and {Szapudi}, Istv{\'a}n and {Szkody}, Paula and {Tanaka}, Masayuki and {Tegmark}, Max and {Teodoro}, Luis F.~A. and {Thakar}, Aniruddha R. and {Tremonti}, Christy A. and {Tucker}, Douglas L. and {Uomoto}, Alan and {Vanden Berk}, Daniel E. and {Vandenberg}, Jan and {Vidrih}, S. and {Vogeley}, Michael S. and {Voges}, Wolfgang and {Vogt}, Nicole P. and {Wadadekar}, Yogesh and {Watters}, Shannon and {Weinberg}, David H. and {West}, Andrew A. and {White}, Simon D.~M. and {Wilhite}, Brian C. and {Wonders}, Alainna C. and {Yanny}, Brian and {Yocum}, D.~R.},
        title = "{The Seventh Data Release of the Sloan Digital Sky Survey}",
      journal = {\apjs},
         year = 2009,
        month = jun,
       volume = {182},
       number = {2},
        pages = {543-558},
          doi = {10.1088/0067-0049/182/2/543},
archivePrefix = {arXiv},
       eprint = {0812.0649},
 primaryClass = {astro-ph},
       adsurl = {https://ui.adsabs.harvard.edu/abs/2009ApJS..182..543A}
}

@ARTICLE{Feldmann2006,
       author = {{Feldmann}, R. and {Carollo}, C.~M. and {Porciani}, C. and {Lilly}, S.~J. and {Capak}, P. and {Taniguchi}, Y. and {Le F{\`e}vre}, O. and {Renzini}, A. and {Scoville}, N. and {Ajiki}, M. and {Aussel}, H. and {Contini}, T. and {McCracken}, H. and {Mobasher}, B. and {Murayama}, T. and {Sanders}, D. and {Sasaki}, S. and {Scarlata}, C. and {Scodeggio}, M. and {Shioya}, Y. and {Silverman}, J. and {Takahashi}, M. and {Thompson}, D. and {Zamorani}, G.},
        title = "{The Zurich Extragalactic Bayesian Redshift Analyzer and its first application: COSMOS}",
      journal = {\mnras},
         year = 2006,
        month = oct,
       volume = {372},
       number = {2},
        pages = {565-577},
          doi = {10.1111/j.1365-2966.2006.10930.x},
archivePrefix = {arXiv},
       eprint = {astro-ph/0609044},
 primaryClass = {astro-ph},
       adsurl = {https://ui.adsabs.harvard.edu/abs/2006MNRAS.372..565F}
}

@ARTICLE{Sheldon2004,
       author = {{Sheldon}, Erin S. and {Johnston}, David E. and {Frieman}, Joshua A. and {Scranton}, Ryan and {McKay}, Timothy A. and {Connolly}, A.~J. and {Budav{\'a}ri}, Tam{\'a}s and {Zehavi}, Idit and {Bahcall}, Neta A. and {Brinkmann}, J. and et al.},
        title = "{The Galaxy-Mass Correlation Function Measured from Weak Lensing in the Sloan Digital Sky Survey}",
      journal = {\aj},
         year = 2004,
        month = may,
       volume = {127},
       number = {5},
        pages = {2544-2564},
          doi = {10.1086/383293},
archivePrefix = {arXiv},
       eprint = {astro-ph/0312036},
 primaryClass = {astro-ph},
       adsurl = {https://ui.adsabs.harvard.edu/abs/2004AJ....127.2544S}
}

@ARTICLE{Mandelbaum2005,
       author = {{Mandelbaum}, Rachel and {Hirata}, Christopher M. and {Seljak}, Uro{\v{s}} and {Guzik}, Jacek and {Padmanabhan}, Nikhil and {Blake}, Cullen and {Blanton}, Michael R. and {Lupton}, Robert and {Brinkmann}, Jonathan},
        title = "{Systematic errors in weak lensing: application to SDSS galaxy-galaxy weak lensing}",
      journal = {\mnras},
         year = 2005,
        month = aug,
       volume = {361},
       number = {4},
        pages = {1287-1322},
          doi = {10.1111/j.1365-2966.2005.09282.x},
archivePrefix = {arXiv},
       eprint = {astro-ph/0501201},
 primaryClass = {astro-ph},
       adsurl = {https://ui.adsabs.harvard.edu/abs/2005MNRAS.361.1287M}
}

@ARTICLE{Singh2017,
       author = {{Singh}, Sukhdeep and {Mandelbaum}, Rachel and {Seljak}, Uro{\v{s}} and {Slosar}, An{\v{z}}e and {Vazquez Gonzalez}, Jose},
        title = "{Galaxy-galaxy lensing estimators and their covariance properties}",
      journal = {\mnras},
         year = 2017,
        month = nov,
       volume = {471},
       number = {4},
        pages = {3827-3844},
          doi = {10.1093/mnras/stx1828},
archivePrefix = {arXiv},
       eprint = {1611.00752},
 primaryClass = {astro-ph.CO},
       adsurl = {https://ui.adsabs.harvard.edu/abs/2017MNRAS.471.3827S}
}

@ARTICLE{Peacock2000,
       author = {{Peacock}, J.~A. and {Smith}, R.~E.},
        title = "{Halo occupation numbers and galaxy bias}",
      journal = {\mnras},
         year = 2000,
        month = nov,
       volume = {318},
       number = {4},
        pages = {1144-1156},
          doi = {10.1046/j.1365-8711.2000.03779.x},
archivePrefix = {arXiv},
       eprint = {astro-ph/0005010},
 primaryClass = {astro-ph},
       adsurl = {https://ui.adsabs.harvard.edu/abs/2000MNRAS.318.1144P}
}

@ARTICLE{Scoccimarro2001,
       author = {{Scoccimarro}, Rom{\'a}n and {Sheth}, Ravi K. and {Hui}, Lam and {Jain}, Bhuvnesh},
        title = "{How Many Galaxies Fit in a Halo? Constraints on Galaxy Formation Efficiency from Spatial Clustering}",
      journal = {\apj},
         year = 2001,
        month = jan,
       volume = {546},
       number = {1},
        pages = {20-34},
          doi = {10.1086/318261},
archivePrefix = {arXiv},
       eprint = {astro-ph/0006319},
 primaryClass = {astro-ph},
       adsurl = {https://ui.adsabs.harvard.edu/abs/2001ApJ...546...20S}
}

@ARTICLE{Berlind2002,
       author = {{Berlind}, Andreas A. and {Weinberg}, David H.},
        title = "{The Halo Occupation Distribution: Toward an Empirical Determination of the Relation between Galaxies and Mass}",
      journal = {\apj},
         year = 2002,
        month = aug,
       volume = {575},
       number = {2},
        pages = {587-616},
          doi = {10.1086/341469},
archivePrefix = {arXiv},
       eprint = {astro-ph/0109001},
 primaryClass = {astro-ph},
       adsurl = {https://ui.adsabs.harvard.edu/abs/2002ApJ...575..587B}
}

@ARTICLE{Zheng2005,
       author = {{Zheng}, Zheng and {Berlind}, Andreas A. and {Weinberg}, David H. and {Benson}, Andrew J. and {Baugh}, Carlton M. and {Cole}, Shaun and {Dav{\'e}}, Romeel and {Frenk}, Carlos S. and {Katz}, Neal and {Lacey}, Cedric G.},
        title = "{Theoretical Models of the Halo Occupation Distribution: Separating Central and Satellite Galaxies}",
      journal = {\apj},
         year = 2005,
        month = nov,
       volume = {633},
       number = {2},
        pages = {791-809},
          doi = {10.1086/466510},
archivePrefix = {arXiv},
       eprint = {astro-ph/0408564},
 primaryClass = {astro-ph},
       adsurl = {https://ui.adsabs.harvard.edu/abs/2005ApJ...633..791Z}
}

@ARTICLE{NFW1997,
       author = {{Navarro}, Julio F. and {Frenk}, Carlos S. and {White}, Simon D.~M.},
        title = "{A Universal Density Profile from Hierarchical Clustering}",
      journal = {\apj},
         year = 1997,
        month = dec,
       volume = {490},
       number = {2},
        pages = {493-508},
          doi = {10.1086/304888},
archivePrefix = {arXiv},
       eprint = {astro-ph/9611107},
 primaryClass = {astro-ph},
       adsurl = {https://ui.adsabs.harvard.edu/abs/1997ApJ...490..493N}
}

@ARTICLE{Zheng2007,
       author = {{Zheng}, Zheng and {Weinberg}, David H.},
        title = "{Breaking the Degeneracies between Cosmology and Galaxy Bias}",
      journal = {\apj},
         year = 2007,
        month = apr,
       volume = {659},
       number = {1},
        pages = {1-28},
          doi = {10.1086/512151},
archivePrefix = {arXiv},
       eprint = {astro-ph/0512071},
 primaryClass = {astro-ph},
       adsurl = {https://ui.adsabs.harvard.edu/abs/2007ApJ...659....1Z}
}

@ARTICLE{Zhao2009,
       author = {{Zhao}, D.~H. and {Jing}, Y.~P. and {Mo}, H.~J. and {B{\"o}rner}, G.},
        title = "{Accurate Universal Models for the Mass Accretion Histories and Concentrations of Dark Matter Halos}",
      journal = {\apj},
         year = 2009,
        month = dec,
       volume = {707},
       number = {1},
        pages = {354-369},
          doi = {10.1088/0004-637X/707/1/354},
archivePrefix = {arXiv},
       eprint = {0811.0828},
 primaryClass = {astro-ph},
       adsurl = {https://ui.adsabs.harvard.edu/abs/2009ApJ...707..354Z}
}

@ARTICLE{Takahashi2012,
       author = {{Takahashi}, Ryuichi and {Sato}, Masanori and {Nishimichi}, Takahiro and {Taruya}, Atsushi and {Oguri}, Masamune},
        title = "{Revising the Halofit Model for the Nonlinear Matter Power Spectrum}",
      journal = {\apj},
         year = 2012,
        month = dec,
       volume = {761},
       number = {2},
          eid = {152},
        pages = {152},
          doi = {10.1088/0004-637X/761/2/152},
archivePrefix = {arXiv},
       eprint = {1208.2701},
 primaryClass = {astro-ph.CO},
       adsurl = {https://ui.adsabs.harvard.edu/abs/2012ApJ...761..152T}
}

@ARTICLE{Tinker2005,
       author = {{Tinker}, Jeremy L. and {Weinberg}, David H. and {Zheng}, Zheng and {Zehavi}, Idit},
        title = "{On the Mass-to-Light Ratio of Large-Scale Structure}",
      journal = {\apj},
         year = 2005,
        month = sep,
       volume = {631},
       number = {1},
        pages = {41-58},
          doi = {10.1086/432084},
archivePrefix = {arXiv},
       eprint = {astro-ph/0411777},
 primaryClass = {astro-ph},
       adsurl = {https://ui.adsabs.harvard.edu/abs/2005ApJ...631...41T}
}

@ARTICLE{Tinker2010,
       author = {{Tinker}, Jeremy L. and {Robertson}, Brant E. and {Kravtsov}, Andrey V. and {Klypin}, Anatoly and {Warren}, Michael S. and {Yepes}, Gustavo and {Gottl{\"o}ber}, Stefan},
        title = "{The Large-scale Bias of Dark Matter Halos: Numerical Calibration and Model Tests}",
      journal = {\apj},
         year = 2010,
        month = dec,
       volume = {724},
       number = {2},
        pages = {878-886},
          doi = {10.1088/0004-637X/724/2/878},
archivePrefix = {arXiv},
       eprint = {1001.3162},
 primaryClass = {astro-ph.CO},
       adsurl = {https://ui.adsabs.harvard.edu/abs/2010ApJ...724..878T}
}

@ARTICLE{Yoo2006,
       author = {{Yoo}, Jaiyul and {Tinker}, Jeremy L. and {Weinberg}, David H. and {Zheng}, Zheng and {Katz}, Neal and {Dav{\'e}}, Romeel},
        title = "{From Galaxy-Galaxy Lensing to Cosmological Parameters}",
      journal = {\apj},
         year = 2006,
        month = nov,
       volume = {652},
       number = {1},
        pages = {26-42},
          doi = {10.1086/507591},
archivePrefix = {arXiv},
       eprint = {astro-ph/0511580},
 primaryClass = {astro-ph},
       adsurl = {https://ui.adsabs.harvard.edu/abs/2006ApJ...652...26Y}
}

@ARTICLE{vdbosch2013,
       author = {{van den Bosch}, Frank C. and {More}, Surhud and {Cacciato}, Marcello and {Mo}, Houjun and {Yang}, Xiaohu},
        title = "{Cosmological constraints from a combination of galaxy clustering and lensing - I. Theoretical framework}",
      journal = {\mnras},
         year = 2013,
        month = apr,
       volume = {430},
       number = {2},
        pages = {725-746},
          doi = {10.1093/mnras/sts006},
archivePrefix = {arXiv},
       eprint = {1206.6890},
 primaryClass = {astro-ph.CO},
       adsurl = {https://ui.adsabs.harvard.edu/abs/2013MNRAS.430..725V}
}

@ARTICLE{Nautilus2023,
       author = {{Lange}, Johannes U.},
        title = "{NAUTILUS: boosting Bayesian importance nested sampling with deep learning}",
      journal = {\mnras},
         year = 2023,
        month = oct,
       volume = {525},
       number = {2},
        pages = {3181-3194},
          doi = {10.1093/mnras/stad2441},
archivePrefix = {arXiv},
       eprint = {2306.16923},
 primaryClass = {astro-ph.IM},
       adsurl = {https://ui.adsabs.harvard.edu/abs/2023MNRAS.525.3181L}
}

@ARTICLE{Treiber2025,
       author = {{Treiber}, Helena and {Amon}, Alexandra and {Wechsler}, Risa H. and {Manwadkar}, Viraj and {Myles}, Justin and {Hahn}, ChangHoon and {Hearin}, Andrew and {Heydenreich}, Sven and {Saintonge}, Am{\'e}lie and {Saraf}, Manasvee and et al.},
        title = "{Dwarf galaxy halo masses from spectroscopic and photometric lensing in DESI and DES}",
      journal = {arXiv e-prints},
         year = 2025,
        month = sep,
          eid = {arXiv:2509.20434},
        pages = {arXiv:2509.20434},
          doi = {10.48550/arXiv.2509.20434},
archivePrefix = {arXiv},
       eprint = {2509.20434},
 primaryClass = {astro-ph.GA},
       adsurl = {https://ui.adsabs.harvard.edu/abs/2025arXiv250920434T}
}

@ARTICLE{Mandelbaum2016,
       author = {{Mandelbaum}, Rachel and {Wang}, Wenting and {Zu}, Ying and {White}, Simon and {Henriques}, Bruno and {More}, Surhud},
        title = "{Strong bimodality in the host halo mass of central galaxies from galaxy-galaxy lensing}",
      journal = {\mnras},
         year = 2016,
        month = apr,
       volume = {457},
       number = {3},
        pages = {3200-3218},
          doi = {10.1093/mnras/stw188},
archivePrefix = {arXiv},
       eprint = {1509.06762},
 primaryClass = {astro-ph.GA},
       adsurl = {https://ui.adsabs.harvard.edu/abs/2016MNRAS.457.3200M}
}

@ARTICLE{Oh2015,
       author = {{Oh}, Se-Heon and {Hunter}, Deidre A. and {Brinks}, Elias and {Elmegreen}, Bruce G. and {Schruba}, Andreas and {Walter}, Fabian and {Rupen}, Michael P. and {Young}, Lisa M. and {Simpson}, Caroline E. and {Johnson}, Megan C. and et al.},
        title = "{High-resolution Mass Models of Dwarf Galaxies from LITTLE THINGS}",
      journal = {\aj},
         year = 2015,
        month = jun,
       volume = {149},
       number = {6},
          eid = {180},
        pages = {180},
          doi = {10.1088/0004-6256/149/6/180},
archivePrefix = {arXiv},
       eprint = {1502.01281},
 primaryClass = {astro-ph.GA},
       adsurl = {https://ui.adsabs.harvard.edu/abs/2015AJ....149..180O}
}

@ARTICLE{Read2017,
       author = {{Read}, J.~I. and {Iorio}, G. and {Agertz}, O. and {Fraternali}, F.},
        title = "{The stellar mass-halo mass relation of isolated field dwarfs: a critical test of {\ensuremath{\Lambda}}CDM at the edge of galaxy formation}",
      journal = {\mnras},
         year = 2017,
        month = may,
       volume = {467},
       number = {2},
        pages = {2019-2038},
          doi = {10.1093/mnras/stx147},
archivePrefix = {arXiv},
       eprint = {1607.03127},
 primaryClass = {astro-ph.GA},
       adsurl = {https://ui.adsabs.harvard.edu/abs/2017MNRAS.467.2019R}
}

@ARTICLE{Kim2026,
       author = {{Kim}, Stacy Y. and {Read}, Justin I. and {Rey}, Martin P. and {Orkney}, Matthew D.~A. and {Nigudkar}, Sushanta and {Pontzen}, Andrew and {Taylor}, Ethan and {Agertz}, Oscar and {Das}, Payel},
        title = "{EDGE: Predictable Scatter in the Stellar Mass-Halo Mass Relation of Dwarf Galaxies}",
      journal = {\mnras},
         year = 2026,
        month = apr,
          doi = {10.1093/mnras/stag825},
archivePrefix = {arXiv},
       eprint = {2408.15214},
 primaryClass = {astro-ph.GA},
       adsurl = {https://ui.adsabs.harvard.edu/abs/2026MNRAS.tmp..763K}
}

@ARTICLE{ManceraPina2025,
       author = {{Mancera Pi{\~n}a}, Pavel E. and {Read}, Justin I. and {Kim}, Stacy and {Marasco}, Antonino and {Benavides}, Jos{\'e} A. and {Glowacki}, Marcin and {Pezzulli}, Gabriele and {Lagos}, Claudia del P.},
        title = "{The galaxy-halo connection of disc galaxies over six orders of magnitude in stellar mass}",
      journal = {\aap},
         year = 2025,
        month = jul,
       volume = {699},
          eid = {A311},
        pages = {A311},
          doi = {10.1051/0004-6361/202554381},
archivePrefix = {arXiv},
       eprint = {2505.22727},
 primaryClass = {astro-ph.GA},
       adsurl = {https://ui.adsabs.harvard.edu/abs/2025A&A...699A.311M}
}

@ARTICLE{Wang2024CLF,
       author = {{Wang}, Yirong and {Yang}, Xiaohu and {Gu}, Yizhou and {Xu}, Xiaoju and {Xu}, Haojie and {Wang}, Yuyu and {Katsianis}, Antonios and {Han}, Jiaxin and {He}, Min and {Zheng}, Yunliang and et al.},
        title = "{Measuring the Conditional Luminosity and Stellar Mass Functions of Galaxies by Combining the Dark Energy Spectroscopic Instrument Legacy Imaging Surveys Data Release 9, Survey Validation 3, and Year 1 Data}",
      journal = {\apj},
         year = 2024,
        month = aug,
       volume = {971},
       number = {1},
          eid = {119},
        pages = {119},
          doi = {10.3847/1538-4357/ad5294},
archivePrefix = {arXiv},
       eprint = {2312.17459},
 primaryClass = {astro-ph.GA},
       adsurl = {https://ui.adsabs.harvard.edu/abs/2024ApJ...971..119W}
}

@ARTICLE{Pillepich2019,
       author = {{Pillepich}, Annalisa and {Nelson}, Dylan and {Springel}, Volker and {Pakmor}, R{\"u}diger and {Torrey}, Paul and {Weinberger}, Rainer and {Vogelsberger}, Mark and {Marinacci}, Federico and {Genel}, Shy and {van der Wel}, Arjen and {Hernquist}, Lars},
        title = "{First results from the TNG50 simulation: the evolution of stellar and gaseous discs across cosmic time}",
      journal = {\mnras},
         year = 2019,
        month = dec,
       volume = {490},
       number = {3},
        pages = {3196-3233},
          doi = {10.1093/mnras/stz2338},
archivePrefix = {arXiv},
       eprint = {1902.05553},
 primaryClass = {astro-ph.GA},
       adsurl = {https://ui.adsabs.harvard.edu/abs/2019MNRAS.490.3196P}
}

@ARTICLE{Schaye2015,
       author = {{Schaye}, Joop and {Crain}, Robert A. and {Bower}, Richard G. and {Furlong}, Michelle and {Schaller}, Matthieu and {Theuns}, Tom and {Dalla Vecchia}, Claudio and {Frenk}, Carlos S. and {McCarthy}, I.~G. and {Helly}, John C. and {Jenkins}, Adrian and {Rosas-Guevara}, Y.~M. and {White}, Simon D.~M. and {Baes}, Maarten and {Booth}, C.~M. and {Camps}, Peter and {Navarro}, Julio F. and {Qu}, Yan and {Rahmati}, Alireza and {Sawala}, Till and {Thomas}, Peter A. and {Trayford}, James},
        title = "{The EAGLE project: simulating the evolution and assembly of galaxies and their environments}",
      journal = {\mnras},
         year = 2015,
        month = jan,
       volume = {446},
       number = {1},
        pages = {521-554},
          doi = {10.1093/mnras/stu2058},
archivePrefix = {arXiv},
       eprint = {1407.7040},
 primaryClass = {astro-ph.GA},
       adsurl = {https://ui.adsabs.harvard.edu/abs/2015MNRAS.446..521S}
}

@ARTICLE{Xu2025,
       author = {{Xu}, Kun and {Jing}, Y.~P. and {Cole}, S. and {Frenk}, C.~S. and {Bose}, S. and {Elbers}, W. and {Wang}, W. and {Wang}, Yirong and {Moore}, S. and {Aguilar}, J. and et al.},
        title = "{PAC in DESI. I. Galaxy stellar mass function into the {}10$^{6}$ M$_{{\ensuremath{\odot}}}$ frontier}",
      journal = {\mnras},
         year = 2025,
        month = jun,
       volume = {540},
       number = {2},
        pages = {1635-1667},
          doi = {10.1093/mnras/staf782},
archivePrefix = {arXiv},
       eprint = {2503.01948},
 primaryClass = {astro-ph.GA},
       adsurl = {https://ui.adsabs.harvard.edu/abs/2025MNRAS.540.1635X}
}

@ARTICLE{Yang2012,
       author = {{Yang}, Xiaohu and {Mo}, H.~J. and {van den Bosch}, Frank C. and {Zhang}, Youcai and {Han}, Jiaxin},
        title = "{Evolution of the Galaxy-Dark Matter Connection and the Assembly of Galaxies in Dark Matter Halos}",
      journal = {\apj},
         year = 2012,
        month = jun,
       volume = {752},
       number = {1},
          eid = {41},
        pages = {41},
          doi = {10.1088/0004-637X/752/1/41},
archivePrefix = {arXiv},
       eprint = {1110.1420},
 primaryClass = {astro-ph.CO},
       adsurl = {https://ui.adsabs.harvard.edu/abs/2012ApJ...752...41Y}
}

@ARTICLE{Moster2018,
       author = {{Moster}, Benjamin P. and {Naab}, Thorsten and {White}, Simon D.~M.},
        title = "{EMERGE - an empirical model for the formation of galaxies since z {\ensuremath{\sim}} 10}",
      journal = {\mnras},
         year = 2018,
        month = jun,
       volume = {477},
       number = {2},
        pages = {1822-1852},
          doi = {10.1093/mnras/sty655},
archivePrefix = {arXiv},
       eprint = {1705.05373},
 primaryClass = {astro-ph.GA},
       adsurl = {https://ui.adsabs.harvard.edu/abs/2018MNRAS.477.1822M}
}

@ARTICLE{Behroozi2019,
       author = {{Behroozi}, Peter and {Wechsler}, Risa H. and {Hearin}, Andrew P. and {Conroy}, Charlie},
        title = "{UNIVERSEMACHINE: The correlation between galaxy growth and dark matter halo assembly from z = 0-10}",
      journal = {\mnras},
         year = 2019,
        month = sep,
       volume = {488},
       number = {3},
        pages = {3143-3194},
          doi = {10.1093/mnras/stz1182},
archivePrefix = {arXiv},
       eprint = {1806.07893},
 primaryClass = {astro-ph.GA},
       adsurl = {https://ui.adsabs.harvard.edu/abs/2019MNRAS.488.3143B}
}

@ARTICLE{Kado-Fong2025,
       author = {{Kado-Fong}, Erin and {Mao}, Yao-Yuan and {Asali}, Yasmeen and {Geha}, Marla and {Wechsler}, Risa H. and {de los Reyes}, Mithi A.~C. and {Wang}, Yunchong and {Nadler}, Ethan O. and {Kallivayalil}, Nitya and {Tollerud}, Erik J. and et al.},
        title = "{SAGAbg III: Environmental Stellar Mass Functions, Self-Quenching, and the Stellar-to-Halo Mass Relation in the Dwarf Galaxy Regime}",
      journal = {arXiv e-prints},
         year = 2025,
        month = sep,
          eid = {arXiv:2509.20444},
        pages = {arXiv:2509.20444},
          doi = {10.48550/arXiv.2509.20444},
archivePrefix = {arXiv},
       eprint = {2509.20444},
 primaryClass = {astro-ph.GA},
       adsurl = {https://ui.adsabs.harvard.edu/abs/2025arXiv250920444K}
}

@ARTICLE{Danieli2023,
       author = {{Danieli}, Shany and {Greene}, Jenny E. and {Carlsten}, Scott and {Jiang}, Fangzhou and {Beaton}, Rachael and {Goulding}, Andy D.},
        title = "{ELVES. IV. The Satellite Stellar-to-halo Mass Relation Beyond the Milky Way}",
      journal = {\apj},
         year = 2023,
        month = oct,
       volume = {956},
       number = {1},
          eid = {6},
        pages = {6},
          doi = {10.3847/1538-4357/acefbd},
archivePrefix = {arXiv},
       eprint = {2210.14233},
 primaryClass = {astro-ph.GA},
       adsurl = {https://ui.adsabs.harvard.edu/abs/2023ApJ...956....6D}
}

@ARTICLE{Manwadkar2022,
       author = {{Manwadkar}, Viraj and {Kravtsov}, Andrey V.},
        title = "{Forward-modelling the luminosity, distance, and size distributions of the Milky Way satellites}",
      journal = {\mnras},
         year = 2022,
        month = nov,
       volume = {516},
       number = {3},
        pages = {3944-3971},
          doi = {10.1093/mnras/stac2452},
archivePrefix = {arXiv},
       eprint = {2112.04511},
 primaryClass = {astro-ph.GA},
       adsurl = {https://ui.adsabs.harvard.edu/abs/2022MNRAS.516.3944M}
}

@ARTICLE{Nadler2020,
       author = {{Nadler}, E.~O. and {Wechsler}, R.~H. and {Bechtol}, K. and {Mao}, Y.-Y. and {Green}, G. and {Drlica-Wagner}, A. and {McNanna}, M. and {Mau}, S. and {Pace}, A.~B. and {Simon}, J.~D. and et al.},
        title = "{Milky Way Satellite Census. II. Galaxy-Halo Connection Constraints Including the Impact of the Large Magellanic Cloud}",
      journal = {\apj},
         year = 2020,
        month = apr,
       volume = {893},
       number = {1},
          eid = {48},
        pages = {48},
          doi = {10.3847/1538-4357/ab846a},
archivePrefix = {arXiv},
       eprint = {1912.03303},
 primaryClass = {astro-ph.GA},
       adsurl = {https://ui.adsabs.harvard.edu/abs/2020ApJ...893...48N}
}

@ARTICLE{Anbajagane2025,
       author = {{Anbajagane}, D. and {Chang}, C. and {Zhang}, Z. and {Tan}, C.~Y. and {Adamow}, M. and {Secco}, L.~F. and {Becker}, M.~R. and {Ferguson}, P.~S. and {Drlica-Wagner}, A. and {Gruendl}, R.~A. and et al.},
        title = "{The DECADE cosmic shear project I: A new weak lensing shape catalog of 107 million galaxies}",
      journal = {arXiv e-prints},
         year = 2025,
        month = feb,
          eid = {arXiv:2502.17674},
        pages = {arXiv:2502.17674},
          doi = {10.48550/arXiv.2502.17674},
archivePrefix = {arXiv},
       eprint = {2502.17674},
 primaryClass = {astro-ph.CO},
       adsurl = {https://ui.adsabs.harvard.edu/abs/2025arXiv250217674A}
}

@ARTICLE{Lagos2011,
       author = {{Lagos}, Claudia Del P. and {Baugh}, Carlton M. and {Lacey}, Cedric G. and {Benson}, Andrew J. and {Kim}, Han-Seek and {Power}, Chris},
        title = "{Cosmic evolution of the atomic and molecular gas contents of galaxies}",
      journal = {\mnras},
         year = 2011,
        month = dec,
       volume = {418},
       number = {3},
        pages = {1649-1667},
          doi = {10.1111/j.1365-2966.2011.19583.x},
archivePrefix = {arXiv},
       eprint = {1105.2294},
 primaryClass = {astro-ph.CO},
       adsurl = {https://ui.adsabs.harvard.edu/abs/2011MNRAS.418.1649L}
}

@ARTICLE{Guo2023,
       author = {{Guo}, Hong and {Wang}, Jing and {Jones}, Michael G. and {Behroozi}, Peter},
        title = "{NeutralUniverseMachine: An Empirical Model for the Evolution of H I and H$_{2}$ Gas in the Universe}",
      journal = {\apj},
         year = 2023,
        month = sep,
       volume = {955},
       number = {1},
          eid = {57},
        pages = {57},
          doi = {10.3847/1538-4357/aced47},
archivePrefix = {arXiv},
       eprint = {2307.07078},
 primaryClass = {astro-ph.GA},
       adsurl = {https://ui.adsabs.harvard.edu/abs/2023ApJ...955...57G}
}

@ARTICLE{Leauthaud2011,
       author = {{Leauthaud}, Alexie and {Tinker}, Jeremy and {Behroozi}, Peter S. and {Busha}, Michael T. and {Wechsler}, Risa H.},
        title = "{A Theoretical Framework for Combining Techniques that Probe the Link Between Galaxies and Dark Matter}",
      journal = {\apj},
         year = 2011,
        month = sep,
       volume = {738},
       number = {1},
          eid = {45},
        pages = {45},
          doi = {10.1088/0004-637X/738/1/45},
archivePrefix = {arXiv},
       eprint = {1103.2077},
 primaryClass = {astro-ph.CO},
       adsurl = {https://ui.adsabs.harvard.edu/abs/2011ApJ...738...45L}
}

@ARTICLE{Munshi2021,
       author = {{Munshi}, Ferah and {Brooks}, Alyson M. and {Applebaum}, Elaad and {Christensen}, Charlotte R. and {Quinn}, T. and {Sligh}, Serena},
        title = "{Quantifying Scatter in Galaxy Formation at the Lowest Masses}",
      journal = {\apj},
         year = 2021,
        month = dec,
       volume = {923},
       number = {1},
          eid = {35},
        pages = {35},
          doi = {10.3847/1538-4357/ac0db6},
archivePrefix = {arXiv},
       eprint = {2101.05822},
 primaryClass = {astro-ph.GA},
       adsurl = {https://ui.adsabs.harvard.edu/abs/2021ApJ...923...35M}
}

@ARTICLE{Carlsten2022,
       author = {{Carlsten}, Scott G. and {Greene}, Jenny E. and {Beaton}, Rachael L. and {Danieli}, Shany and {Greco}, Johnny P.},
        title = "{The Exploration of Local VolumE Satellites (ELVES) Survey: A Nearly Volume-limited Sample of Nearby Dwarf Satellite Systems}",
      journal = {\apj},
         year = 2022,
        month = jul,
       volume = {933},
       number = {1},
          eid = {47},
        pages = {47},
          doi = {10.3847/1538-4357/ac6fd7},
archivePrefix = {arXiv},
       eprint = {2203.00014},
 primaryClass = {astro-ph.GA},
       adsurl = {https://ui.adsabs.harvard.edu/abs/2022ApJ...933...47C}
}

@ARTICLE{Drlica-Wagner2020,
       author = {{Drlica-Wagner}, A. and {Bechtol}, K. and {Mau}, S. and {McNanna}, M. and {Nadler}, E.~O. and {Pace}, A.~B. and {Li}, T.~S. and {Pieres}, A. and {Rozo}, E. and {Simon}, J.~D. and et al.},
        title = "{Milky Way Satellite Census. I. The Observational Selection Function for Milky Way Satellites in DES Y3 and Pan-STARRS DR1}",
      journal = {\apj},
         year = 2020,
        month = apr,
       volume = {893},
       number = {1},
          eid = {47},
        pages = {47},
          doi = {10.3847/1538-4357/ab7eb9},
archivePrefix = {arXiv},
       eprint = {1912.03302},
 primaryClass = {astro-ph.GA},
       adsurl = {https://ui.adsabs.harvard.edu/abs/2020ApJ...893...47D}
}

@ARTICLE{Monzon2024,
       author = {{Monzon}, J. Sebastian and {van den Bosch}, Frank C. and {Mitra}, Kaustav},
        title = "{Constraining the Low-mass End of the Stellar-to-halo Mass Relation with Surveys of Satellite Galaxies}",
      journal = {\apj},
         year = 2024,
        month = dec,
       volume = {976},
       number = {2},
          eid = {197},
        pages = {197},
          doi = {10.3847/1538-4357/ad834e},
archivePrefix = {arXiv},
       eprint = {2410.02873},
 primaryClass = {astro-ph.GA},
       adsurl = {https://ui.adsabs.harvard.edu/abs/2024ApJ...976..197M}
}

@ARTICLE{Garrison-Kimmel2017,
       author = {{Garrison-Kimmel}, Shea and {Bullock}, James S. and {Boylan-Kolchin}, Michael and {Bardwell}, Emma},
        title = "{Organized chaos: scatter in the relation between stellar mass and halo mass in small galaxies}",
      journal = {\mnras},
         year = 2017,
        month = jan,
       volume = {464},
       number = {3},
        pages = {3108-3120},
          doi = {10.1093/mnras/stw2564},
archivePrefix = {arXiv},
       eprint = {1603.04855},
 primaryClass = {astro-ph.GA},
       adsurl = {https://ui.adsabs.harvard.edu/abs/2017MNRAS.464.3108G}
}

@ARTICLE{Bullock2017,
       author = {{Bullock}, James S. and {Boylan-Kolchin}, Michael},
        title = "{Small-Scale Challenges to the {\ensuremath{\Lambda}}CDM Paradigm}",
      journal = {\araa},
         year = 2017,
        month = aug,
       volume = {55},
       number = {1},
        pages = {343-387},
          doi = {10.1146/annurev-astro-091916-055313},
archivePrefix = {arXiv},
       eprint = {1707.04256},
 primaryClass = {astro-ph.CO},
       adsurl = {https://ui.adsabs.harvard.edu/abs/2017ARA&A..55..343B}
}

@ARTICLE{Brook2014,
       author = {{Brook}, C.~B. and {Di Cintio}, A. and {Knebe}, A. and {Gottl{\"o}ber}, S. and {Hoffman}, Y. and {Yepes}, G. and {Garrison-Kimmel}, S.},
        title = "{The Stellar-to-halo Mass Relation for Local Group Galaxies}",
      journal = {\apjl},
         year = 2014,
        month = mar,
       volume = {784},
       number = {1},
          eid = {L14},
        pages = {L14},
          doi = {10.1088/2041-8205/784/1/L14},
archivePrefix = {arXiv},
       eprint = {1311.5492},
 primaryClass = {astro-ph.CO},
       adsurl = {https://ui.adsabs.harvard.edu/abs/2014ApJ...784L..14B}
}

@ARTICLE{Jethwa2018,
       author = {{Jethwa}, P. and {Erkal}, D. and {Belokurov}, V.},
        title = "{The upper bound on the lowest mass halo}",
      journal = {\mnras},
         year = 2018,
        month = jan,
       volume = {473},
       number = {2},
        pages = {2060-2083},
          doi = {10.1093/mnras/stx2330},
archivePrefix = {arXiv},
       eprint = {1612.07834},
 primaryClass = {astro-ph.GA},
       adsurl = {https://ui.adsabs.harvard.edu/abs/2018MNRAS.473.2060J}
}

@ARTICLE{Wang2024SAGAUM,
       author = {{Wang}, Yunchong and {Nadler}, Ethan O. and {Mao}, Yao-Yuan and {Wechsler}, Risa H. and {Abel}, Tom and {Behroozi}, Peter and {Geha}, Marla and {Asali}, Yasmeen and {de los Reyes}, Mithi A.~C. and {Kado-Fong}, Erin and {Kallivayalil}, Nitya and {Tollerud}, Erik J. and {Weiner}, Benjamin and {Wu}, John F.},
        title = "{The SAGA Survey. V. Modeling Satellite Systems around Milky Way{\textendash}Mass Galaxies with Updated UNIVERSEMACHINE}",
      journal = {\apj},
         year = 2024,
        month = nov,
       volume = {976},
       number = {1},
          eid = {119},
        pages = {119},
          doi = {10.3847/1538-4357/ad7f4c},
archivePrefix = {arXiv},
       eprint = {2404.14500},
 primaryClass = {astro-ph.GA},
       adsurl = {https://ui.adsabs.harvard.edu/abs/2024ApJ...976..119W}
}

@ARTICLE{Leauthaud2012,
       author = {{Leauthaud}, Alexie and {Tinker}, Jeremy and {Bundy}, Kevin and {Behroozi}, Peter S. and {Massey}, Richard and {Rhodes}, Jason and {George}, Matthew R. and {Kneib}, Jean-Paul and {Benson}, Andrew and {Wechsler}, Risa H. and et al.},
        title = "{New Constraints on the Evolution of the Stellar-to-dark Matter Connection: A Combined Analysis of Galaxy-Galaxy Lensing, Clustering, and Stellar Mass Functions from z = 0.2 to z =1}",
      journal = {\apj},
         year = 2012,
        month = jan,
       volume = {744},
       number = {2},
          eid = {159},
        pages = {159},
          doi = {10.1088/0004-637X/744/2/159},
archivePrefix = {arXiv},
       eprint = {1104.0928},
 primaryClass = {astro-ph.CO},
       adsurl = {https://ui.adsabs.harvard.edu/abs/2012ApJ...744..159L}
}

@ARTICLE{Posti2019,
       author = {{Posti}, Lorenzo and {Fraternali}, Filippo and {Marasco}, Antonino},
        title = "{Peak star formation efficiency and no missing baryons in massive spirals}",
      journal = {\aap},
         year = 2019,
        month = jun,
       volume = {626},
          eid = {A56},
        pages = {A56},
          doi = {10.1051/0004-6361/201935553},
archivePrefix = {arXiv},
       eprint = {1812.05099},
 primaryClass = {astro-ph.GA},
       adsurl = {https://ui.adsabs.harvard.edu/abs/2019A&A...626A..56P}
}

@ARTICLE{Thornton2024,
       author = {{Thornton}, Joseph and {Amon}, Alexandra and {Wechsler}, Risa H. and {Adhikari}, Susmita and {Mao}, Yao-Yuan and {Myles}, Justin and {Geha}, Marla and {Kallivayalil}, Nitya and {Tollerud}, Erik and {Weiner}, Benjamin},
        title = "{The mass profiles of dwarf galaxies from Dark Energy Survey lensing}",
      journal = {\mnras},
         year = 2024,
        month = nov,
       volume = {535},
       number = {1},
        pages = {1-20},
          doi = {10.1093/mnras/stae2040},
archivePrefix = {arXiv},
       eprint = {2311.14659},
 primaryClass = {astro-ph.GA},
       adsurl = {https://ui.adsabs.harvard.edu/abs/2024MNRAS.535....1T}
}

@ARTICLE{Mandelbaum2006,
       author = {{Mandelbaum}, Rachel and {Seljak}, Uro{\v{s}} and {Kauffmann}, Guinevere and {Hirata}, Christopher M. and {Brinkmann}, Jonathan},
        title = "{Galaxy halo masses and satellite fractions from galaxy-galaxy lensing in the Sloan Digital Sky Survey: stellar mass, luminosity, morphology and environment dependencies}",
      journal = {\mnras},
         year = 2006,
        month = may,
       volume = {368},
       number = {2},
        pages = {715-731},
          doi = {10.1111/j.1365-2966.2006.10156.x},
archivePrefix = {arXiv},
       eprint = {astro-ph/0511164},
 primaryClass = {astro-ph},
       adsurl = {https://ui.adsabs.harvard.edu/abs/2006MNRAS.368..715M}
}

@ARTICLE{Jing1998,
       author = {{Jing}, Y.~P. and {Mo}, H.~J. and {B{\"o}rner}, G.},
        title = "{Spatial Correlation Function and Pairwise Velocity Dispersion of Galaxies: Cold Dark Matter Models versus the Las Campanas Survey}",
      journal = {\apj},
         year = 1998,
        month = feb,
       volume = {494},
       number = {1},
        pages = {1-12},
          doi = {10.1086/305209},
archivePrefix = {arXiv},
       eprint = {astro-ph/9707106},
 primaryClass = {astro-ph},
       adsurl = {https://ui.adsabs.harvard.edu/abs/1998ApJ...494....1J}
}

@ARTICLE{Buckley2018,
       author = {{Buckley}, Matthew R. and {Peter}, Annika H.~G.},
        title = "{Gravitational probes of dark matter physics}",
      journal = {\physrep},
         year = 2018,
        month = oct,
       volume = {761},
        pages = {1-60},
          doi = {10.1016/j.physrep.2018.07.003},
archivePrefix = {arXiv},
       eprint = {1712.06615},
 primaryClass = {astro-ph.CO},
       adsurl = {https://ui.adsabs.harvard.edu/abs/2018PhR...761....1B}
}

@ARTICLE{Nadler2021,
       author = {{Nadler}, E.~O. and {Drlica-Wagner}, A. and {Bechtol}, K. and {Mau}, S. and {Wechsler}, R.~H. and {Gluscevic}, V. and {Boddy}, K. and {Pace}, A.~B. and {Li}, T.~S. and {McNanna}, M. and et al.},
        title = "{Constraints on Dark Matter Properties from Observations of Milky Way Satellite Galaxies}",
      journal = {\prl},
         year = 2021,
        month = mar,
       volume = {126},
       number = {9},
          eid = {091101},
        pages = {091101},
          doi = {10.1103/PhysRevLett.126.091101},
archivePrefix = {arXiv},
       eprint = {2008.00022},
 primaryClass = {astro-ph.CO},
       adsurl = {https://ui.adsabs.harvard.edu/abs/2021PhRvL.126i1101N}
}

@ARTICLE{Zacharegkas2025,
       author = {{Zacharegkas}, G. and {Chang}, C. and {Prat}, J. and {Hartley}, W. and {Mucesh}, S. and {Alarcon}, A. and {Alves}, O. and {Amon}, A. and {Bechtol}, K. and {Becker}, M.~R. and et al.},
        title = "{Constraining the Stellar-to-Halo Mass Relation with Galaxy Clustering and Weak Lensing from DES Year 3 Data}",
      journal = {arXiv e-prints},
         year = 2025,
        month = jun,
          eid = {arXiv:2506.22367},
        pages = {arXiv:2506.22367},
          doi = {10.48550/arXiv.2506.22367},
archivePrefix = {arXiv},
       eprint = {2506.22367},
 primaryClass = {astro-ph.GA},
       adsurl = {https://ui.adsabs.harvard.edu/abs/2025arXiv250622367Z}
}

@ARTICLE{Dvornik2020,
       author = {{Dvornik}, Andrej and {Hoekstra}, Henk and {Kuijken}, Konrad and {Wright}, Angus H. and {Asgari}, Marika and {Bilicki}, Maciej and {Erben}, Thomas and {Giblin}, Benjamin and {Graham}, Alister W. and {Heymans}, Catherine and et al.},
        title = "{KiDS+GAMA: The weak lensing calibrated stellar-to-halo mass relation of central and satellite galaxies}",
      journal = {\aap},
         year = 2020,
        month = oct,
       volume = {642},
          eid = {A83},
        pages = {A83},
          doi = {10.1051/0004-6361/202038693},
archivePrefix = {arXiv},
       eprint = {2006.10777},
 primaryClass = {astro-ph.CO},
       adsurl = {https://ui.adsabs.harvard.edu/abs/2020A&A...642A..83D}
}

@ARTICLE{Coupon2015,
       author = {{Coupon}, J. and {Arnouts}, S. and {van Waerbeke}, L. and {Moutard}, T. and {Ilbert}, O. and {van Uitert}, E. and {Erben}, T. and {Garilli}, B. and {Guzzo}, L. and {Heymans}, C. and et al.},
        title = "{The galaxy-halo connection from a joint lensing, clustering and abundance analysis in the CFHTLenS/VIPERS field}",
      journal = {\mnras},
         year = 2015,
        month = may,
       volume = {449},
       number = {2},
        pages = {1352-1379},
          doi = {10.1093/mnras/stv276},
archivePrefix = {arXiv},
       eprint = {1502.02867},
 primaryClass = {astro-ph.CO},
       adsurl = {https://ui.adsabs.harvard.edu/abs/2015MNRAS.449.1352C}
}

@ARTICLE{Rodriguez2017,
       author = {{Rodr{\'\i}guez-Puebla}, Aldo and {Primack}, Joel R. and {Avila-Reese}, Vladimir and {Faber}, S.~M.},
        title = "{Constraining the galaxy-halo connection over the last 13.3 Gyr: star formation histories, galaxy mergers and structural properties}",
      journal = {\mnras},
         year = 2017,
        month = sep,
       volume = {470},
       number = {1},
        pages = {651-687},
          doi = {10.1093/mnras/stx1172},
archivePrefix = {arXiv},
       eprint = {1703.04542},
 primaryClass = {astro-ph.GA},
       adsurl = {https://ui.adsabs.harvard.edu/abs/2017MNRAS.470..651R}
}

@ARTICLE{Wechsler2018,
       author = {{Wechsler}, Risa H. and {Tinker}, Jeremy L.},
        title = "{The Connection Between Galaxies and Their Dark Matter Halos}",
      journal = {\araa},
         year = 2018,
        month = sep,
       volume = {56},
        pages = {435-487},
          doi = {10.1146/annurev-astro-081817-051756},
archivePrefix = {arXiv},
       eprint = {1804.03097},
 primaryClass = {astro-ph.GA},
       adsurl = {https://ui.adsabs.harvard.edu/abs/2018ARA&A..56..435W}
}

@ARTICLE{Kravtsov2022,
       author = {{Kravtsov}, Andrey and {Manwadkar}, Viraj},
        title = "{GRUMPY: a simple framework for realistic forward modelling of dwarf galaxies}",
      journal = {\mnras},
         year = 2022,
        month = aug,
       volume = {514},
       number = {2},
        pages = {2667-2691},
          doi = {10.1093/mnras/stac1439},
archivePrefix = {arXiv},
       eprint = {2106.09724},
 primaryClass = {astro-ph.GA},
       adsurl = {https://ui.adsabs.harvard.edu/abs/2022MNRAS.514.2667K}
}

@ARTICLE{Wheeler2017,
       author = {{Wheeler}, Coral and {Pace}, Andrew B. and {Bullock}, James S. and {Boylan-Kolchin}, Michael and {O{\~n}orbe}, Jose and {Elbert}, Oliver D. and {Fitts}, Alex and {Hopkins}, Philip F. and {Kere{\v{s}}}, Du{\v{s}}an},
        title = "{The no-spin zone: rotation versus dispersion support in observed and simulated dwarf galaxies}",
      journal = {\mnras},
         year = 2017,
        month = feb,
       volume = {465},
       number = {2},
        pages = {2420-2431},
          doi = {10.1093/mnras/stw2583},
archivePrefix = {arXiv},
       eprint = {1511.01095},
 primaryClass = {astro-ph.GA},
       adsurl = {https://ui.adsabs.harvard.edu/abs/2017MNRAS.465.2420W}
}

@ARTICLE{Koposov2009,
       author = {{Koposov}, Sergey E. and {Yoo}, Jaiyul and {Rix}, Hans-Walter and {Weinberg}, David H. and {Macci{\`o}}, Andrea V. and {Escud{\'e}}, Jordi Miralda},
        title = "{A Quantitative Explanation of the Observed Population of Milky Way Satellite Galaxies}",
      journal = {\apj},
         year = 2009,
        month = may,
       volume = {696},
       number = {2},
        pages = {2179-2194},
          doi = {10.1088/0004-637X/696/2/2179},
archivePrefix = {arXiv},
       eprint = {0901.2116},
 primaryClass = {astro-ph.GA},
       adsurl = {https://ui.adsabs.harvard.edu/abs/2009ApJ...696.2179K}
}

@ARTICLE{York2000,
       author = {{York}, Donald G. and {Adelman}, J. and {Anderson}, Jr., John E. and {Anderson}, Scott F. and {Annis}, James and {Bahcall}, Neta A. and {Bakken}, J.~A. and {Barkhouser}, Robert and {Bastian}, Steven and {Berman}, Eileen and et al.},
        title = "{The Sloan Digital Sky Survey: Technical Summary}",
      journal = {\aj},
         year = 2000,
        month = sep,
       volume = {120},
       number = {3},
        pages = {1579-1587},
          doi = {10.1086/301513},
archivePrefix = {arXiv},
       eprint = {astro-ph/0006396},
 primaryClass = {astro-ph},
       adsurl = {https://ui.adsabs.harvard.edu/abs/2000AJ....120.1579Y}
}

@ARTICLE{Golden-Marx2023,
       author = {{Golden-Marx}, Jesse B. and {Zu}, Ying and {Wang}, Jiaqi and {Li}, Hekun and {Zhang}, Jun and {Yang}, Xiaohu},
        title = "{Satellite content and halo mass of galaxy clusters: comparison between red-sequence and halo-based optical cluster finders}",
      journal = {\mnras},
         year = 2023,
        month = sep,
       volume = {524},
       number = {3},
        pages = {4455-4471},
          doi = {10.1093/mnras/stad2174},
archivePrefix = {arXiv},
       eprint = {2212.13270},
 primaryClass = {astro-ph.GA},
       adsurl = {https://ui.adsabs.harvard.edu/abs/2023MNRAS.524.4455G}
}

@ARTICLE{Li2022HSC,
       author = {{Li}, Xiangchong and {Miyatake}, Hironao and {Luo}, Wentao and {More}, Surhud and {Oguri}, Masamune and {Hamana}, Takashi and {Mandelbaum}, Rachel and {Shirasaki}, Masato and {Takada}, Masahiro and {Armstrong}, Robert and et al.},
        title = "{The three-year shear catalog of the Subaru Hyper Suprime-Cam SSP Survey}",
      journal = {\pasj},
         year = 2022,
        month = apr,
       volume = {74},
       number = {2},
        pages = {421-459},
          doi = {10.1093/pasj/psac006},
archivePrefix = {arXiv},
       eprint = {2107.00136},
 primaryClass = {astro-ph.CO},
       adsurl = {https://ui.adsabs.harvard.edu/abs/2022PASJ...74..421L}
}

@ARTICLE{Chen2019,
       author = {{Chen}, Yangyao and {Mo}, H.~J. and {Li}, Cheng and {Wang}, Huiyuan and {Yang}, Xiaohu and {Zhou}, Shuang and {Zhang}, Youcai},
        title = "{ELUCID. VI. Cosmic Variance of the Galaxy Distribution in the Local Universe}",
      journal = {\apj},
         year = 2019,
        month = feb,
       volume = {872},
       number = {2},
          eid = {180},
        pages = {180},
          doi = {10.3847/1538-4357/ab0208},
archivePrefix = {arXiv},
       eprint = {1809.00523},
 primaryClass = {astro-ph.GA},
       adsurl = {https://ui.adsabs.harvard.edu/abs/2019ApJ...872..180C}
}

@ARTICLE{Moore2025,
       author = {{Moore}, Samuel G. and {Cole}, Shaun and {Wilson}, Michael and {Norberg}, Peder and {Moustakas}, John and {Aguilar}, J. and {Ahlen}, S. and {Anand}, A. and {Bianchi}, D. and {Brooks}, D. and et al.},
        title = "{DESI DR2 Galaxy Luminosity Functions}",
      journal = {arXiv e-prints},
         year = 2025,
        month = nov,
          eid = {arXiv:2511.01803},
        pages = {arXiv:2511.01803},
          doi = {10.48550/arXiv.2511.01803},
archivePrefix = {arXiv},
       eprint = {2511.01803},
 primaryClass = {astro-ph.GA},
       adsurl = {https://ui.adsabs.harvard.edu/abs/2025arXiv251101803M}
}

@ARTICLE{DES2018,
       author = {{Abbott}, T.~M.~C. and {Abdalla}, F.~B. and {Allam}, S. and {Amara}, A. and {Annis}, J. and {Asorey}, J. and {Avila}, S. and {Ballester}, O. and {Banerji}, M. and {Barkhouse}, W. and et al.},
        title = "{The Dark Energy Survey: Data Release 1}",
      journal = {\apjs},
         year = 2018,
        month = dec,
       volume = {239},
       number = {2},
          eid = {18},
        pages = {18},
          doi = {10.3847/1538-4365/aae9f0},
archivePrefix = {arXiv},
       eprint = {1801.03181},
 primaryClass = {astro-ph.IM},
       adsurl = {https://ui.adsabs.harvard.edu/abs/2018ApJS..239...18A}
}

@ARTICLE{DES2021,
       author = {{Abbott}, T.~M.~C. and {Adam{\'o}w}, M. and {Aguena}, M. and {Allam}, S. and {Amon}, A. and {Annis}, J. and {Avila}, S. and {Bacon}, D. and {Banerji}, M. and {Bechtol}, K. and et al.},
        title = "{The Dark Energy Survey Data Release 2}",
      journal = {\apjs},
         year = 2021,
        month = aug,
       volume = {255},
       number = {2},
          eid = {20},
        pages = {20},
          doi = {10.3847/1538-4365/ac00b3},
archivePrefix = {arXiv},
       eprint = {2101.05765},
 primaryClass = {astro-ph.IM},
       adsurl = {https://ui.adsabs.harvard.edu/abs/2021ApJS..255...20A}
}

@ARTICLE{Tinker2021,
       author = {{Tinker}, Jeremy L.},
        title = "{A Self-Calibrating Halo-Based Group Finder: Application to SDSS}",
      journal = {\apj},
         year = 2021,
        month = dec,
       volume = {923},
       number = {2},
          eid = {154},
        pages = {154},
          doi = {10.3847/1538-4357/ac2aaa},
archivePrefix = {arXiv},
       eprint = {2010.02946},
 primaryClass = {astro-ph.CO},
       adsurl = {https://ui.adsabs.harvard.edu/abs/2021ApJ...923..154T}
}

@ARTICLE{Geha2017,
       author = {{Geha}, Marla and {Wechsler}, Risa H. and {Mao}, Yao-Yuan and {Tollerud}, Erik J. and {Weiner}, Benjamin and {Bernstein}, Rebecca and {Hoyle}, Ben and {Marchi}, Sebastian and {Marshall}, Phil J. and {Mu{\~n}oz}, Ricardo and {Lu}, Yu},
        title = "{The SAGA Survey. I. Satellite Galaxy Populations around Eight Milky Way Analogs}",
      journal = {\apj},
         year = 2017,
        month = sep,
       volume = {847},
       number = {1},
          eid = {4},
        pages = {4},
          doi = {10.3847/1538-4357/aa8626},
archivePrefix = {arXiv},
       eprint = {1705.06743},
 primaryClass = {astro-ph.GA},
       adsurl = {https://ui.adsabs.harvard.edu/abs/2017ApJ...847....4G}
}

@ARTICLE{Guo2017,
       author = {{Guo}, Hong and {Li}, Cheng and {Zheng}, Zheng and {Mo}, H.~J. and {Jing}, Y.~P. and {Zu}, Ying and {Lim}, S.~H. and {Xu}, Haojie},
        title = "{Constraining the H I-Halo Mass Relation from Galaxy Clustering}",
      journal = {\apj},
         year = 2017,
        month = sep,
       volume = {846},
       number = {1},
          eid = {61},
        pages = {61},
          doi = {10.3847/1538-4357/aa85e7},
archivePrefix = {arXiv},
       eprint = {1707.01999},
 primaryClass = {astro-ph.GA},
       adsurl = {https://ui.adsabs.harvard.edu/abs/2017ApJ...846...61G}
}

@ARTICLE{Zu2021,
       author = {{Zu}, Ying and {Shan}, Huanyuan and {Zhang}, Jun and {Singh}, Sukhdeep and {Shao}, Zhiwei and {Chen}, Xiaokai and {Yao}, Ji and {Golden-Marx}, Jesse B. and {Cui}, Weiguang and {Jullo}, Eric and et al.},
        title = "{Does concentration drive the scatter in the stellar-to-halo mass relation of galaxy clusters?}",
      journal = {\mnras},
         year = 2021,
        month = aug,
       volume = {505},
       number = {4},
        pages = {5117-5128},
          doi = {10.1093/mnras/stab1712},
archivePrefix = {arXiv},
       eprint = {2012.08629},
 primaryClass = {astro-ph.CO},
       adsurl = {https://ui.adsabs.harvard.edu/abs/2021MNRAS.505.5117Z}
}

@ARTICLE{Kravtsov2004,
       author = {{Kravtsov}, Andrey V. and {Berlind}, Andreas A. and {Wechsler}, Risa H. and {Klypin}, Anatoly A. and {Gottl{\"o}ber}, Stefan and {Allgood}, Brandon and {Primack}, Joel R.},
        title = "{The Dark Side of the Halo Occupation Distribution}",
      journal = {\apj},
         year = 2004,
        month = jul,
       volume = {609},
       number = {1},
        pages = {35-49},
          doi = {10.1086/420959},
archivePrefix = {arXiv},
       eprint = {astro-ph/0308519},
 primaryClass = {astro-ph},
       adsurl = {https://ui.adsabs.harvard.edu/abs/2004ApJ...609...35K}
}

@ARTICLE{SpecS5,
       author = {{Besuner}, Robert and {Dey}, Arjun and {Drlica-Wagner}, Alex and {Ebina}, Haruki and {Fernandez Moroni}, Guillermo and {Ferraro}, Simone and {Forero-Romero}, Jaime and {Honscheid}, Klaus and {Jelinsky}, Pat and {Lang}, Dustin and et al.},
        title = "{The Spectroscopic Stage-5 Experiment}",
      journal = {arXiv e-prints},
         year = 2025,
        month = mar,
          eid = {arXiv:2503.07923},
        pages = {arXiv:2503.07923},
          doi = {10.48550/arXiv.2503.07923},
archivePrefix = {arXiv},
       eprint = {2503.07923},
 primaryClass = {astro-ph.CO},
       adsurl = {https://ui.adsabs.harvard.edu/abs/2025arXiv250307923B}
}

@ARTICLE{DESI2,
       author = {{Schlegel}, David J. and {Ferraro}, Simone and {Aldering}, Greg and {Baltay}, Charles and {BenZvi}, Segev and {Besuner}, Robert and {Blanc}, Guillermo A. and {Bolton}, Adam S. and {Bonaca}, Ana and {Brooks}, David and et al.},
        title = "{A Spectroscopic Road Map for Cosmic Frontier: DESI, DESI-II, Stage-5}",
      journal = {arXiv e-prints},
         year = 2022,
        month = sep,
          eid = {arXiv:2209.03585},
        pages = {arXiv:2209.03585},
          doi = {10.48550/arXiv.2209.03585},
archivePrefix = {arXiv},
       eprint = {2209.03585},
 primaryClass = {astro-ph.CO},
       adsurl = {https://ui.adsabs.harvard.edu/abs/2022arXiv220903585S}
}

@ARTICLE{Euclid2025,
       author = {{Euclid Collaboration} and {Mellier}, Y. and {Abdurro'uf} and {Acevedo Barroso}, J.~A. and {Ach{\'u}carro}, A. and {Adamek}, J. and {Adam}, R. and {Addison}, G.~E. and {Aghanim}, N. and {Aguena}, M. and et al.},
        title = "{Euclid: I. Overview of the Euclid mission}",
      journal = {\aap},
         year = 2025,
        month = may,
       volume = {697},
          eid = {A1},
        pages = {A1},
          doi = {10.1051/0004-6361/202450810},
archivePrefix = {arXiv},
       eprint = {2405.13491},
 primaryClass = {astro-ph.CO},
       adsurl = {https://ui.adsabs.harvard.edu/abs/2025A&A...697A...1E}
}

@ARTICLE{CSST2025,
       author = {{CSST Collaboration} and {Gong}, Yan and {Miao}, Haitao and {Zhan}, Hu and {Li}, Zhao-Yu and {Shangguan}, Jinyi and {Li}, Haining and {Liu}, Chao and {Chen}, Xuefei and {Yuan}, Haibo and et al.},
        title = "{Introduction to the Chinese Space Station Survey Telescope (CSST)}",
      journal = {arXiv e-prints},
         year = 2025,
        month = jul,
          eid = {arXiv:2507.04618},
        pages = {arXiv:2507.04618},
          doi = {10.48550/arXiv.2507.04618},
archivePrefix = {arXiv},
       eprint = {2507.04618},
 primaryClass = {astro-ph.IM},
       adsurl = {https://ui.adsabs.harvard.edu/abs/2025arXiv250704618C}
}

@ARTICLE{LSST2019,
       author = {{Ivezi{\'c}}, {\v{Z}}eljko and {Kahn}, Steven M. and {Tyson}, J. Anthony and {Abel}, Bob and {Acosta}, Emily and {Allsman}, Robyn and {Alonso}, David and {AlSayyad}, Yusra and {Anderson}, Scott F. and {Andrew}, John and et al.},
        title = "{LSST: From Science Drivers to Reference Design and Anticipated Data Products}",
      journal = {\apj},
         year = 2019,
        month = mar,
       volume = {873},
       number = {2},
          eid = {111},
        pages = {111},
          doi = {10.3847/1538-4357/ab042c},
archivePrefix = {arXiv},
       eprint = {0805.2366},
 primaryClass = {astro-ph},
       adsurl = {https://ui.adsabs.harvard.edu/abs/2019ApJ...873..111I}
}

@ARTICLE{Roman2019,
       author = {{Akeson}, Rachel and {Armus}, Lee and {Bachelet}, Etienne and {Bailey}, Vanessa and {Bartusek}, Lisa and {Bellini}, Andrea and {Benford}, Dominic and {Bennett}, David and {Bhattacharya}, Aparna and {Bohlin}, Ralph and et al.},
        title = "{The Wide Field Infrared Survey Telescope: 100 Hubbles for the 2020s}",
      journal = {arXiv e-prints},
         year = 2019,
        month = feb,
          eid = {arXiv:1902.05569},
        pages = {arXiv:1902.05569},
          doi = {10.48550/arXiv.1902.05569},
archivePrefix = {arXiv},
       eprint = {1902.05569},
 primaryClass = {astro-ph.IM},
       adsurl = {https://ui.adsabs.harvard.edu/abs/2019arXiv190205569A}
}

@ARTICLE{Xu2026,
       author = {{Xu}, Kun and {Frenk}, Carlos S. and {Jing}, Y.~P. and {Cole}, Shaun and {Bose}, Sownak and {Aguilar}, J. and {Ahlen}, S. and {Bianchi}, D. and {Brooks}, D. and {Castander}, F.~J. and et al.},
        title = "{PAC in DESI. II. Galaxy-halo connection into the $10^{6}{\rm M}_{\odot}$ frontier}",
      journal = {arXiv e-prints},
         year = 2026,
        month = mar,
          eid = {arXiv:2603.29331},
        pages = {arXiv:2603.29331},
          doi = {10.48550/arXiv.2603.29331},
archivePrefix = {arXiv},
       eprint = {2603.29331},
 primaryClass = {astro-ph.GA},
       adsurl = {https://ui.adsabs.harvard.edu/abs/2026arXiv260329331X}
}

@ARTICLE{Planck18,
       author = {{Planck Collaboration} and {Aghanim}, N. and {Akrami}, Y. and {Ashdown}, M. and {Aumont}, J. and {Baccigalupi}, C. and {Ballardini}, M. and {Banday}, A.~J. and {Barreiro}, R.~B. and {Bartolo}, N. and et al.},
        title = "{Planck 2018 results. VI. Cosmological parameters}",
      journal = {\aap},
         year = 2020,
        month = sep,
       volume = {641},
          eid = {A6},
        pages = {A6},
          doi = {10.1051/0004-6361/201833910},
archivePrefix = {arXiv},
       eprint = {1807.06209},
 primaryClass = {astro-ph.CO},
       adsurl = {https://ui.adsabs.harvard.edu/abs/2020A&A...641A...6P}
}

@unpublished{Manwadkar2026,
  author       = {Manwadkar, Viraj and Wechsler, R.~H. and others},
  title        = {When Galaxies Fall Apart: Addressing Photometric Shredding with Color-Based Aperture
2 Photometry in DESI DR1},
  note         = {in preparation},
  year         = {in prep}
}
\bibliographystyle{aasjournalv7}



\end{document}